\documentclass{memoir}
\usepackage{pdfpages}
\usepackage{hyperref}
\usepackage{amsmath} 
\usepackage{palatino}
\usepackage{listings}
\usepackage[font={footnotesize,sf},labelfont=bf]{caption}
\usepackage{color}
\usepackage{xcolor}
\usepackage{latexsym}
\definecolor{light-gray}{gray}{0.95}
\colorlet{shadecolor}{light-gray} 
\usepackage{framed}
\usepackage{booktabs}

\usepackage{longtable}
\usepackage{array}
\usepackage{wrapfig}
\usepackage{calc}
\usepackage{upgreek}

\definecolor{navyblue}{rgb}{0,0,0.55}     
\definecolor{darkgreen}{rgb}{0,0.3,0.3}
\definecolor{verylightgray}{gray}{0.95}   

\hypersetup{colorlinks=true, linkcolor=navyblue, urlcolor=navyblue, pdfnewwindow=true, citecolor=darkgreen}

\setsecnumdepth{subsection} 
\abnormalparskip{0.75em}

\newcommand*{\mytitlepage}{
  \begingroup
  \vfill
  \hbox{
    \hspace*{0.05\textwidth}
    \rule{2pt}{0.95\textheight}
    \hspace*{0.05\textwidth}
    \parbox[t]{0.9\textwidth}{
      \vbox{
          {\noindent\HUGE\bfseries Dissecting the \\[0.33\baselineskip]
          Graphcore IPU \\[0.4\baselineskip]
          Architecture \\[0.5\baselineskip]
          via Microbenchmarking}\\[4\baselineskip]

          {\Large\itshape Technical Report \\[1.6\baselineskip]
          December 7, 2019
          }\\[3\baselineskip]

          {\Large Zhe Jia\\[0.35\baselineskip]
          Blake Tillman\\[0.35\baselineskip]
          Marco Maggioni\\[0.35\baselineskip]
          Daniele P. Scarpazza
          }\par

          \vspace{0.08\textheight}
          \includegraphics[trim=12.5cm 13.65cm 13.5cm 13.25cm, clip=true, totalheight=0.9cm]{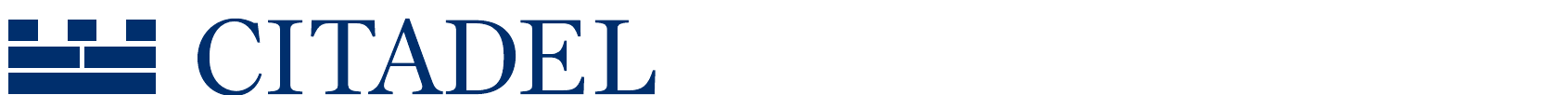}
          \\[0.4em]
          {High Performance Computing R\&D Team\\[0.1em]
          Citadel, 131~S.~Dearborn St., Chicago.
          }\par
          \vspace{0.02\textheight}
      }
    }
  }
  \vfill
  \null
  \endgroup}

\begin{document}
\clearpage\mytitlepage
\thispagestyle{empty}
\sloppy

%
%


  \noindent
  Copyright \copyright \ 2019, Citadel Enterprise Americas, LLC. All rights
  reserved.

  \noindent
  Chicago, United States of America.

  \noindent
  \vspace{1ex}
  \noindent
  \makebox[\textwidth]{\centering
    \begin{tabular}{ll}
      \toprule
      Edition  & Date \\
      \midrule
      First    &  December 7, 2019 \\
      \bottomrule
    \end{tabular}
  }
  \vspace{1ex}

  \noindent
  The authors make this document available on the
  \href{http://arXiv.org}{arXiv.org} e-print service owned and
  operated by Cornell University.  Citadel grants arXiv.org a
  perpetual, non-exclusive license to distribute this manuscript.

  \noindent
  \textbf{Disclaimer.} This presentation solely reflects the analysis
  and views of the authors.  No recipient should interpret this
  presentation to represent the general views of Citadel or its
  personnel.  Facts, analyses, and views presented herein have not
  been reviewed by, and may not reflect information known to other
  Citadel professionals.

  \noindent
  \textbf{Objectivity.} This research aims at being objective and
  impartial. The authors strive to ensure that each claim is based on
  solid evidence. Results published here are meant to be reproducible
  by anybody who recreates the experimental conditions described by
  the authors, with the possible exception of good faith mistakes. No
  such mistakes are known to the authors at the time of publication.

  \noindent
  \textbf{Conflicts of interest.} This research was supported
  entirely by Citadel. Citadel received no financial support from
  Graphcore or any other entity in connection with this work.
  Graphcore helped this work by providing an early-availability test
  system to Citadel, and by reviewing the contents of this manuscript
  for mistakes and omissions.  The authors did not receive any
  consulting fees, research funding or reimbursement in conjunction
  with this work, and are not employed by Graphcore.

  \noindent
  \textbf{Acknowledgments.} The authors thank their colleague Jeffrey
  Smith for his valuable input and assistance during the preparation
  of this manuscript.

  \noindent
  All product names, trademarks and registered trademarks are property
  of their respective owners.

\thispagestyle{empty}

\setcounter{page}{2}
\newpage
\setcounter{tocdepth}{4}
\tableofcontents

%
\chapter*{Abstract}

This report focuses on the architecture and performance of the
Intelligence Processing Unit (IPU), a novel, massively parallel
platform recently introduced by Graphcore and aimed at Artificial
Intelligence/Machine Learning (AI/ML) workloads.

We dissect the IPU's performance behavior using microbenchmarks that
we crafted for the purpose. We study the IPU's memory organization and
performance. We study the latency and bandwidth that the on-chip and
off-chip interconnects offer, both in point-to-point transfers and in a
spectrum of collective operations, under diverse loads.  We evaluate
the IPU's compute power over matrix multiplication, convolution, and AI/ML
primitives. We discuss actual performance in comparison with its
theoretical limits.

Our findings reveal how the IPU's architectural design affects its
performance. Moreover, they offer simple mental models to predict an
application's performance on the IPU, on the basis of the computation
and communication steps it involves.

This report is the natural extension to a novel architecture of a
continuing effort of ours~\cite{zhe2018,zhe2019} that focuses on the
microbenchmark-based discovery of massively parallel architectures.

%

\chapter{Architecture}
\label{chap:arch}

In this chapter, we introduce the reader to the fundamentals of the
IPU's architecture and its programming paradigm.

\section{Design Philosophy}
\label{sec:philosophy}

The IPU architecture and its compute paradigm were co-designed from
the ground up specifically to tackle machine intelligence workloads.
For that reason, they incarnate certain design choices that depart
radically from more common architectures like CPUs and GPUs, and might
be less familiar to the reader. This section and the following ones
discuss these choices and how they affect applications and application
designers.

\begin{figure}
  \includegraphics[width=1.05\columnwidth, trim=0cm 0cm 1cm 0cm]{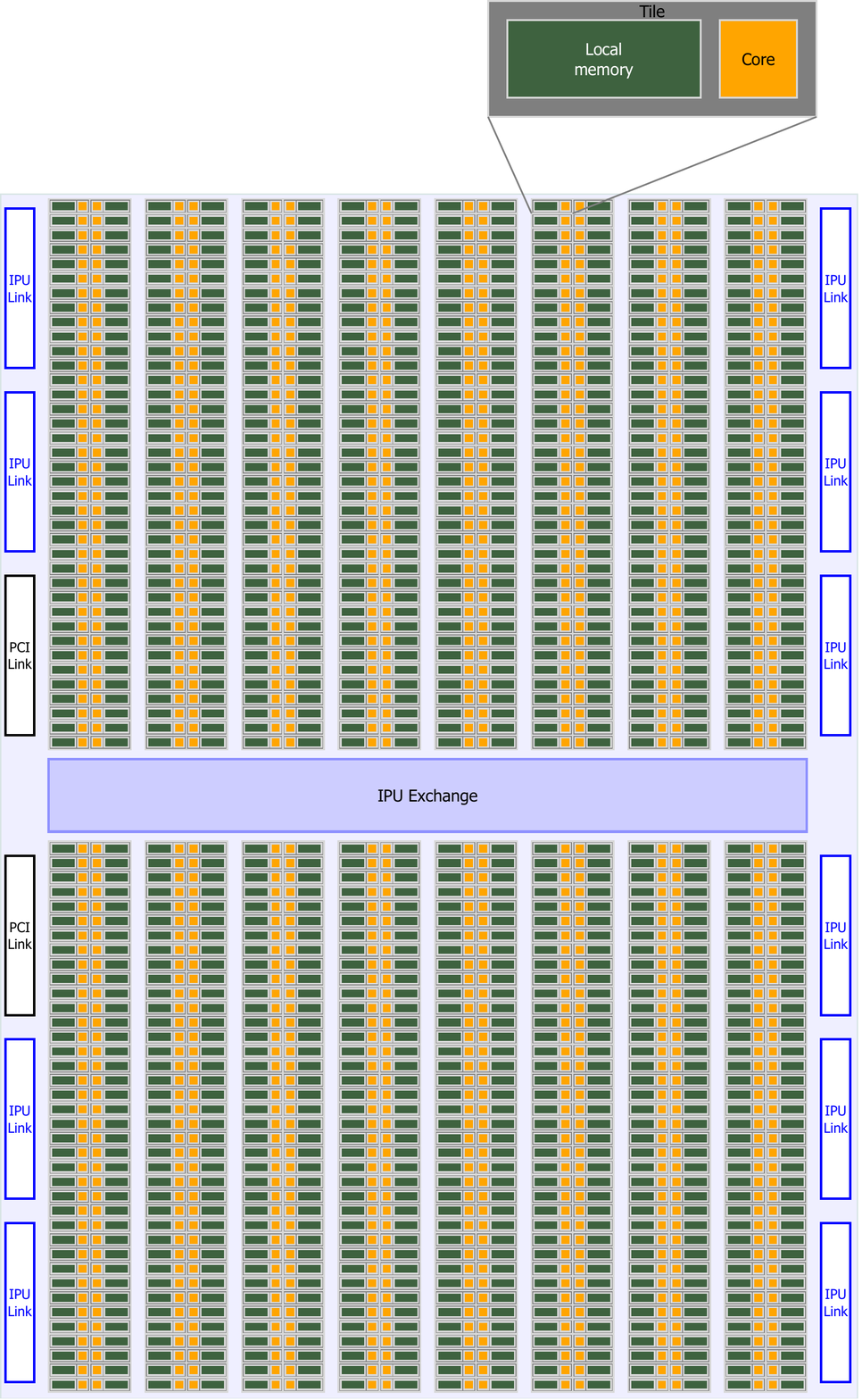}
  \caption{Simplified block diagram of an IPU processor: the processor
    features 1,216 tiles (each containing one core and its local
    memory), the exchange (an on-chip interconnect), IPU link
    interfaces that provide connectivity to other IPU chips, and PCIe
    interfaces for host connectivity.}
  \label{fig:basic-IPU-topology}
\end{figure}

At the heart of any IPU-based system is the IPU processor, of which we
offer a simplified block diagram in
Figure~\ref{fig:basic-IPU-topology}. Its design aim is the
\textbf{efficient execution of fine-grained operations across a
  relatively large number of parallel threads}.  This means that the
IPU, unlike other massively parallel architectures (e.g., the GPU)
adapts well to fine-grained, irregular computation that exhibits
irregular data accesses.  The IPU offers true MIMD (Multiple
Instruction, Multiple Data) parallelism and has distributed, local
memory as its only form of memory on the device.

Each IPU contains 1,216 processing elements called \emph{tiles}; a
tile consists of one computing core plus 256 KiB of local
memory. Except for the register file, the IPU offers no memory other
than the distributed memories local to each tile.

In addition to the tiles, the IPU processor contains the
\emph{exchange}, an on-chip interconnect that allows for
high-bandwidth, low-latency communication among tiles.

Each IPU also contains ten \emph{IPU link} interfaces; the IPU link is
a Graphcore-proprietary interconnect that enables low latency,
high-throughput communication between IPU processors. Because IPU
links make transfers between remote tiles as simple to the programmer
as between local tiles, they are the linchpin to the IPU paradigm's
scalability.  Finally, the IPU contains two PCIe links for
communication with CPU-based hosts.

\begin{shaded}
\small
\noindent\textbf{Terminology.} While Graphcore's commercial literature uses the
terms IPU-Tiles\texttrademark, IPU-Core\texttrademark,
IPU-Exchange\texttrademark and IPU-Links\texttrademark, we refer to
the same components as tiles, core, exchange and IPU links,
respectively, with no risk of confusion.

\noindent\textbf{Source.} Information we report here on IPU
architecture derives from Graphcore's technical literature or from
direct correspondence with Graphcore, and is republished with
permission.
\end{shaded}

\section{Fine-grained Parallelism}
\label{sec:fine-grained-parallelism}

The IPU's emphasis on \emph{fine-grained} parallelism means that the
IPU can efficiently run applications that have irregular and sparse
data access patterns and control flow. Unlike SIMD/SIMT architectures,
IPUs don't need large warps of threads consuming contiguous vector
data to achieve high efficiency. Instead, IPUs can run individual
processing threads on smaller data blocks, in a highly parallel MIMD
fashion. Each thread can have completely distinct code and execution
flow without incurring performance penalties.

Architecturally, IPUs differ significantly from platforms commonly
used to execute ML/AI workloads, namely CPUs and GPUs. We discuss
those differences by comparing the fundamental approaches behind the
design of CPUs and GPUs with those behind IPUs.

\noindent\textbf{CPUs}

CPUs tend to offer complex cores in relatively small counts.  CPU
cores feature sophisticated latency-reducing techniques like branch
prediction, branch speculation, and out-of-order execution. These
optimizations make CPUs excel at single-thread performance and
control-dominated code, possibly at the expense of energy efficiency
and aggregate arithmetic throughput per area of silicon.

Even if most modern CPUs offer vectorization (SIMD, Single
Instruction, Multiple Data), they can't match GPUs in aggregate
floating-point arithmetic or in energy efficiency (performance per
Watt) on large, regular, array-based workloads.

To hide memory latency, CPUs typically employ a deep memory hierarchy
containing multiple levels of caches, together with prefetching and
sophisticated prefetch predictors.

\noindent\textbf{GPUs}

GPUs, in contrast, feature smaller cores in a significantly higher
count per device (e.g., thousands).  GPU cores are architecturally
simpler than those of CPUs and do not typically offer branch
speculation, sophisticated branch prediction, out-of-order execution,
or hardware prefetching.

GPUs arrange their cores into clusters that operate in lockstep; all
cores in a cluster execute the same instruction at any point in
time. Threads are grouped together in warps; warps are scheduled
together to clusters. Threads in a warp will perform the same
operation on independent data.  This execution model is referred to as
SIMT: Single Instruction, Multiple Threads.

Even though GPUs have evolved deep memory hierarchies featuring both
cache and scratchpad memories~\cite{zhe2018,zhe2019}, their
fundamental approach to hiding memory latency remains the
oversubscription of threads to cores and the ability to inexpensively
switch among threads. In this approach, when a warp of threads is
awaiting operands from main memory, the hardware can suspend them and
switch to another warp that has received its operands from memory and
is ready to continue.  The programming model explicitly encourages
developers to expose sufficient thread parallelism so that a portion
of the threads are always ready to execute, while the remainder await
operands.

GPUs only access main memory at peak throughput when load/store
operations within each warp involve contiguous regions of memory; that
access is said to be \emph{coalesced}.  Uncoalesced memory accesses
only achieve a fraction of the theoretical peak memory bandwidth.  For
best performance, developers must instantiate large blocks of threads
that share the same program, attempt to consume input data (and produce
output) in a coordinated, coalesced manner, and use control flow
sparingly. Because of the SIMT model, GPUs pay a performance penalty
when threads diverge in their control flow.

Because of their organization, GPUs excel at regular, dense,
numerical, data-flow-dominated workloads that naturally lead to
coalesced accesses and a coherent control flow. On these workloads,
GPUs also tend to be more energy efficient than CPUs because they
dedicate a higher fraction of their silicon to arithmetic units,
rather than caches and latency-oriented features. Moreover, GPUs
refactor instruction decoding circuitry outside of individual cores;
this is possible because clusters of cores operate on the same
instruction.

\noindent\textbf{IPUs}

IPUs provide large core counts (1,216 per processor) and offer cores
complex enough to be capable of executing completely distinct
programs.  The IPU's approach to reducing memory latency is
radical---it does away with shared memory entirely. The IPU only offers
small, distributed memories that are local and tightly coupled to each
core. Each tile contains 256 KiB of memory, totaling 304 MiB per
device.

IPU memories are scratchpads, not caches.
They are implemented as SRAM and therefore offer much higher
bandwidth (45 TB/s, aggregate) and lower latency (6 clock cycles)
than DRAMs. Their performance is comparable to L2 CPU caches and
superior to GPU shared memories (or L1 caches).
Section~\ref{sec:memory-arch} extends our discussion on memory.

IPU cores pay no penalty when their control flows diverge or when the
addresses of their memory accesses diverge. In fact, they pay no
penalty for running disjoint instruction flows that exhibit
uncorrelated memory accesses.  Cores access data from their respective
local memory at a fixed cost that is independent of access patterns.
This makes IPUs more efficient than GPUs at executing applications
with irregular or random data access patterns and/or applications
that are control-flow dominated, provided that working sets fit in
IPU memory.

Similarly to CPUs and GPUs, IPUs achieve higher efficiency by
oversubscribing threads to cores. Specifically, each IPU tile offers
hardware support for 6 threads in a manner that is functionally
similar to the SMT technique (Simultaneous MultiThreading, or Intel's
Hyper-Threading) commonly found on CPUs.  Each IPU tile maintains 6
resident execution contexts and multiplexes them onto shared
resources, thus hiding instruction latencies (dependency, memory
access and branch latencies), reducing corresponding pipeline stalls,
and increasing aggregate throughput. Each tile rotates among threads
according to a static, round-robin schedule. The entire IPU supports
therefore 6 $\times$ 1,216 $=$ 7,296 threads. For maximum occupancy,
software designers are encouraged to instantiate that many threads.

On the IPU, efficiency on irregular workloads does not come at the
expenses of regular, numerical, array- or matrix-based workloads,
which are known to run well on GPUs. In the next section we show that
IPUs outperform GPUs on a per-board comparison when operands fit in
memory.

\section{Arithmetic Throughput}

The IPU offers an impressive arithmetic throughput, up to 31.1
TFlops/s in single precision and 124.5 TFlops/s in mixed
precision\footnote{With \emph{mixed precision} we denote operations in
  which multiplicands are in half precision (FP16) and their products
  are accumulated onto a single precision result (FP32).} per chip,
surpassing contemporary GPUs in a comparison of theoretical limits.

In a per-board comparison, the IPU's theoretical advantage over the
GPU grows roughly by a factor of two, and so it does in an energy
efficiency comparison.

This level of throughput is made possible by the use of specialized
pipelines called \emph{Accumulating Matrix Product} (AMP) units that
are present in each IPU tile.  AMP unit are used to accelerate matrix
multiplication and convolution operations. An AMP unit can finish 64
mixed-precision or 16 single-precision floating point operations per
clock cycle.

However, actual arithmetic performance, both on IPUs and GPU, depends
dramatically on the properties of the specific numerical workload at
hand, and may significantly differ from theoretical limits.
Evaluating the IPU's arithmetic performance on a meaningful basket of
real-world numerical workloads is far outside of the scope of a
microbenchmarking report like this. However, we benchmark matrix
multiplication performance as offered by Poplar's linear algebra
library (poplin) and we compare them with theoretical limits, and with
respective performance numbers associated with NVidia's V100 GPU.

The actual performance we measured shows the IPU as a clear winner in
single precision against NVidia's V100 GPU (per-chip comparison).  In
mixed precision, the comparison does not yield a clear winner and
requires a more nuanced discussion. We postpone this discussion,
together with experimental setup details and quantitative results to
Section~\ref{sec:matmul}, which is entirely dedicated to the matrix
multiplication workload.

\section{Memory Architecture}
\label{sec:memory-arch}

In the IPU's fine-grained processing philosophy, the role of local
memories is fundamental.  In fact, the very choice to adopt
distributed SRAM memories located next to the cores is what allows
threads to access data efficiently even when access patterns are
irregular, sparse and incoherent at a fine grain.

Each tile contains 256 KiB, totaling 304 MiB on the entire
processor. Each tile owns an independent, contiguous 21-bit address
space that is shared among the 6 hardware execution contexts, where
code executed locally and data processed locally must fit.  The
nominal aggregate bandwidth for the entire IPU memory is 45 TB/s,
while the latency is 6 clock cycles.

While the IPU's aggregate capacity is lower than the typical DRAM
memory on a GPU (e.g., 32 GiB), IPU memories make up in speed what
they lack in capacity; they have shorter latency than both L1 caches
and shared memories on the NVidia Turing T4 GPUs, and comparable
latency with L2 caches on Intel Skylake/Kaby Lake/Coffee Lake CPUs.
See Table~\ref{tab:memory-latency-comparison} for a quantitative
comparison between IPU memory and SRAM memories with comparable
latency on GPUs and CPUs.

\begin{table}[h]
  \center
  \footnotesize
  \begin{tabular}{llcccc}
    \toprule
    Architecture     & Memory     & Per-chip        & \multicolumn{1}{c}{Latency} & Latency  & Clock \\
    \,               &            & Capacity        & \multicolumn{1}{c}{(ns)}    & (cycles) & Frequency \\
    \,               &            &   (MiB)          &                             &          & (GHz) \\
    \midrule
    Graphcore IPU    & Tile-local &         304     &     3.75   &       6 & 1.60  \\
    \midrule
    NVidia T4 GPU    & Shared     &  1.25 ... 2.5   &  11.94  &      19 & 1.59  \\
                     & L1         &  1.25 ... 2.5   &  20.13  &      32 &       \\
    \midrule
    Intel *-Lake CPU & L1D        &  0.25 ... 0.875 & 0.93 ... 1.92    & 4 ... 5 & 2.60 ... 4.30 \\
                     & L2         &     4 ... 28    & 2.79 ... 4.62    &      12 &           \\
    \bottomrule
  \end{tabular}
  \caption{Size and latency comparison between IPU memories and
    similar SRAM-based memory hierarchy levels on contemporary GPUs
    and CPUs.  The IPU's local memory has lower latency than the
    fastest memories on the Turing GPUs and is on par with the L2 cache
    in modern Intel CPUs. However, the IPU's local memory is vastly
    larger than those memories in a per-chip comparison. (Intel data:
    from public sources; intervals range over the product offering at
    the time of the writing.  GPU data: from our prior
    work~\cite{zhe2019}.)}
  \label{tab:memory-latency-comparison}
\end{table}

As far as capacity is concerned, the aggregate IPU memory is larger
than memory layers of equivalent latency on CPUs and GPUs,
surpassing them by one to two orders of magnitude (see column
\emph{Per-chip Capacity} in
Table~\ref{tab:memory-latency-comparison}). The aggregate size of the
IPU's memory removes the need for a cache hierarchy similar to those
found on GPUs and CPUs.

We dedicate the entirety of Chapter~\ref{chap:local-memory} to
studying the performance of the IPU's local memory, which we measure
via microbenchmarks.

The IPU's memory organization requires software designers to partition
their working set across the tiles' memories appropriately and make
tiles exchange data with each other when they need non-local
operands. The programming model, along with Graphcore's Poplar
language and associated compiler, allows for the automatic
orchestration of these data transfers. Designers describe the
operands' flow but need not worry about explicit variable placement
and transfer scheduling. Transient data that is only consumed once
should be streamed into the device via PCI from the host.

Graphcore's optimized machine-learning and linear algebra libraries
adopt the data partitioning approach we described.
Performance-sensitive software designers are encouraged to follow it
as well.

The cumulative memory of 304 MiB per IPU (608 MiB per board) is
typically sufficient for many models used in contemporary AI/ML
applications to reside entirely on-chip. Models that fit entirely on
chip benefit from the high bandwidth and low latency offered by local
memory. Models of larger size can be sharded across IPU processors and
IPU boards, thanks to the architectural and programming paradigm
features described in the next section.

\section{Interconnect Architecture}

The IPU interconnect is what allows tiles on an IPU system to work
tightly together and exchange data efficiently with each other.  It is
also what truly makes a system with multiple IPUs act as a single,
coherent machine.

A system with multiple IPUs exposes the single IPU devices
independently, but it also exposes \emph{Multi-IPUs}. A Multi-IPU is a
virtual IPU device that is comprised of multiple physical IPUs and
offers all their memory and compute resources as if they belonged to a
single device.  The ability to federate multiple physical IPUs into a
virtual, monolithic device is precisely what allows users to train and
infer models larger than a single IPU's capacity, while also taking
advantage of the cumulative compute power. Software designers can
scale their applications to multiple IPUs with no additional
development effort because the same APIs can target physical IPUs
or Multi-IPUs indifferently. The tight cooperation between on-chip
exchanges and IPU links is the crucial factor that allows Multi-IPUs
to exist and put the cumulative memory capacity and compute resources
of its participants at the developer's disposal.

To study the IPU's on-chip and off-chip interconnect performance, we
adopt the terms, the methods and the interests of classic research
focusing on the characterization of parallel systems and
high-performance
networks~\cite{culler1993,alexandrov1995,panda2004,kistler2006}.

On chip, the interconnect exhibits an impressive aggregate throughput
of 7.7 TB/s (actual). We measured this throughput under load, with
all tiles transferring data concurrently to randomly selected
destinations, with a benchmark that is representative of all-to-all
exchanges (Section~\ref{sec:peak-bw}). On a per-tile basis, each of
the 1,216 tiles can simultaneously use 6.3 GB/s of bandwidth to
transfer data to an arbitrary destination on chip.  The latency of an
on-chip tile-to-tile exchange is 165 nanoseconds or lower and does
not degrade under load above that value.

In multiprocessor systems, the exchange and the IPU links work
together to support tile-to-tile communication transparently to the
user, regardless of where in the system the two endpoints are located;
it is as easy to program a Multi-IPU as it is a single, physical IPU.

Each board is connected to peer boards via IPU links and to its host
system via PCIe interfaces. The 2 IPUs on each board are connected by
three links with a nominal bidirectional bandwidth of 64 GB/s each, two
of which reserved for intra-board transfers. Our benchmarks achieved an
actual bandwidth of 108 GB/s.

Connecting boards in an IPU-link network is the key to building larger
IPU-based systems. While our study is limited to a single-host
configuration featuring 8 boards, much larger systems can be built,
including hostless systems. They are beyond the scope of this report.
Although different network topologies are possible, our experimental
evaluation focuses on the concrete configuration adopted in the test
system provided to us by Graphcore; we describe it in detail in
Chapter~\ref{chap:experimental-setup}.

Then, we dedicate the entirety of Chapter~\ref{chap:interconnect} to
the study of the IPU's interconnect performance, both on chip and
across chips.

In summary, we remark two general observations:
\begin{itemize}
\item \textbf{performance:} the aggregate arithmetic resources of the
  virtual IPU scale linearly, and the overall interconnect performance
  scales relatively well. For example, communication latencies degrade
  gracefully with system diameter, as our benchmarking results show in
  Sections~\ref{sec:congestion-free-latency} and
  \ref{sec:ipu-proximity-min-latency};
\item \textbf{programmability:} the Multi-IPU programming model is
  transparent to the developer. The underlying hardware makes the
  abstraction efficient and, in practice, no extra development effort is
  needed to scale applications onto large IPU systems.  In contrast,
  CUDA applications do require extra development effort and complexity
  to parallelize across multiple GPUs, and even more to parallelize
  across hosts. The same is true for CPU parallel applications, especially
  across hosts.
\end{itemize}

\section{The Bulk Synchronous Parallel Model}

An important design factor that underlies the IPU programming paradigm
is the Bulk Synchronous Parallel (BSP) model~\cite{bsp1990}. BSP is
the very approach that IPUs use to organize their compute and exchange
operations.

Proposed in the 1980s, the BSP model is an abstraction for parallel
computation that facilitates expressing parallel algorithms and
reasoning on the performance they achieve as they execute on parallel
computers.

The BSP model organizes computation in multiple sequential
\emph{supersteps}; a superstep is composed of a local computation
phase, followed by a communication phase and  a barrier
synchronization:

\begin{itemize}
\item in the \textbf{local computation} phase, every process performs
  computation that operates solely on local memory. No communication
  between processes occurs in this phase;
\item in the \textbf{communication} phase, processes exchange data;
  each process may send a message to each desired destination
  counterpart (all-to-all personalized exchange). No computation
  occurs in this phase;
\item a \textbf{barrier} synchronization phase follows; no process
  continues to the next superstep till all processes have reached the
  barrier. Neither computation nor communication occurs in this phase
  except for that strictly required by the barrier itself.
\end{itemize}

Processes can use the communication phase not only to send each other
intermediate computation results, but also to request and (at a later
communications stage) receive data from remote memories. This
mechanism allows each process to use any other's local memory as a
remote memory and to ultimately access the entire aggregate system
memory as one larger store.

Parallel algorithms of arbitrary complexity can be described in the
BSP model without restriction of generality.

\emph{The IPU is a true BSP machine.} It faithfully incarnates
hardware support, enforcement and optimization for the three phases of
each BSP superstep. Its programming model lets programmers specify
processes in terms of graph vertices that compute on local data (and
local data \emph{only}).

Input operands are brought to each process by the run-time system
before the computation begins, in the communication step associated
with the previous superstep.

The programming model and the hardware jointly enforce
the separation between phases:
\begin{itemize}
\item IPU cores can only access directly local memories; this
  organization naturally enforces the \emph{local} restriction of the
  computation phase;
\item the on-chip exchange provides native hardware support and
  acceleration for the all-to-all exchanges of the communication phase
  ...
\item and for the barrier synchronization as well.
\end{itemize}

In the IPU paradigm supported by the Poplar SDK, programmers describe
computation as vertices, data as tensors and data exchanges as static
edges, without having to worry about allocation of data at rest in local
memories, allocation of input/output transfer buffers, or scheduling of
transfers. All these tasks are carried out by the Poplar compiler.  The
compiler organizes the processing of machine intelligence models
exclusively using this BSP paradigm.

Because the IPU implements the BSP model using its exchange and IPU
links, the interconnect's performance is the primary factor that
determines the performance of BSP communication and synchronization
phases and ultimately affects overall application performance. We
dedicate Chapter~\ref{chap:interconnect} to study the performance of
intra- and inter-IPU transfers via benchmarks. Those benchmarks are,
in fact, compiled into and executed as BSP communication and barrier
synchronization phases.

\chapter{Experimental Setup}
\label{chap:experimental-setup}

In this chapter, we specify the experimental system setup we use in
our benchmarks. Of particular interest to the reader is the IPU link
network topology, which determines the IPUs' relative proximity to
each other and influences the performance of data transfers across
IPUs.

\section{System Configuration}

All results in this report derive from benchmarks that we ran on a
test system provided by Graphcore and equipped with 8 Graphcore C2
PCIe boards. Each C2 board hosts 2 IPU processors running at 1.6
GHz. While this test system was configured for 1.6 GHz IPU operation,
production systems may differ in configuration and performance;
contact Graphcore directly for inquiries.

\begin{figure}
  \includegraphics[width=\columnwidth, trim=0cm 3cm 0cm 0cm]{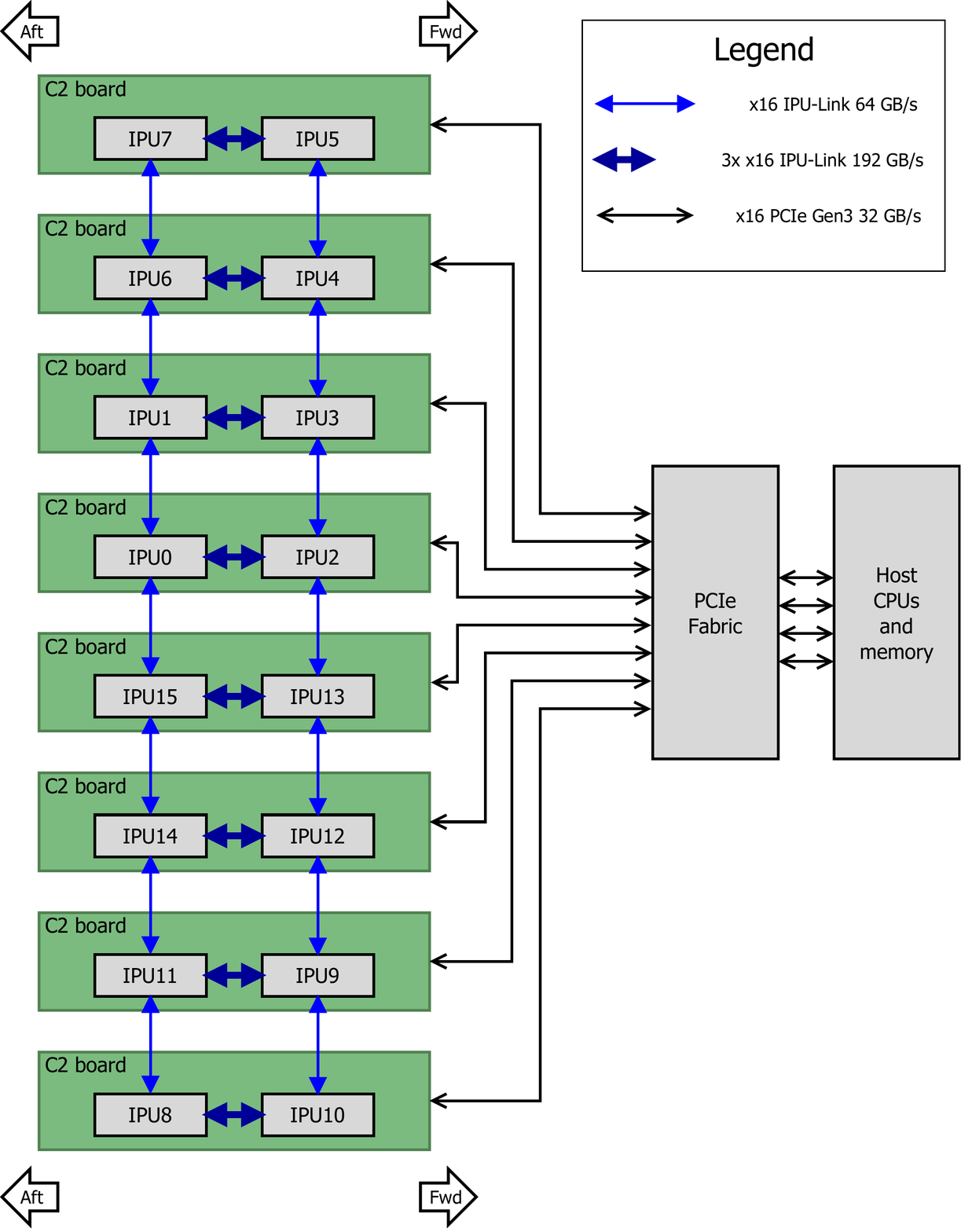}
  \caption{Topology and interconnection of C2 boards and IPU
    processors in the system that we employed in all our
    experiments. All IPU numbers in this figure are \emph{Device
      IDs}. Board placement in the chassis is depicted accurately (see
    Fwd and Aft arrows).}
  \label{fig:system-topology}
\end{figure}

We show a simplified representation of the server's topology and its
connections in Figure~\ref{fig:system-topology}.  The system
features two Intel Xeon Platinum 8168 CPUs, each containing 24 cores,
with 33 MiB of L3 cache and a clock frequency of 2.70 GHz.

IPUs are connected to each other via IPU links, organized in a ladder
network (thin and thick blue arrows in the figure). The network
topology directly explains the interconnect performance that we
measured (Chapter~\ref{chap:interconnect}).

In this ladder-shaped topology,
\begin{itemize}
\item each vertical ``side rail'' represents one single-link chain
  connecting either all even or all odd IPUs; each such link has a
  nominal 64 GB/s bandwidth;
\item each horizontal ``rung'' (thick, horizontal, blue arrow)
  represents a triple-link, bundled connection between the two IPUs
  on each C2 board. The three links offer a nominal bidirectional
  bandwidth of 64 GB/s each, or 192 GB/s aggregate.
\end{itemize}

The routing of traffic along the bundled on-board IPU links is
statically configured so that:
\begin{itemize}
  \item two links are reserved for data transfers where the two IPUs
    on a board are the source and destination; this yields a 128 GB/s
    nominal aggregate bidirectional bandwidth between the two IPUs of
    each board;
  \item one link is reserved for \emph{pass-through} transfers, where
    either the source or the destination IPU belongs to another board
    and has a different parity.
\end{itemize}

In the topology, each board has exactly two neighbors, with the
exception of the first and last boards, which only have one neighbor.
Each IPU is connected directly to the other IPU on the same C2 board
and up to two other IPUs located on neighboring boards. For
example, IPU1 is connected directly to IPU3 (on the same board), and
to IPU0 and IPU6 (on different boards). IPU0 is not directly
connected to IPU3; each data transfer from IPU0 to IPU3 uses either
IPU1 or IPU2 as a relay.

For the avoidance of doubt, the ladder structure is not a torus, i.e.,
it does not wrap around at the edges. The card hosting IPUs 7 and 5 is
not a neighbor of the card hosting IPUs 8 and 10. A transfer from IPU7
to IPU8 must traverse the entire ladder vertically.

All the IPU numbers we use here are \emph{Device ID} numbers as
exposed and used by the Poplar SDK and by the IPU's command-line tools.
Device IDs reflect the lexicographic order of the respective devices'
PCI domains, not their physical placement in the server. For these
reasons, Device IDs do not respect proximity in the network topology.
IPUs with consecutive Device IDs are not on the same board or, in
general, neighbors. For example, consider IPU pair 7 and 8. On the
other hand, Poplar provides a mapping (DNC IDs) that accounts for
the IPU's network proximity. We discuss that in the next section.

Our microbenchmarks reveal that proximity matters. Specifically, tile
proximity directly affects the latency between pairs of tiles in
on-chip communication, and network proximity between IPUs directly
affects inter-IPU latency.

Neighboring IPUs experience the lowest round-trip latency
when communicating with each other, whereas marginal latency
increases progressively when the endpoint IPUs are farther and farther
away.  For example, the two IPUs on a board can typically perform a
minimum data transfer in 0.63...0.83 microseconds, and two neighboring
IPUs of same parity can perform a transfer in 0.54...0.77 microseconds;
any addition hop costs on average 0.16 microseconds (varying between
0.145 and 0.174); we present these results in
Section~\ref{sec:ipu-proximity-min-latency} and
Figure~\ref{fig:ipu-latency-topo}. IPUs 7 and 10 are the farthest
pair and experience a latency of approximately 1.76 microseconds.

On the other hand, peak bandwidth between two endpoint IPUs is not
affected by their proximity (Figure~\ref{fig:ipu-bandwidth-topo}) as
the topology suggests. Also in accordance with topology are our
findings on bidirectional communications between the IPUs on each board:
we achieved 108 GB/s, or 84\% of the nominal peak bandwidth
(Section~\ref{sec:peak-bw},
Table~\ref{tab:point-to-point-bandwidth-long}).  We find that
monodirectional bandwidth achieves almost exactly half that much
bandwidth.

The cursory results we just listed are only examples of our findings;
the next chapter is entirely dedicated to studying the IPU's
interconnect performance in diverse communication patterns and under
different loads, as measured with our benchmarks.

\begin{figure}
  \center
  \includegraphics[width=0.7\columnwidth, angle=270]{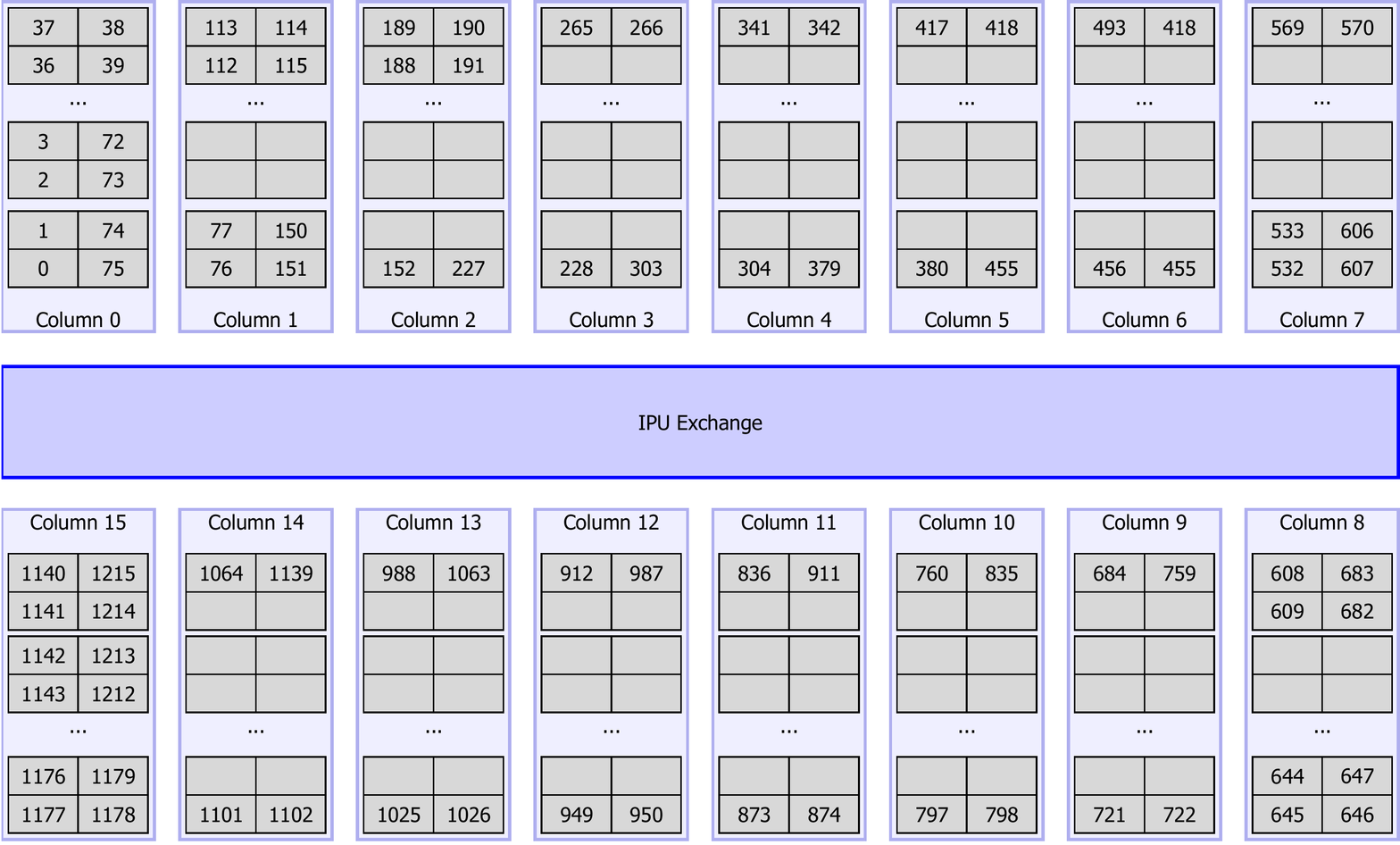}
  \vspace{-1cm}
  \caption{Detailed tile layout within an IPU processor, including
    tile numbering.  The logical tile IDs depicted are those exposed
    to the programmer.  Tiles topologically close to each other
    actually experience shorter communication latencies than tiles
    farther away (see
    Section~\ref{sec:tile-latency-proximity}). Source: direct
    communication with Graphcore.}
  \label{fig:ipu-topology}
\end{figure}

\textbf{Tile numbering.}  The on-chip interconnect among tiles within
an IPU is represented in Figure~\ref{fig:ipu-topology}, where the actual
logical tile ID numbering is shown.  Tiles are arranged in columns
containing 76 total tiles.  These 76 tiles are organized in 19 \emph{islands}
containing four tiles each.  Our benchmarks shows that proximity
affects latency and that neighboring tiles within the same island
experience a marginally shorter latency between each other than tiles
further away within the same column. Tiles in different columns
experience a marginally higher latency.

A developer who is strongly focused on latency could, at the extreme,
manually place compute vertices on tiles in order to take advantage of
the respective proximities.

\begin{figure}
  \includegraphics[width=\columnwidth]{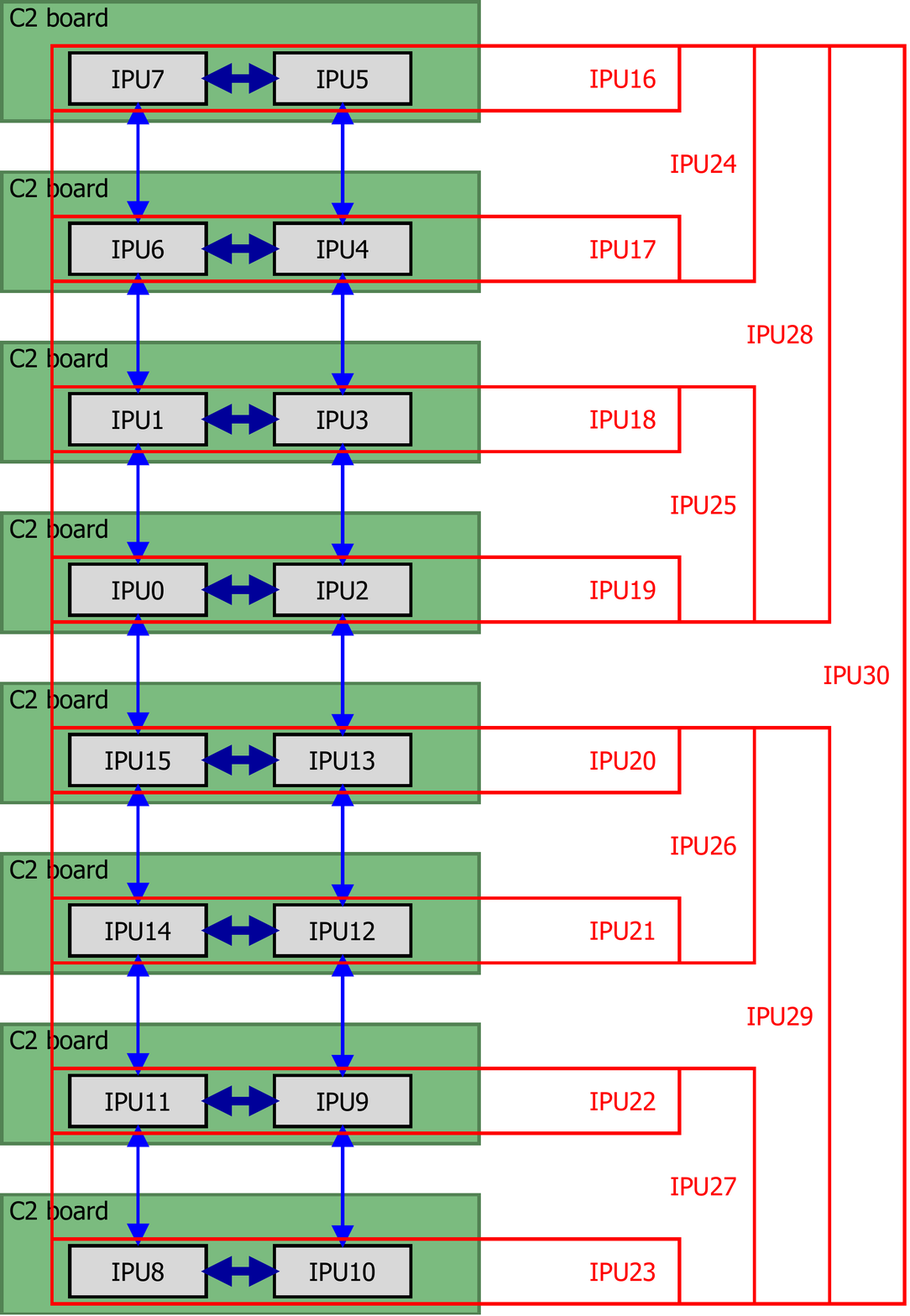}
  \caption{Topology of Multi-IPU virtual devices (depicted in red)
    with respect to the physical IPUs (in black). All numbers are
    Device IDs. For example, IPU16 is a virtual device containing
    physical IPUs 5 and 7.  IPU30 is a virtual device containing all
    16 physical IPUs in the server.}
  \label{fig:multi-ipu-topology}
\end{figure}

\section{Multi-IPUs}
\label{sec:multi-ipus}

This section discusses how the platform groups physical IPUs into
virtual Multi-IPU devices and how IPU tiles are numbered within each
Multi-IPU device. All details reflect the concrete configuration
of the Graphcore evaluation system we benchmarked.

On our evaluation system, the drivers and SDK expose the exact set of
Multi-IPU devices of Figure~\ref{fig:multi-ipu-topology}.  Each
Multi-IPU possesses a unique Device ID; their numbering resumes from
where physical ID numbering ends.  On our system, the highest-numbered
physical IPU has Device ID 15, and the first Multi-IPU (virtual) has
Device ID 16.

For performance reasons, Multi-IPU are constructed only out of
neighboring IPUs and only in powers of two (2, 4, 8, and 16 IPUs).
All 2-IPU devices contain a pair of IPUs residing on the same
board. For example, IPU16 contains IPU7 and IPU5.  All 4-IPU devices
contains the IPU from a pair of C2 cards that are neighbors in the
ladder network topology, and so on.

Multi-IPUs correspond to partitions of the following sequence of
Device IDs into equally-sized, contiguous sub-sequences with power-of-two
lengths:\\

\begin{center}
  5, 7, 4, 6, 3, 1, 2, 0, 13, 15, 12, 14, 9, 11, 10, 8.
\end{center}

\noindent Trivially, this sequence corresponds to a front-to-back,
left-to-right enumeration of IPUs in the chassis. For the complete
avoidance of doubt, chassis front and back are denoted in the picture
by the Fwd and Aft arrows, whereas left and right are as seeon by an
observer placed at the back of the chassis and looking forward (Fwd
direction).

Within each Multi-IPU device, the run-time numbers IPUs according to the
sequence just reported. This 0-based index of each physical IPU
within a Multi-IPU is called a \emph{DNC ID}.  For example, IPU16 is
composed of IPUs 5 and 7 (5 and 7 are their Device IDs); their DNC IDs
are trivially 0 and 1, respectively.  Consider IPU30, the device
containing all IPUs in the system, whose Device IDs are 5, 7, 4, 6
... 10; their respective DNC IDs are 0,1,2,3 ... 15.

To avoid confusion between Device IDs and DNC IDs, the reader only needs to
remember that:
\begin{itemize}
\item \textbf{Device IDs} reflect PCI device numeration; consecutive
  Device IDs do not indicate IPU proximity;
\item \textbf{DNC IDs} reflect actual IPU proximity in the IPU Link
  topology; consecutive DNC IDs denote IPUs that are close to each other.
\end{itemize}

DNC IDs matter to the final user because they also reflect the order
in which tiles are ordered within the Multi-IPU. Tiles are ordered
sequentially according to the DNC IDs: tiles 0...1,215 belong to IPU
with DNC ID 0, tiles 1,216 ... 2,431 belong to DNC ID 1, and so on.

A user employing a different system than the one we evaluated can
obtain an explicit enumeration of all IPU devices (physical, and
virtual) and their topology, including DNC numbering, by using command
line tool \lstinline|gcinfo| \lstinline|--list-all-devices|.

\section{Methods}
\label{sec:methods}

\noindent\textbf{Software SDK version.}  At the time this report is
written, the IPU is a novel architecture and Graphcore is refreshing
its Poplar SDK and its IPU drivers with relative frequency. Because
subsequent releases include incremental optimizations, the choice of
any one SDK version in conjunction with a benchmark affects the
benchmark's performance results. We used SDK version
1.0.49. Researchers intending to duplicate our results should employ
the same version.

Whenever meaningful, we timed the latency of each operation with care
for the following concerns:
\begin{itemize}
\item \textbf{Single-IPU measurements.} In benchmarks involving a
  single IPU, we timed operations on the IPU, via the cycle-accurate
  primitive \lstinline|popsys::cycleStamp()|. The use of this
  primitive isolates measurements from CPU-IPU communication overheads
  and IPU program launch overheads.
\item \textbf{Multi-IPU measurements.} Benchmarks involving multiple
  IPUs can not practically use \lstinline|popsys::cycleStamp()|.  We
  time them instead from the host system. We time multiple iterations
  of each benchmark in order to amortize overheads, making their
  impact negligible.  To remove spurious overhead between one
  iteration and the next, we use the Poplar primitive
  \lstinline|program::Repeat()|, which excludes the host from having
  any role in the repetition.
\item\textbf{Warm up.} Whenever we desire steady-state measurements,
  we take appropriate countermeasures to exclude warm-up overheads
  from the results. For example, whenever benchmarks involve the host,
  we typically execute an untimed warm-up iteration of the
  benchmark before we start timing.
\end{itemize}

\noindent\textbf{Units.} In this paper we prefer harmonized ISO/IEC
80000-13:2008 standard to denote data sizes and capacities: 1 KiB =
1,024 bytes; 1 MiB = 1,024 KiB; 1 GiB = 1,024 MiB; 1 TiB = 1,024 GiB.
When units of capacity are used to express throughputs, we use
customary 1,000-based prefixes (k,M,G,T ---thus, kB/s, MB/s, GB/s,
TB/s) for consistency with the literature and ease of calculation.

\chapter{Local Memory}
\label{chap:local-memory}

We start our analysis from the basic constituents of an IPU system and
proceed outward. In this chapter, we focus on the performance of the
memories located within each tile.

On an IPU, each tile possesses 256 KiB of memory that it can access
directly via instructions.  (For a tile to access memory that is local
to another tile, it must use the exchange; we characterize the
exchange's performance in the next chapter.)

We find that the performance of each local memory is fixed (6 cycles
latency, 31.1 TB/s aggregate peak read bandwidth); local memories are
completely decoupled from each other, in function and performance, as
architectural considerations suggest.  Pressure on the local memory in
one tile does not affect memory performance in any other tile.

\section{Latency}

The latency experienced by a tile reading a value from its local
memory is 6 clock cycles. Our experiments show that this latency is
fixed and does not depend on access patterns, stride, working set size,
number of threads used on each tile (1 ... 6), number of concurrent
tiles running the same benchmark simultaneously, or size of the target
working set accessed by the benchmark.

A minimal benchmark suitable to demonstrate latency invariance
follows, courtesy of Graphcore. The code implements a pointer chase
that scans an array of configurable size. The array is pre-populated
with indices that realize a linked-list visit with configurable
stride.

\begin{lstlisting}[basicstyle={\scriptsize\ttfamily}]
template <const int unroll_factor>
class PChaseVertex : public poplar::Vertex {
public:
  poplar::Input<Vector<uint32_t>> in;
  poplar::Input<uint32_t>         start;
  poplar::Input<bool>             flag;

  poplar::Output<uint32_t>        out;

  bool compute() {
    uint32_t index = start;
    for (int i = 0; i < unroll_factor; i += 1)
      index = in[index];
    if (flag)
      *out = index;
    return true;
  }
};

template class PchaseVertex<1000>; // explicit instantiation
\end{lstlisting}

In the listing, variable \lstinline|flag| presents the compiler with a
possible side effect; its purpose is to prevent the compiler from
detecting the entire benchmark as dead code and optimizing it out.

Experimental results confirm that neither size nor stride affect
memory access latency.

\section{Bandwidth}

We measure the aggregate read bandwidth available to user code on the
entire IPU and compare it with theoretical limits derived from
hardware specifications (31.1 TB/s). We achieved bandwidths closely
matching theoretical values only with a benchmark written in assembly
language containing a zero-overhead loop of 128-bit loads. Experiments
show that Poplar load-dense code with narrower loads, 32- and 64-bit
wide, access roughly a quarter and half of the theoretical bandwidth,
respectively. Code with lower-than-perfect load instruction density
may achieve even lower bandwidth. However, naive code consisting of
array-based, single-precision read loops without hand optimizations
emits relatively dense 32-bit loads, achieving a quarter of the
theoretical limit. Finally, we show how developers can use the
\lstinline|float2| vector types to increase access width, roughly
doubling their bandwidth, without resorting to assembly code. All
these results are presented in detail in this section.

\noindent\textbf{Theoretical limit.} The theoretical aggregate read bandwidth
derives from the following assumptions: each tile can read 16 bytes
per clock cycle, the clock frequency is 1.6 GHz, and the tile count is
1,216. The result, 31.1 TB/s, is the product of these three
factors. (Top row in Table~\ref{tab:memory-bw}.)

\noindent\textbf{Multi-threading.} All benchmarks in this section use six
identical threads per tile in order to achieve complete hardware
thread occupancy. Using a smaller amount of threads yields
proportionally lower bandwidth.

\begin{table}[h]
  \center
  \footnotesize
  \begin{tabular}{lcrrrc}
    \toprule
    Approach                      & Language & Load width &  Bandwidth &  Fraction of \\
                                  &          & (bits)     &   (TB/s)   &  Theoretical \\
    \midrule
    Theoretical limit             & --       &            &      31.13 &  100 \% \\
    \midrule
    Best actual                   & assembly &        128 &      30.70 &  98.6\% \\
    \midrule
    float2                        & Poplar   &         64 &      15.30 &  49.2\% \\
    64-bit loads                  & assembly &         64 &      15.26 &  49.0\% \\
    float4                        & Poplar   &         64 &      15.02 &  48.3\% \\
    \midrule
    Naive, float (upper limit)    & Poplar   &         32 &       7.59 &  24.4\% \\
    \bottomrule
  \end{tabular}
  \caption{Theoretical and actual aggregate read bandwidths available
    on the entire IPU chip, as measured via diverse benchmarks written
    in assembly or in Poplar.}
  \label{tab:memory-bw}
\end{table}

\noindent\textbf{Naive code.} To measure the performance limits of loop-based
code written in Poplar/C++, we use a benchmark specifically designed
to generate long sequences of load instructions.  This benchmark
delivers 7.59 TB/s, roughly a quarter of the theoretical
limit. (Bottom row in Table~\ref{tab:memory-bw}.)  Real-world user
code will see even lower performance than 7.59 TB/s if it exhibits a
lower load instruction density.  An excerpt of the benchmark source
code follows, courtesy of Graphcore.

\begin{lstlisting}[basicstyle={\scriptsize\ttfamily}]
#define UNROLL 256

class AccaddVertex32 : public poplar::Vertex {
public:
  Input<Vector<float>>         in;
  poplar::Input<uint32_t>      size;

  bool compute() {
    for (int i = 0, e = size ; i < e; i += UNROLL) {
      float* base = &in[i];

      #pragma unroll UNROLL
      for (int j = 0; j < UNROLL; j++)
        asm volatile("" :: "r"(base[j]));
    }
    return true;
  }
};
\end{lstlisting}

The listing contains the array-based loop, with an inner loop
containing an inline assembly block accompanied by an unroll
pragma directive.  The inner loop instructs the compiler to generate
an unrolled sequence of load instructions.

We verify by inspection that the corresponding emitted code contains
an unrolled sequence of 256 load instructions, each 32-bit wide.



\noindent\textbf{Best actual bandwidth.} To determine the absolute
highest bandwidth achievable by user code, we benchmark hand-written
assembly code that features zero-overhead loops of 128-bit wide memory
accesses. The details of this code are beyond the scope of this
document.  It achieves a bandwidth closely matching the theoretical:
30.7 TB/s, or 98.6\% of the theoretical limit. (``Best actual'' in
Table~\ref{tab:memory-bw}.) We know of no other means to emit 128-bit
load instructions than resorting to assembly.


\noindent\textbf{Vector types.} Applications loading 32-bit words can
achieve the higher 64-bit bandwidth by loading two words at a time if
the source data is contiguous. This avenue benefits developers who
desire higher performance but are not willing to write assembly code.
They can instruct the compiler to generate wider loads by using vector
type \lstinline|float2| and explicitly aligning their input arrays, as
we illustrate in the listing below. Vector type \lstinline|float2|
expresses an array of two 32-bit floating point values intended to be
handled together by the hardware.  This code will emit 64-bit load
instructions and achieve the same performance as a hand-written
assembly loop of 64-bit loads (rows labeled ``float2'' and ``64-bit
loads'' in Table~\ref{tab:memory-bw}), which is roughly half of the
theoretical limit.

\begin{lstlisting}[basicstyle={\scriptsize\ttfamily}]
#define UNROLL 256

class AccaddVertex64 : public poplar::Vertex {
public:
  Input<Vector<float, VectorLayout::SPAN, 8>> in;
  poplar::Input<uint32_t>                     size;

  bool compute() {
    float2 *f2in = reinterpret_cast<float2 *>(&in[0]);
    for (int i = 0, e = size / 2; i < e; i += UNROLL) {
      float2* base = &f2in[i];
      #pragma unroll UNROLL
      for (int j = 0; j<UNROLL; j++)
        asm volatile("" :: "r"(base[j]));
    }
    return true;
  }
};
\end{lstlisting}
The crucial aspects of this listing are the alignment directives
associated with input tensor \lstinline|in| and the
\lstinline|reinterpret_cast<>| syntax necessary to access
\lstinline|in| via vector type \lstinline|float2|.  The inline
assembly loop at the bottom is not part of the technique we propose,
and it is just intended to achieve perfect load instruction density
(forcing the compiler to generate an unrolled sequence of load
instructions), as in an example we presented earlier.

This vector type approach does not extend to \lstinline|float4|, which
does not cause the compiler to emit 128-bit loads. We find that a
benchmark based on \lstinline|float4| roughly achieves the same
performance as the code just illustrated (``float4'' in
Table~\ref{tab:memory-bw}). We omit its listing.

\begin{figure}
  \includegraphics[width=\columnwidth]{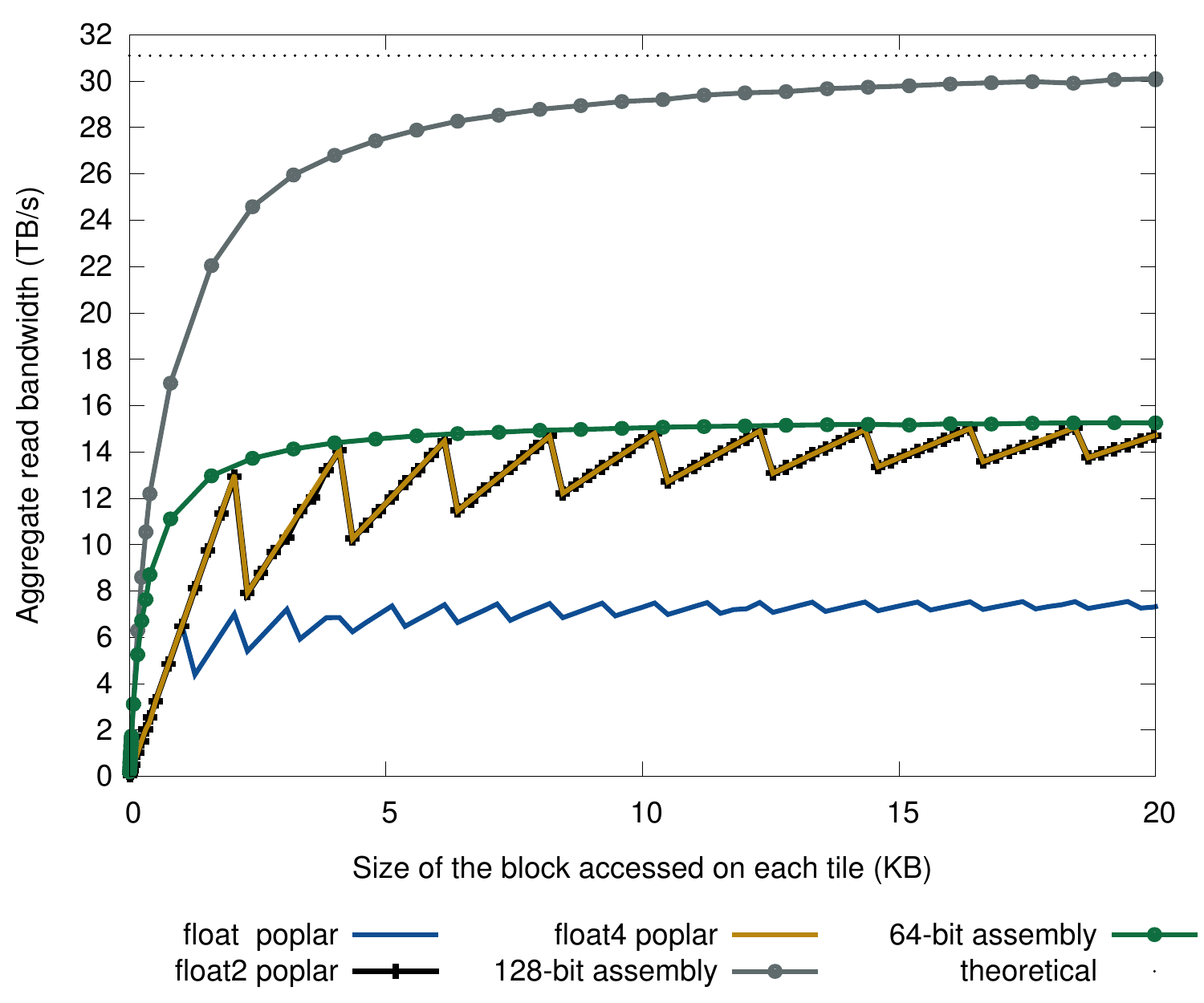}
  \caption{Aggregate local memory bandwidth on the entire IPU as a
    function of block size.}
  \label{fig:tile_bandwidth}
\end{figure}

\noindent\textbf{Sensitivity to block size.} We show how the aggregate
local memory bandwidth seen by a workload varies according to the size
of blocks it accesses.  Our benchmark scans a block of configurable
size. We instantiate the workload six times per tile to optimally
occupy the six hardware threads. We vary block sizes from a single
word to 200 KiB, which is close to the user-available capacity.  Our
results are in Figure~\ref{fig:tile_bandwidth}; for sufficiently large
blocks, the bandwidth saturates smoothly.  Accessing memory in blocks
of 8 KiB is sufficient to achieve 95\% of peak bandwidth.

No congestion occurs as more and more tiles use the respective
memories. This finding is consistent with the operation of completely
distributed local memories.


\noindent\textbf{Write bandwidth.} The theoretical limit is 15.5 TB/s,
derived from each tile's ability to write 8 bytes per clock cycle.
Naive, array-based, single-precision write loops achieve the same
performance as the corresponding read loops (``float poplar'') in
Figure~\ref{fig:tile_bandwidth}. That code benefits from the use of
vector types in the same manner as illustrated above and with the same
performance benefits. The following listing, courtesy of Graphcore,
illustrates how to instruct the compiler to emit 64-bit wide, aligned
write instructions for single precision (32-bit) elements.

\begin{lstlisting}[basicstyle={\scriptsize\ttfamily}]
#define UNROLL 8

using namespace poplar;

class AccaddVertex : public poplar::Vertex {
public:
  Input<Vector<float, VectorLayout::SPAN, 8>> in;
  // Request 8-byte alignment (SPAN) for variable `in'

  poplar::Input<uint32_t>      size;
  poplar::Input<bool>          flag;
  poplar::Output<float>        out;

  bool compute() {
    float2 *f2in = reinterpret_cast<float2 *>(&in[0]);
    float tmp = 0.0;

    // loop limit is size/2 because each element consists of 2 floats
    for (int i = 0; i < size/2; i+= UNROLL ){
      float2 tmps[UNROLL];
      #pragma unroll UNROLL
      for (int j = 0; j < UNROLL; j++)
        tmps[j] = f2in[i+j];
      if (flag)
        for (int j = 0; j < UNROLL; j++)
          tmp += tmps[j][0] + tmps[j][1];
    }

    if (flag)
      *out = tmp;
    return true;
  }
};
\end{lstlisting}

\noindent Using an unroll factor equal to 8 achieves the highest
bandwidth. This code achieves the same bandwidth as the corresponding
code that we illustrated above for array reads.


\chapter{Interconnect}
\label{chap:interconnect}

We evaluate the IPU interconnect's performance by benchmarking
point-to-point and collective operations.  We analyze whether and how
latency and bandwidth degrade as the scale of the communication
operation increases, involving on-chip and off-chip interconnects.

We selected a sufficiently broad set of primitives for benchmarking that
represent communication patterns commonly found in parallel
applications. Our results are intended to help software designers
derive early performance estimates for their applications.

Our choice of microbenchmarks and metrics conforms with publicly
available benchmarking suites for parallel computing, such as the
\emph{OSU Micro-benchmarks}\cite{omb2019} by the Ohio State
University's Network-based Computing Laboratory.

The first section of this chapter focuses on point-to-point transfers;
the remaining sections each focus on one collective operation.

\section{Point-to-point Transfers}

In this section, we study the performance of IPU systems when engaged
in point-to-point communications (data transfers from one source tile
to one destination tile) under diverse load conditions.

Our results draw a map of the relative proximity of the constituents
of an IPU system (specifically, how quickly tiles can reach each other
in terms of latency and bandwidth, depending on relative distance in
the system) that is predictable and consistent with the known
topology of the system presented in Figures~\ref{fig:system-topology}
and \ref{fig:ipu-topology}.

Software designers can take advantage of our results to gain an
understanding of the latency penalty and bandwidth associated with
traversing the on-chip interconnect, crossing IPU links on a board,
and crossing IPU links across boards.

We benchmark latency and bandwidth in a spectrum of congestion
conditions ranging from global silence (the system is idle, with the
only exception of the one transfer benchmarked) to full load (all
tiles are engaged in communication with the same pattern as the one
benchmarked).

\begin{figure}[h]
  \hspace{-1.5cm}\includegraphics[width=1.2\columnwidth, trim=0cm 20.2cm 1cm 0cm]{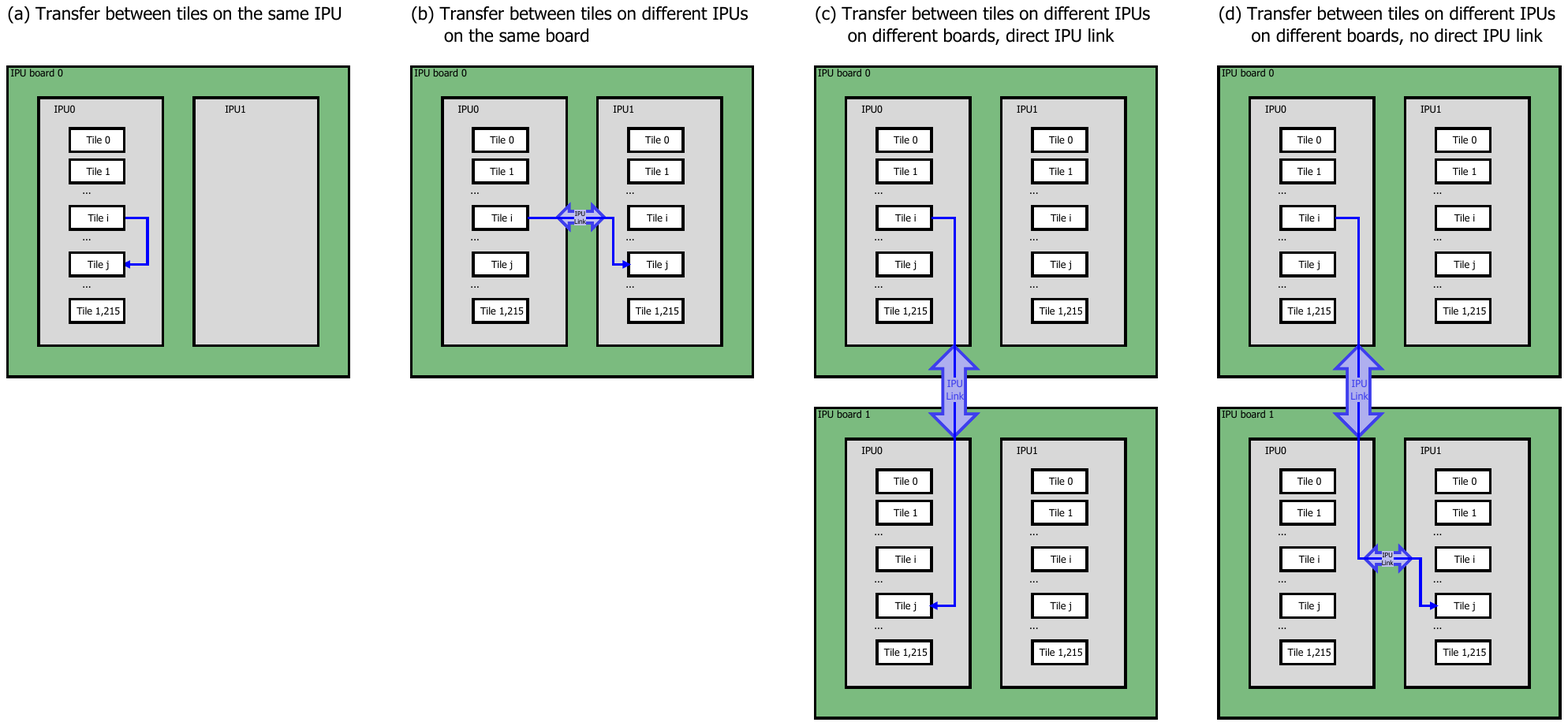}
  \caption{Communication topologies we evaluate in point-to-point latency
    benchmarks. Data transfers between the source and the destination
    tiles are depicted with blue arrows. The experiments are designed
    to exercise both on-chip (a) and off-chip (b,c,d) interconnects.}
  \label{fig:point-to-point-setup}
\end{figure}


\subsection{Congestion-free Latency}
\label{sec:congestion-free-latency}
We study the latency incurred by a single, minimum-size transfer
between two tiles while the rest of the system is idle (global
silence on the interconnects).

The latencies we measure are consistent with the topology of the
experiment.  On chip, a source tile can reach any destination tile on
the same IPU within 0.13 microseconds on average.  As soon as a message
needs to cross an IPU link to reach its destination, a penalty of
approximately 0.5 microseconds applies.  Surprisingly, reaching a tile
on the second IPU located on the same board as the source is
marginally more expensive than reaching a directly connected IPU on a
different board. Detailed results are in
Table~\ref{tab:point-to-point-latency-no-load}.

Detailed descriptions of the experimental topologies we benchmarked
follow, matching the corresponding illustrations in
Figure~\ref{fig:point-to-point-setup} and corresponding
results of Table~\ref{tab:point-to-point-latency-no-load}:
\begin{itemize}
\item in experiment (a) we exercise the on-chip interconnect; the
  source and destination tiles ($i$ and $j$) reside on the same IPU.
  We obtain the average latency value over a large number of
  experiments. In each such experiment, $i$ and $j$ are chosen
  randomly;
\item in experiment (b) we exercise the off-chip IPU Link interconnect
  between two IPUs on the same board;
\item in experiment (c) we exercise the off-chip IPU Link interconnect
  between two IPUs on different boards when those IPUs are directly
  connected via an IPU link. Surprisingly, this latency is marginally
  lower than that of experiment (b);
\item in experiment (d) we exercise the off-chip IPU Link interconnect
  in conditions where the two IPUs involved are not directly connected
  via an IPU link and communication requires traversing more than one
  IPU link.
\end{itemize}

\begin{table}[h]
  \center
  \small
  \begin{tabular}{lcrr}
  \toprule
  Experiment                            &     &  Latency           \\
  \midrule
  On-chip                               & (a) &  0.133 $\upmu$s    \\
  Off-chip, on board                    & (b) &  0.633 $\upmu$s    \\
  Off-board, direct IPU link            & (c) &  0.524 $\upmu$s    \\
  Off-board, indirect IPU link          & (d) &  0.779 $\upmu$s    \\
  \bottomrule
  \end{tabular}
  \caption{Point-to-point latency for small messages under no
    load. Experiments labeled (a)...(d) correspond to the
    illustration in Figure~\ref{fig:point-to-point-setup} and
    are described above.}
  \label{tab:point-to-point-latency-no-load}
\end{table}

\subsection{Latency between Tiles on a Chip by Proximity}
\label{sec:tile-latency-proximity}

\begin{figure}
  \includegraphics[width=\columnwidth]{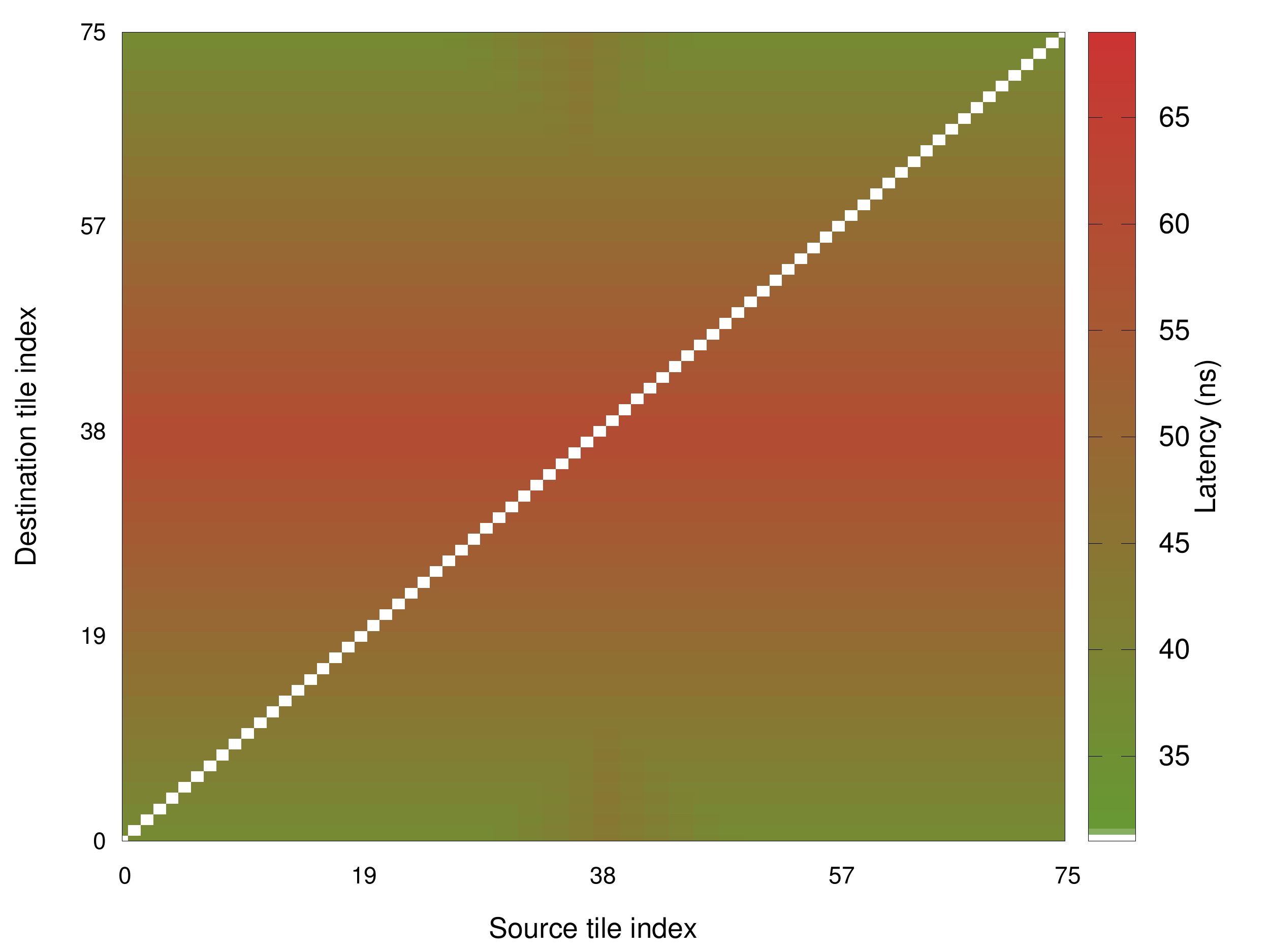}
        \caption{Minimum latency between all tiles belonging to the same
    Column on an IPU processor (we first presented columns in
    Figure~\ref{fig:ipu-topology}).}
  \label{fig:tile-latency-topo-sub}
\end{figure}

\begin{figure}
  \includegraphics[width=\columnwidth]{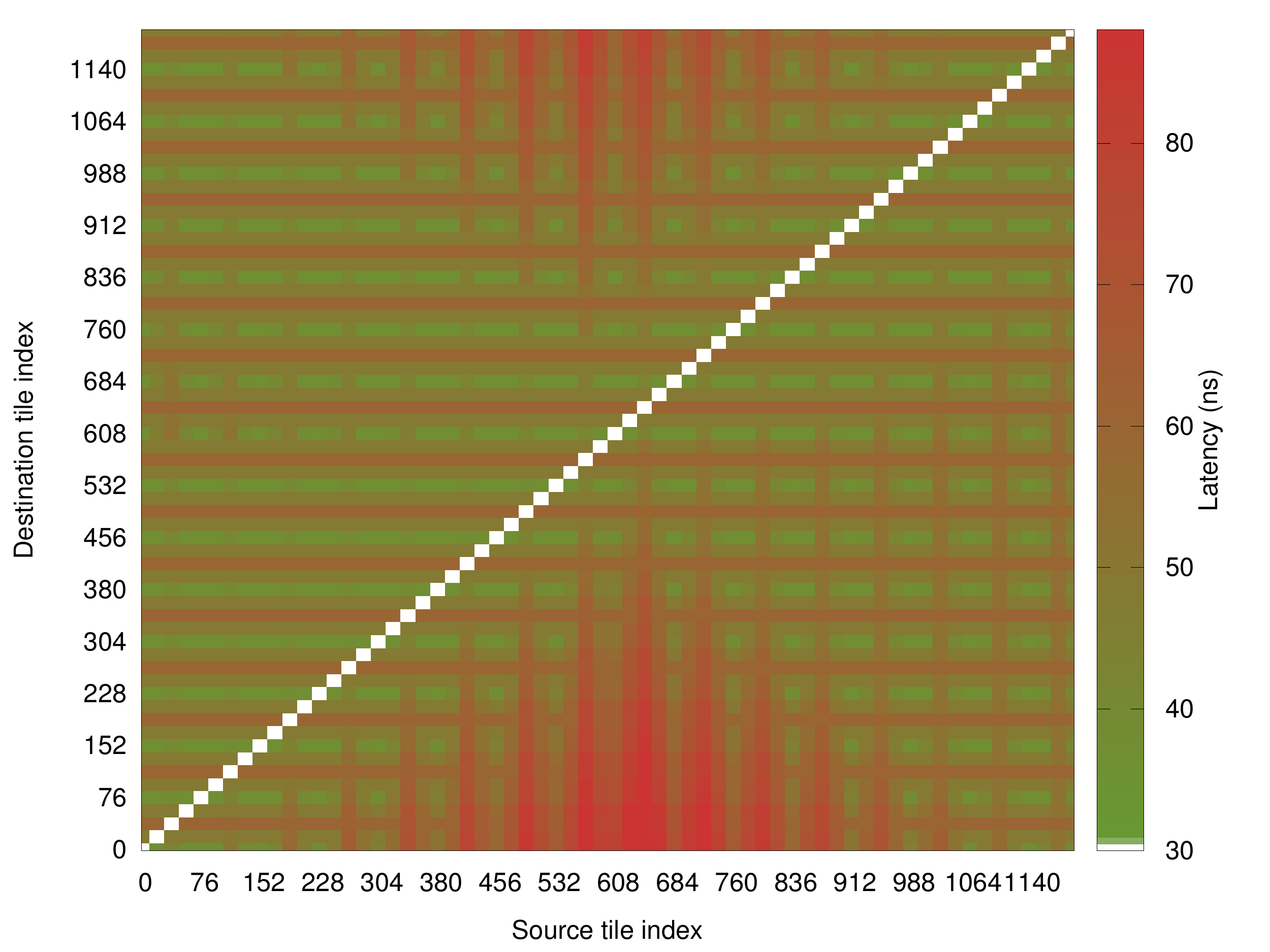}
  \caption{Minimum latency between all pairs of tiles on an IPU processor.}
  \label{fig:tile-latency-topo}
\end{figure}

We study how the physical proximity between pairs of tiles on a chip
affects their communication latency in congestion-free conditions.

We measure latency between all pairs of tiles and depict our results
in Figures~\ref{fig:tile-latency-topo-sub} and
\ref{fig:tile-latency-topo}.  The first figure focuses on tiles within
a column; the second shows the entire IPU.  In both figures, the main
diagonal (in white) corresponds to local transfers, which are carried
out in local memory and do not involve the exchange.  Latencies
reflect tile topology (Figure~\ref{fig:ipu-topology}).

\noindent\textbf{Latency measurements.} The reader should pay
attention to the fact that the latency measurements presented within
this section may differ from those of other sections, e.g.,
Table~\ref{tab:point-to-point-latency-no-load}.  The discrepancy is
due to the use of different timing methods. Specifically, experiments
in this section use the on-IPU, fine-grained profiling facilities
offered with the Poplar SDK (that we described under ``Single-IPU
measurements'' in Section~\ref{sec:methods}) which do not include the
cost of the on-IPU synchronization phase that precedes the transfer,
whereas other sections typically use host-based ``Multi-IPU
measurements'' that include that cost. Both measurements are
meaningful when appropriately characterized.

\noindent\textbf{Within a column.} Data transfers within the same
column take 37...59 ns (59...95 clock cycles); see
Figure~\ref{fig:tile-latency-topo-sub}. Intra-column latency primarily
depends on what island the destination tile belongs to; transfers to
the same island have same latency, no matter the source island.
Latency is minimal when transferring to a tile belonging to the island
closest to the IPU Exchange (e.g., for Column 0, that's island of
tiles 0, 1, 74 and 75). Latency increases by 1.25 ns (2 clock cycles)
for every island the destination moves away from the exchange.

\noindent\textbf{Across columns.} Latency is 98 ns (the highest) in
transfers from Column 8 to 0 (tile 646 to 0). Transfers from the
rightmost columns (column 7 and 8) to the leftmost columns (0 and 15)
take more than 63 ns. Transfers in the opposite direction have lower
latency.

\subsection{Latency between IPUs by Proximity}
\label{sec:ipu-proximity-min-latency}

\begin{figure}
  \hspace{-5mm}
  \includegraphics[width=1.1\columnwidth]{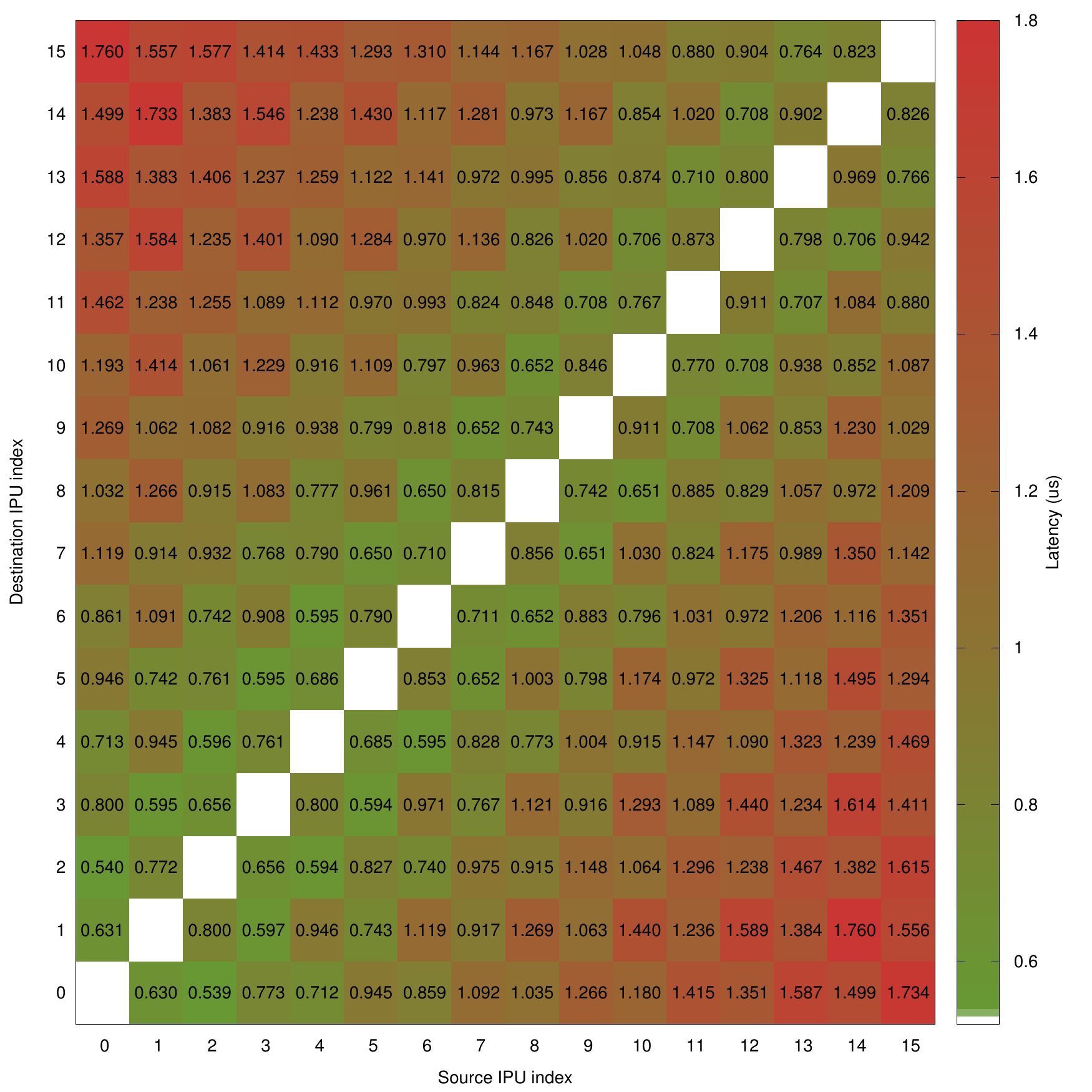}
  \caption{Minimum latency between each pair of IPUs in our
    experimental 16-IPU system, measured in zero-congestion
    conditions. IPUs are numbered according to their \emph{DNC IDs},
    as discussed in Section~\ref{sec:multi-ipus}.}
  \label{fig:ipu-latency-topo}
\end{figure}

We measure and chart the minimum latency between pairs of IPUs across
the entire 16-IPU experimental system.  Latencies correlate directly
with the distance between source and destination IPUs along the network
topology of Figure~\ref{fig:system-topology}.  We chart our results in
Figure~\ref{fig:ipu-latency-topo}.

\noindent\textbf{Numbering.} In this section we only refer to IPUs by \emph{DNC
  IDs} (as opposed to \emph{Device IDs}---recall the distinction from
Section~\ref{sec:multi-ipus}). This is precisely because DNC IDs
account for the proximity between pairs of IPUs, whereas Device IDs
don't.

\noindent\textbf{Results.} We find that:
\begin{itemize}
\item the lowest latency is between IPU with DNC IDs $i$ and $i \pm
  2$, i.e., each even-numbered IPU reaches the fastest those
  even-numbered IPUs that are facing it on the neighboring boards. The
  same applies to odd-numbered IPUs. These links are visible in the
  matrix as the two diagonal lines with the most intense shade of
  green; they are two cells above the main diagonal and two cells
  below the main diagonal;
\item pairs of IPUs located on the same board see the second-best
  latencies across the system;
\item there is no wrap-around. The IPUs with DNC IDs 0 and 14 are not
  neighbors; neither are 1 and 15.  These four IPUs suffer from edge
  effects in the sense that they only have one neighbor of same
  parity.  This reflects the IPU Link network shape (i.e., not a
  torus).
\end{itemize}

\clearpage

\subsection{Latency under Load}

In this section, we investigate the impact of congestion on
point-to-point latency.  We simulate traffic conditions by
instantiating identical point-to-point transfers that occur
concurrently.

We compare latencies in the congestion-free scenario and under load in
Table~\ref{tab:point-to-point-latency-load-short}.  In the table,
experimental conditions (a)...(d) refer to same topologies we
considered in our congestion-free study
(Section~\ref{sec:congestion-free-latency}). In summary, congestion
increases on-chip latency only marginally (+24\%), but it affects
off-chip IPU link latency significantly, with slowdowns of
4.0...7.7$\times$.

We also find that the average per-message latency in a multi-IPU system
is remarkably scalable: randomized concurrent transfers across a
16-IPU system experience a latency that is minimally higher (1.93 ns/message)
than on a single-board 2-IPU system (1.76 ns/message).

The methods we used to put the interconnect under load are described
below.

\begin{table}[h]
  \center
  \small
  \begin{tabular}{lcrrr}
  \toprule
  Experiment                            &     &  Congestion-free &  Latency    & Congestion \\
                                        &     &  Latency         &  under load & degradation\\
  \midrule
  On-chip                               & (a) &  0.133 $\upmu$s  &  0.165 $\upmu$s & 1.2$\times$\\
  Off-chip, on board                    & (b) &  0.633 $\upmu$s  &  2.521 $\upmu$s & 4.0$\times$\\
  Off-board, direct IPU link            & (c) &  0.524 $\upmu$s  &  2.524 $\upmu$s & 4.8$\times$\\
  Off-board, indirect IPU link          & (d) &  0.779 $\upmu$s  &  5.989 $\upmu$s & 7.7$\times$\\
  \bottomrule
  \end{tabular}
  \caption{Effect of congestion on point-to-point latency. This is a
    short summary that compares congestion and no-congestion results
    in conditions (a)...(d) corresponding to the previous
    section. More detailed experiments are in the next table.}
  \label{tab:point-to-point-latency-load-short}
\end{table}

Experiments (a)...(d) in presence of congestion match the
corresponding congestion-free experiments (described in the previous
section), with the following additional details:
\begin{itemize}
\item in experiment (a), each tile on an IPU performs one transfer to
  one randomly selected tile on the same IPU; as many transfers occur
  concurrently as there are tiles on one IPU (1,216). Each tile
  participates in exactly one transfer as a source and in exactly one
  (other) transfer as a destination;
\item in experiment (b), we exercise the off-chip IPU Link
  interconnect between two IPUs on the same board. Each tile on one
  IPU performs one transfer to one randomly selected tile on the other
  IPU on the same board. As many transfers occur concurrently as there
  are tiles on one IPU. Each tile on the first IPU participates in
  exactly one transfer as a source, and each tile on the second IPU
  participates as a destination. All cross-IPU traffic is
  monodirectional;
\item in experiment (c), we exercise the off-chip IPU Link
  interconnect between two IPUs on different boards when those IPUs
  are directly connected via an IPU link (e.g., each tile on IPU5
  performs one transfer to one randomly selected tile on IPU4; see
  Figure~\ref{fig:system-topology}). As many transfers occur
  concurrently as there are tiles on one IPU. Each tile on the source
  IPU participates in exactly one transfer as a source, and each tile
  on the destination IPU participates in exactly one transfer as a destination.
  All cross-IPU traffic is monodirectional;
\item in experiment (d) we exercise the off-chip IPU Link interconnect
  in conditions where the two IPUs involved are not directly connected
  via an IPU link and communication requires traversing more than one
  IPU link (e.g., each tile on IPU5 performs one transfer to one
  randomly selected tile on IPU6; see
  Figure~\ref{fig:system-topology}). All other conditions are the same
  as in experiment (c).
\end{itemize}

For a finer-grained characterization of congestion impact, see the
augmented results of Table~\ref{tab:point-to-point-latency-long} and
our observations that follow.

\begin{table}[h]
  \center
  \footnotesize
  \begin{tabular}{clcrrrr}
  \toprule
  Scale  & Experiment                                      &  & Concurrent & Total         & Avg. latency \\
  (IPUs) &                                                 &  & transfers  & latency       & per message \\
  \midrule
        & \multicolumn{4}{l}{On-chip: source and destination tiles are on the same IPU}\\
        &                                                  &     &         1 &  0.123 $\upmu$s  & 122.565  ns \\
        &                                                  &     &         2 &  0.130 $\upmu$s  &  65.040  ns \\
        &                                                  &     &         4 &  0.157 $\upmu$s  &  39.253  ns \\
        &                                                  &     &         8 &  0.152 $\upmu$s  &  18.994  ns \\
        & ...                                              &     &        16 &  0.160 $\upmu$s  &  10.009  ns \\
        & ...                                              &     &        38 &  0.148 $\upmu$s  &  3.884   ns \\
        & ...                                              &     &        76 &  0.165 $\upmu$s  &  2.173   ns \\
   1/8  & ...                                              &     &       152 &  0.163 $\upmu$s  &  1.070   ns \\
   1/4  & ... a quarter of one IPU                         &     &       304 &  0.165 $\upmu$s  &  0.541   ns \\
   1/2  & ... half IPU                                     &     &       608 &  0.165 $\upmu$s  &  0.272   ns \\
   1    & ... the entire IPU                               & (a) &     1,216 &  0.165 $\upmu$s  &  0.136   ns \\
   \midrule
   2    & \multicolumn{5}{l}{Cross-IPU experiments - 2 IPUs - monodirectional traffic} \\
        & ... both IPUs on the same board                  & (b) &    1,216 & 2.521 $\upmu$s  & 2.073 ns \\
        & ... across boards, direct IPU link               & (c) &    1,216 & 2.524 $\upmu$s  & 2.076 ns \\
        & ... across boards, no direct IPU link            & (d) &    1,216 & 5.980 $\upmu$s  & 4.917 ns \\
  \midrule
        & \multicolumn{5}{l}{Cross-IPU experiments - random system-wide destinations } \\
   2    & ... both IPUs on the same board                  &     &     2,432 &  4.282 $\upmu$s  &  1.761 ns \\
   2    & ... across boards, direct IPU link               &     &     2,432 &  2.572 $\upmu$s  &  1.058 ns \\
   2    & ... across boards, no direct IPU link            &     &     2,432 &  6.030 $\upmu$s  &  2.480 ns \\
   4    & 4 IPUs on two boards                             &     &     4,864 &  5.981 $\upmu$s  &  1.230 ns \\
   8    & 8 IPUs on four boards                            &     &     9,728 & 16.532 $\upmu$s  &  1.699 ns \\
   16   & 16 IPUs, entire system                           &     &    19,456 & 37.611 $\upmu$s  &  1.933 ns \\
  \bottomrule
  \end{tabular}
  \caption{Fine-grained and larger-scale detailed results on the
    effect of congestion on latency.}
  \label{tab:point-to-point-latency-long}
\end{table}

In the on-chip experiments, we vary congestion by varying the number
of concurrent transfers from 1 to 1,216. In each transfer, a randomly
selected set of tiles (of count 2, 4, 8 ... 1,216) transfer to as many
destination tiles on the same IPU. At the upper extreme, this
coincides with experiment (a) as already described.

The middle band of the table contains results of experiments (b), (c)
and (d) involving 2 IPUs where the source tiles are segregated on one
IPU and all the destination tiles are on another. All cross-IPU
traffic in these experiments is monodirectional.  We have described
these experiments earlier in this section.

In the bottom band of the table, we present results corresponding to
desegregated destinations. We consider varying system scales (from 2
to 16 IPUs) and, at every scale, all the tiles in the system are the
source of one transfer.  The destination of each transfer is chosen
randomly among all tiles in the system. The destination tile can be on
the same IPU as the source or on any other IPU.  This communication
scheme is representative of parallel applications using a domain
decomposition that requires uniformly spread communication.  Results
show good scalability across system sizes.  The average per-message
latency does not seem to grow significantly in a 16-IPU system
compared to smaller systems.

\subsection{Congestion-free Peak Bandwidth}

We study the peak bandwidth available to a single point-to-point
transfer between pairs of tiles in different topologies in
congestion-free conditions (no other operations occurring in the
rest of the system; global silence on all interconnects except for the
transfer being benchmarked).

\noindent\textbf{Peak.} Everywhere in this paper, the term \emph{peak
  bandwidth} denotes the bandwidth seen by transfers of sufficiently
large messages (we use the terms \emph{block} and \emph{message}
equivalently).  On most interconnect architectures, peak bandwidth
values are achieved with larger messages.  That happens because
transfers both of large and small blocks tend to incur similar
communication setup overheads, but with larger blocks those overheads
are amortized on higher byte counts, thus yielding higher throughput
values.  This property holds on the IPU too, for both the on-chip IPU
interconnect and the off-chip IPU Link network. We verify this
claim experimentally in
Section~\ref{sec:bw-as-function-of-block-size}.

Our experiments cover the same topologies already described in
Section~\ref{sec:congestion-free-latency} and depicted in
Figure~\ref{fig:point-to-point-setup}.

All results follow in
Table~\ref{tab:point-to-point-bandwidth-no-load}.  There is a 14\%
decrease in point-to-point bandwidth when moving from on-chip to
off-chip, on-board communication. There is an additional 9\% penalty
when moving from on-board to inter-board communication with source and
destination IPU linked directly. When IPUs are not connected directly,
bandwidth decrease marginally (less than 2\%).

\begin{table}[h]
  \center
  \small
  \begin{tabular}{lcrr}
  \toprule
  Experiment                            &     &  Peak Bandwidth \\
  \midrule
  On chip                               & (a) &  6.34  GB/s   \\
  Off chip, on board                    & (b) &  5.46  GB/s   \\
  Off board, direct IPU link            & (c) &  4.99  GB/s   \\
  Off board, indirect IPU link          & (d) &  4.91  GB/s   \\
  \bottomrule
  \end{tabular}
  \caption{Point-to-point peak bandwidth under no load. Experiments
    labeled (a)...(d) correspond to the topologies in
    Figure~\ref{fig:point-to-point-setup}.}
  \label{tab:point-to-point-bandwidth-no-load}
\end{table}

\begin{figure}[h]
  \includegraphics[width=\columnwidth]{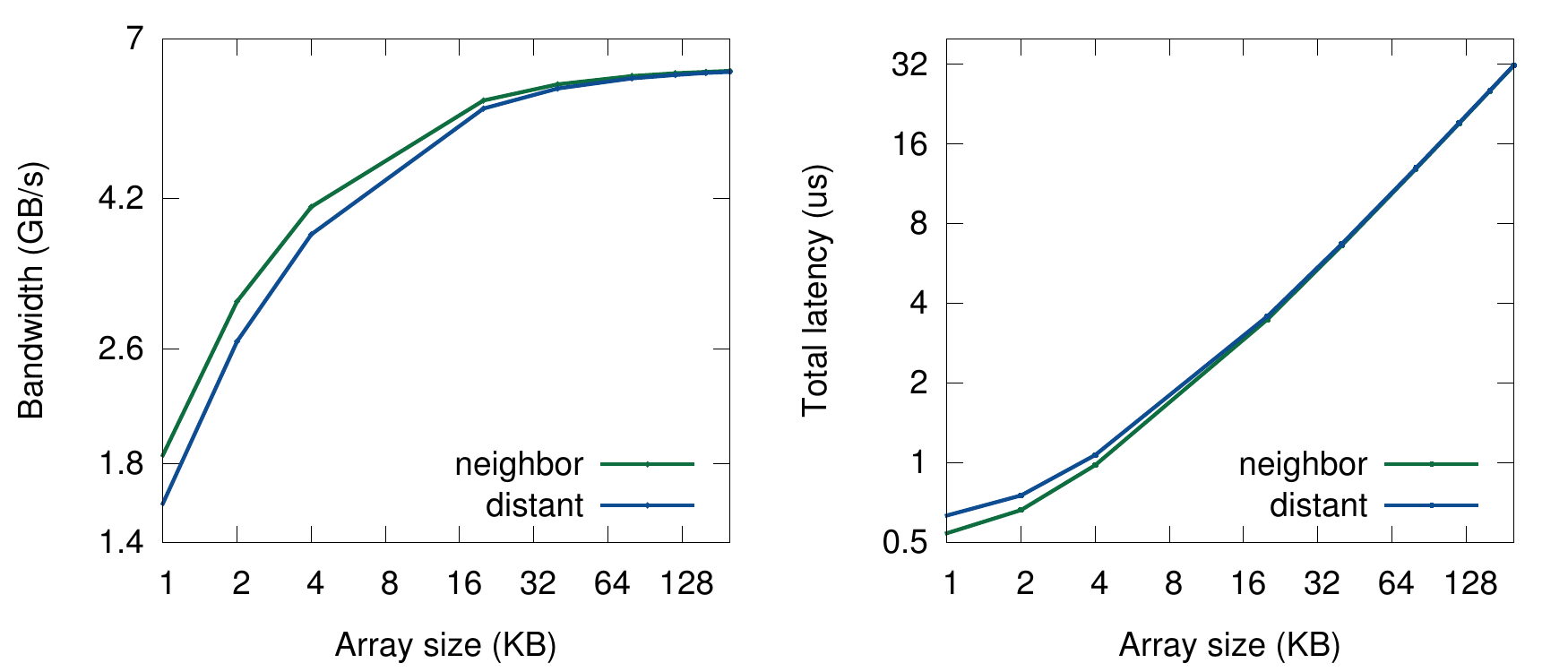}
  \caption{The effect of physical proximity on tile-to-tile transfer
    bandwidth within an IPU is negligible, especially with large
    messages. In our experiments, the chosen pair of neighboring tiles
    is (0,1) and the chosen pair of far tiles is (0,644),
    consistent with the tile enumeration of
    Figure~\ref{fig:ipu-topology}.}
  \label{fig:tile_to_tile_bandwidth_near_far}
\end{figure}

\noindent\textbf{Tile proximity on chip}. Physical proximity between
the source tile and destination tile on a chip does not affect the peak
bandwidth available between the two tiles.  We compare the bandwidth
accessible between physically near and far tiles at various block
sizes; results are in
Figure~\ref{fig:tile_to_tile_bandwidth_near_far}.  Performance is
indistinguishable for sufficiently large data blocks.  When blocks are
smaller than 32 KiB, far tile pairs suffer a slightly longer transfer
setup time.

\clearpage

\begin{table}[h]
  \center
  \small
  \begin{tabular}{lcrrr}
  \toprule
  Experiment                            &     &  Congestion-free &  Bandwidth    & Degradation \\
                                        &     &  Bandwidth       &  under load   & \\
                                        &     &  (GB/s)          &  (GB/s)       & \\
  \midrule
  On chip                               & (a) &  6.34            & 6.21          & 1.02  $\times$\\
  Off chip, on board                    & (b) &  5.46            & 0.0436        & 125.2 $\times$\\
  Off board, direct IPU link            & (c) &  4.99            & 0.0224        & 222.8 $\times$\\
  Off board, indirect IPU link          & (d) &  4.91            & 0.0224        & 219.2 $\times$\\
  \bottomrule
  \end{tabular}
  \caption{Effect of congestion on point-to-point bandwidth. This is a
    short summary that compares congestion and no-congestion results
    in conditions (a)...(d) corresponding to the previous
    section. Extended results are available in the next table.}
  \label{tab:point-to-point-bandwidth-load-short}
\end{table}

\subsection{Peak Bandwidth under Load}
\label{sec:peak-bw}

We study the peak bandwidth available to concurrent point-to-point
transfers between pairs of tiles in different topologies and under
different loads. We consider the same topologies as in the previous
sections (Figure~\ref{fig:point-to-point-setup}). We compare our
results against congestion-free ones
(Table~\ref{tab:point-to-point-bandwidth-load-short}).

\noindent\textbf{On chip.} The aggregate bandwidth available on the
on-chip interconnect scales virtually perfectly with the number of
concurrent transfers. As concurrent transfers grow in number, the
bandwidth seen by each transfer remains virtually constant at 6.3
GB/s; see the first band of
Table~\ref{tab:point-to-point-bandwidth-long}, culminating in
experiment (a).

\noindent\textbf{Off chip.} The off-chip bandwidth offered by IPU
links is lower than that offered on chip by the exchange.  An
intra-board IPU link connection offers 55 GB/s in each direction.  IPU
Link connections between different boards offer approximately half
that much bandwidth per direction, 28 GB/s, regardless of whether
the pair of IPUs is directly or indirectly connected via IPU links.
Bidirectional bandwidth under load effectively doubles the
monodirectional values.

Our extended results of Table~\ref{tab:point-to-point-bandwidth-long}
provide a finer-grained characterization of congestion.

\begin{table}[h]
  \center \footnotesize
  \begin{tabular}{clcrrrr}
  \toprule
  Scale  & Experiment                                                                    & \multicolumn{2}{c}{Concurrent} &  \multicolumn{1}{c}{Aggregate} & \multicolumn{1}{c}{Bandwidth} \\
  (IPUs) &                                                                               & \multicolumn{2}{c}{transfers}  &  \multicolumn{1}{c}{bandwidth} & \multicolumn{1}{c}{per transfer} \\
         &                                                                               &     &                          &  \multicolumn{1}{c}{(GB/s)}    & \multicolumn{1}{c}{(MB/s)} \\
  \midrule
      & \multicolumn{4}{l}{On-chip: source and destination tiles are on the same IPU} \\
      & One tile to one tile                                                             &     &     1   &      6.34   &   6,341.\phantom{00} \\
      & 2 tile to 2 tiles                                                                &     &     2   &     12.65   &   6,323.\phantom{00} \\
      & ...                                                                              &     &     4   &     25.29   &   6,323.\phantom{00} \\
      & ...                                                                              &     &     8   &     50.58   &   6,323.\phantom{00} \\
      & ...                                                                              &     &    16   &    101.06   &   6,316.\phantom{00} \\
      & ...                                                                              &     &    38   &    240.12   &   6,319.\phantom{00} \\
      & ...                                                                              &     &    76   &    480.32   &   6,320.\phantom{00} \\
  1/8 & ...                                                                              &     &   152   &    959.89   &   6,315.\phantom{00} \\
  1/4 & ... a quarter of one IPU                                                         &     &   304   &  1,919.68   &   6,315.\phantom{00} \\
  1/2 & ... half IPU                                                                     &     &   608   &  3,839.22   &   6,315.\phantom{00} \\
  1   & ... the entire IPU                                                               & (a) & 1,216   &  7,679.01   &   6,315.\phantom{00} \\
  \midrule
  2    &  \multicolumn{5}{l}{Cross-IPU experiments - 2 IPUs - monodirectional} \\
       & ... both IPUs on the same board                                                 & (b) &  1,216 &   55.00      &  45.23    \\
       & ... across boards, direct IPU link                                              & (c) &  1,216 &   27.72      &  22.79    \\
       & ... across boards, no direct IPU link                                           & (d) &  1,216 &   27.71      &  22.79    \\
  \midrule
  2    &  \multicolumn{5}{l}{Cross-IPU experiments - 2 IPUs - bidirectional} \\
       & ... both IPUs on the same board                                                 & (b*) &  2,432 &   108.09     &  44.44    \\
       & ... across boards, direct IPU link                                              & (c*) &  2,432 &   54.86      &  22.56    \\
       & ... across boards, no direct IPU link                                           & (d*) &  2,432 &   55.02      &  22.62    \\
  \midrule
     & \multicolumn{5}{l}{Cross-IPU experiments - randomized system-wide destinations - no segregation} \\
  2  & to 2 IPUs \\
  2  & ... both IPUs on the same board                                                   &     &  2,432 &   109.57     &  45.05     \\
  2  & ... across boards, direct IPU link                                                &     &  2,432 &   54.86      &  22.56     \\
  2  & ... across boards, no direct IPU link                                             &     &  2,432 &   55.02      &  22.62     \\
  4  & to 4 IPUs on two boards                                                           &     &  4,864 &   109.72     &  22.56     \\
  8  & to 8 IPUs on four boards                                                          &     &  9,728 &   111.13     &  11.42     \\
  16 & to 16 IPUs, entire system                                                         &     & 19,456 &   111.28     &  5.72      \\
  \bottomrule
  \end{tabular}
  \caption{Point-to-point peak bandwidth: bandwidth available
    to concurrent transfers under load.}
  \label{tab:point-to-point-bandwidth-long}
\end{table}

Experiments (b),(c) and (d) focus on the measurement of
monodirectional inter-IPU bandwidth in the same topologies depicted
in Figure~\ref{fig:point-to-point-setup}.  They all involve 1,216
concurrent transfers, arranged so that each tile on the source IPU is
the source of exactly one transfer directed at exactly one randomly
selected tile located on the destination IPU.

Results show that the on-board, inter-IPU monodirectional bandwidth is
roughly twice as high as across boards (55 vs 28 GB/s).

Additionally, we measure bidirectional inter-IPU bandwidth with
experiments (b*),(c*) and (d*), which are the bidirectional extension
of (b), (c), and (d), respectively.  Experiments (b*),(c*) and (d*)
all involve 2,432 transfers, of which 1,216 go from one IPU to
another and 1,216 go in the opposite direction.  Each tile on the
first IPU is involved in exactly two transfers: in one as a source
tile, and in the other as a destination tile. In both such transfers,
the other endpoint is on the other tile. The same is true for the
second IPU. All transfers are across IPUs. No transfers are IPU-local.

These experiments' results show that bidirectional aggregate bandwidth
is twice as high as the corresponding monodirectional values (108 GB/s
vs. 55, and 55 GB/s vs 28).

The bottom band in the table studies per-transfer bandwidth
degradation as the system size scales up. We instantiate Multi-IPUs
comprising growing physical IPU counts (2, 4, 8, and 16), and in each
experiment we originate exactly one transfer from each tile in the
Multi-IPU system toward a randomly selected tile in the system
(uniform distribution). Source and destination tiles may or may not be
on the same IPU or C2 board.

In a 16-IPU system, the average per-tile bandwidth degrades to
approximately 5.72 MB/s, which is 25\% of the monodirectional
aggregate per-tile bandwidth available when the system involves a pair
of IPUs (not on the same board).

%

\subsection{Peak Bandwidth between IPUs by Proximity}

\begin{figure}
  \includegraphics[width=\columnwidth]{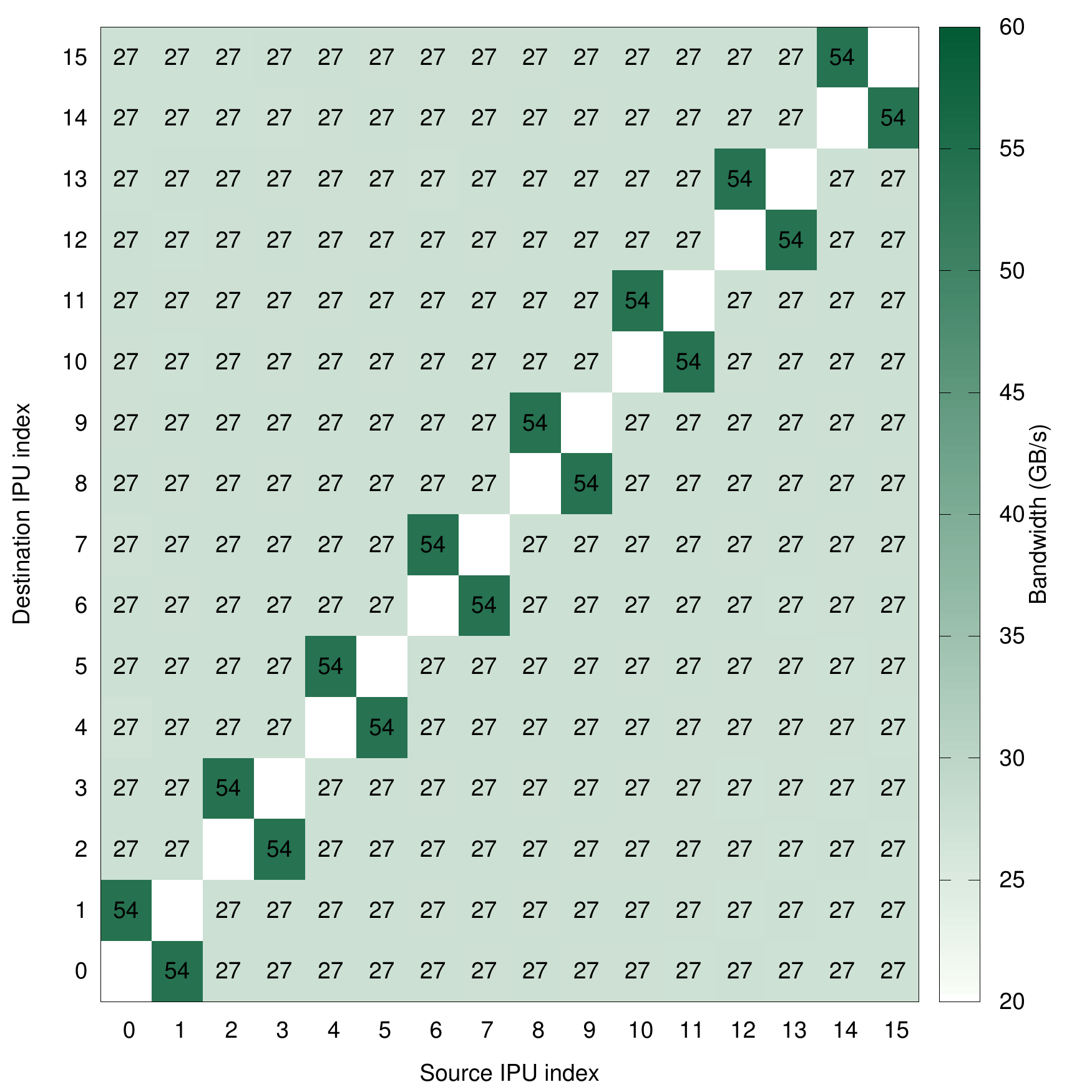}
  \caption{Peak monodirectional bandwidth between pairs of IPUs, in
    congestion-free conditions. IPUs are numbered according to their
    \emph{DNC IDs}, as discussed in
    Section~\ref{sec:multi-ipus}.}
  \label{fig:ipu-bandwidth-topo}
\end{figure}

We study how IPU proximity in a multi-IPU system affects
monodirectional peak bandwidth in congestion-free conditions.  We
find that it does not---a randomly selected pair of IPUs not located on the same
board can transfer data at 27 GB/s (monodirectional) regardless of
their respective position in the ladder network
(Figure~\ref{fig:ipu-bandwidth-topo}).

\noindent\textbf{Benchmark.} Our benchmark considers all
(source,destination) pairs of IPUs.  We number IPUs according to their
DNC IDs. In each experiment, we create exactly 1,216 transfers.  Each
tile in the source IPU transfers a 4 KiB message to a distinct,
randomly selected tile on the destination IPU. No other transfers are
occurring in the system.

\begin{table}[b]
  \center
  \scriptsize
  \begin{tabular}{lrrrrrrrrrrrrrr}
  \toprule
                                                     & \multicolumn{6}{c}{Aggregate bandwidth (GB/s)}                         \\
  Experiment                                         & 1 thread & 2 threads & 3 threads & 4 threads & 5 threads & 6 threads   \\
  \midrule
  \multicolumn{6}{l}{On-chip: source and destination tiles are on the same IPU} \\
   One tile to one tile                               & 6.14      & 6.13      & 6.13      & 6.13      & 6.13      & 6.13      \\
   2 tiles to 2 tiles                                 & 12.22     & 12.27     & 12.24     & 12.26     & 12.07     & 12.29     \\
   4 tiles to 4 tiles                                 & 24.45     & 24.52     & 24.48     & 24.54     & 24.51     & 24.49     \\
   8 tiles to 8 tiles                                 & 48.96     & 48.95     & 48.92     & 48.94     & 48.92     & 48.93     \\
   16 tiles to 16 tiles                               & 97.85     & 97.86     & 97.87     & 97.94     & 97.89     & 97.76     \\
   38 tiles to 38 tiles                               & 232.54    & 232.44    & 232.37    & 231.96    & 232.03    & 232.15    \\
   76 tiles to 76 tiles                               & 464.62    & 464.59    & 464.32    & 464.76    & 464.60    & 464.49    \\
   152 tiles to 152 tiles                             & 928.64    & 927.85    & 928.49    & 928.49    & 928.41    & 928.19    \\
   304 tiles to 304 tiles                             & 1,857.74   & 1,855.70   & 1,856.45   & 1,856.45   & 1,857.36   & 1,856.30   \\
   608 tiles to 608 tiles                             & 3,713.81   & 3,712.90   & 3,712.75   & 3,712.45   & 3,712.90   & 3,702.61   \\
   1,216 tiles to 1,216 tiles                         & 7,425.80   & 7,427.92   & 7,425.80   & 7,425.80   & 7,426.41   & 7,428.22   \\
  \midrule
  \multicolumn{5}{l}{Cross-IPU experiments - 2 IPUs - monodirectional} \\
   ... both IPUs on the same board \hspace{-8ex}      & 54.99 & 54.99 & 54.99 & 54.90 & 54.88 & 54.91\\
   ... across boards, direct IPU link\hspace{-8ex}    & 27.71 & 27.71 & 27.71 & 27.67 & 27.66 & 27.67\\
   ... across boards, no direct IPU link\hspace{-6ex} & 27.71 & 27.71 & 27.71 & 27.66 & 27.63 & 27.66\\
  \bottomrule
  \end{tabular}
  \caption{Multi-threaded point-to-point peak bandwidth: the use of
    different threads per tile does not affect the bandwidth available
    to concurrent transfers.}
    \label{tab:point-to-point-bandwidth-long-thread-impact}
\end{table}

\subsection{Multi-threaded Peak Bandwidth}

We study the effect of multi-threading on the peak bandwidth available
on-chip and off-chip. We find the following:
\begin{itemize}
\item a single thread is sufficient to achieve full bandwidth;
\item the use of multiple threads is not necessary to achieve peak values;
\item if used (potentially for other purposes), multi-threading will
  not cause any bandwidth degradation.
\end{itemize}

This investigation is motivated by the fact that tiles support
executing instructions from up to six concurrent thread contexts in
an SMT-like fashion. It is legitimate for a software designer to
wonder whether the use of multiple threads to initiate concurrent
transfers will achieve a higher aggregate bandwidth.

Results show that the use of additional threads causes no material
change in performance, neither on chip nor across chips.

\clearpage
\begin{figure}
  \hspace{-1cm}\includegraphics[width=0.9\columnwidth, trim=0cm 3.5cm 5cm 0cm]{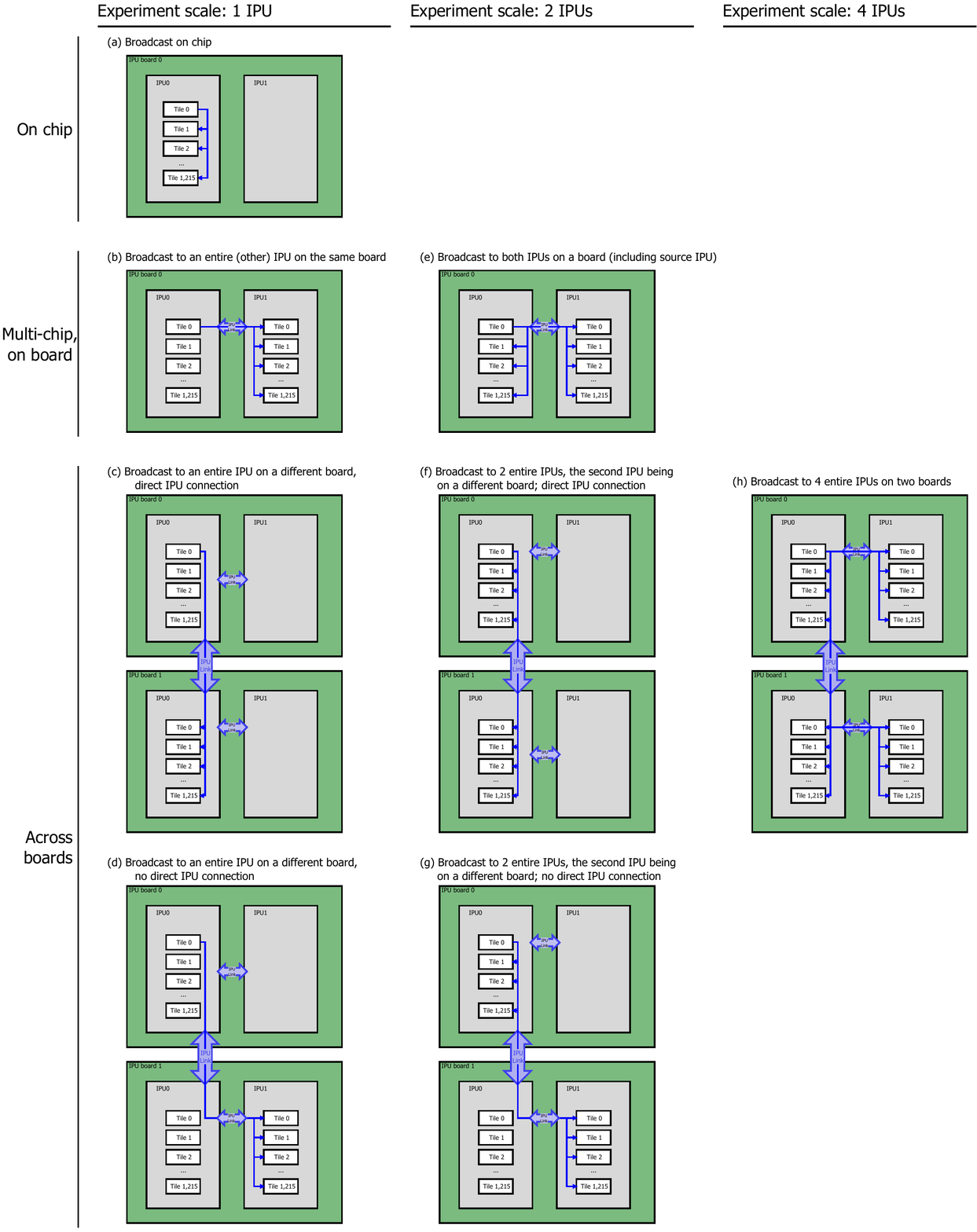}
  \caption{Topologies we benchmark in our broadcast and scatter
    experiments. Different topologies have different connectivity
    between source and destination tiles and perform
    differently.  We depict data flows from the source to the
    destination tiles with blue arrows.  Reversing
    all blue arrows depicts the data flows in the gather
    experiments. }
  \label{fig:broadcast-setup}
\end{figure}
\thispagestyle{empty}
\clearpage

\section{Broadcast}
\label{sec:broadcast}

This is the first section in this chapter dedicated to studying the
performance of collective operations. Specifically, this section
focuses on latency and bandwidth available to broadcast operations at
different scales and for different topologies.

\noindent\textbf{Broadcast.} In a broadcast operation, one tile sends
the same message to multiple destination tiles. Each destination tile
receives an identical copy of the data. The source tile maintains a
source buffer available to the operation, in local memory, for the
entire duration of the operation. Correspondingly, each destination
tile maintains a local destination buffer available to the
operation. In benchmarks involving whole IPUs, the source tile is also
among the destinations, and maintains both a source and a destination
buffer in its local memory.  This limits the largest block size
available for this benchmark to approximately 100 KiB.

\noindent\textbf{Scale.} By scale we mean the destination tile count,
or the IPU count if whole IPUs are involved
(Table~\ref{fig:broadcast-setup}). We vary scale from one tile to the
entire test system, which includes 16 IPUs on 8 boards.  When using 2
IPUs, we study the impact of topology and source-destinations
proximity: we place the source and destinations on the same board
(experiment (b)) and on different boards. We consider pairs of boards
directly or indirectly connected with IPU Links (experiments (c) and
(d), respectively).

\noindent\textbf{No load.} Experiments in this section study load
conditions in which no other operation is in progress. Each transfer's
performance is affected only by the load caused by the remainder of
the collective operation.

\noindent\textbf{Topologies.} We benchmark the topologies of
Figure~\ref{fig:broadcast-setup}, where thin blue arrows depict the
data flows from the source to the destination tiles. Experiments at
different scales are illustrated in different columns of the figure:
subfigures in the first column illustrate experiments where the
destination tiles belong to 1 IPU; the second column illustrates
experiments where destination tiles span 2 IPUs; and the third column
illustrates experiments where the destination tiles span 4 IPUs. We
omit experiments involving 8 and 16 IPUs for brevity.

In the figure, experiments on different rows involve different
portions of the interconnect. In the first row, only the on-chip
interconnect is involved. Experiments in the second row also involve
the IPU Link connection between two IPUs located on the same
board. Experiments in the lower rows involve IPU Links across multiple
boards.

Experiments (b) and (e) study the cost of performing a broadcast
between IPUs when both IPUs reside in the same tile. The difference
between (b) and (e) is that in (e), the tiles on the first IPU are
also broadcast destinations, while in (b), no tiles of the first IPU
are destinations.

The same difference appears between experiments (c) and (f). Again,
the same difference appears between experiments (d) and (g). This is
symbolized in the picture by additional blue arrowheads in (f) and (g)
that point to tiles on the source IPU.

Experiments (c) and (f) involve 2 IPUs that are not on the same board
but are connected directly via IPU Links.  In contrast, experiments
(d) and (g) involve 2 IPUs that are not connected directly via IPU
Links.


\subsection{Congestion-free Broadcast Latency}

We focus first on the minimum latency associated with a broadcast
operation. To do that, we transfer a message of minimum size, i.e., a
word of 32 bits.  We vary the experiment's scale from one tile to the
whole system.  We find that the system displays remarkable
scalability, with a 16-IPU broadcast taking less than 2
microseconds. Our results are in Table~\ref{tab:broadcast-latency}.

\begin{table}[h]
  \center
  \small
  \begin{tabular}{clcrrrr}
  \toprule
  Scale  & Experiment                               & \multicolumn{2}{c}{Destination} & Total         \\ 
  (IPUs) &                                          & \multicolumn{2}{c}{tile count}  & latency       \\ 
         &                                          &     &                           & \hspace{10ex} \\ 
  \midrule
         & Transfer to self (via local memory)             &     &         1 & 0.012 $\upmu$s  \\ 
  \midrule
        & \multicolumn{4}{l}{On-chip: source and destination tiles are on the same IPU} \\
        & One tile to $n$ tiles                            &     &         1 & 0.094 $\upmu$s  \\ 
        & ...                                              &     &         2 & 0.094 $\upmu$s  \\ 
        & ...                                              &     &         4 & 0.094 $\upmu$s  \\ 
        & ...                                              &     &         8 & 0.094 $\upmu$s  \\ 
        & ...                                              &     &        16 & 0.094 $\upmu$s  \\ 
        & ...                                              &     &        38 & 0.134 $\upmu$s  \\ 
        & ...                                              &     &        76 & 0.136 $\upmu$s  \\ 
   1/8  & ...                                              &     &       152 & 0.142 $\upmu$s  \\ 
   1/4  & ... a quarter of one IPU                         &     &       304 & 0.153 $\upmu$s  \\ 
   1/2  & ... half IPU                                     &     &       608 & 0.176 $\upmu$s  \\ 
   1    & ... the entire IPU                               & (a) &     1,216 & 0.194 $\upmu$s  \\ 
  \midrule
  1  & \multicolumn{4}{l}{to an entire IPU, different than that of source tile} \\
     & ... both IPUs on the same board                  & (b) &     1,216 & 0.747 $\upmu$s  \\ 
     & ... across boards, direct IPU link               & (c) &     1,216 & 0.637 $\upmu$s  \\ 
     & ... across boards, no direct IPU link            & (d) &     1,216 & 0.896 $\upmu$s  \\ 
  \midrule
  2  & to 2 IPUs \\
     & ... both IPUs on the same board                  & (e) &     2,432 & 0.747 $\upmu$s  \\ 
     & ... across boards, direct IPU link               & (f) &     2,432 & 0.637 $\upmu$s  \\ 
     & ... across boards, no direct IPU link            & (g) &     2,432 & 0.896 $\upmu$s  \\ 
   \midrule
  4  & to 4 IPUs on two boards                         & (h) &     4,864 & 0.900 $\upmu$s  \\ 
  8  & to 8 IPUs on four boards                        & (i) &     9,728 & 1.231 $\upmu$s  \\ 
  16 & to 16 IPUs, entire system                       & (j) &    19,456 & 1.921 $\upmu$s  \\ 
  \bottomrule
  \end{tabular}
  \caption{Broadcast minimum latency: latency necessary to
    broadcast one value from one tile to a set of tiles. We study
    destination sets of varying size and location.  System scale and
    respective location of source and destination tiles affect
    latency.  Experiments labeled (a)-(j) benchmark the topologies of
    Figure~\ref{fig:broadcast-setup}.}
  \label{tab:broadcast-latency}
\end{table}

The first experiment in the table (not labeled) is a mere transfer
from a tile to itself.  It results in a local memory copy that doesn't
involve the exchange or IPU links. We report it (12 nanoseconds)
only for reference; the reader can subtract this value from subsequent
result values to separate the local and interconnect contributions to
latency.

In the experiments described in the following rows, the set of
destination tiles grows till it reaches the entire IPU.

One tile can broadcast one word to the entirety of its IPU in less
than 0.2 microseconds. The additional penalty to perform an off-chip
broadcast involving 2 IPUs is approximately 0.5 microseconds. The
total latency of a broadcast targeting all 16 IPUs in the test system
is below 2 microseconds.

\begin{figure}
  \includegraphics[width=0.49\columnwidth, trim=2mm 0mm 3mm 0mm]{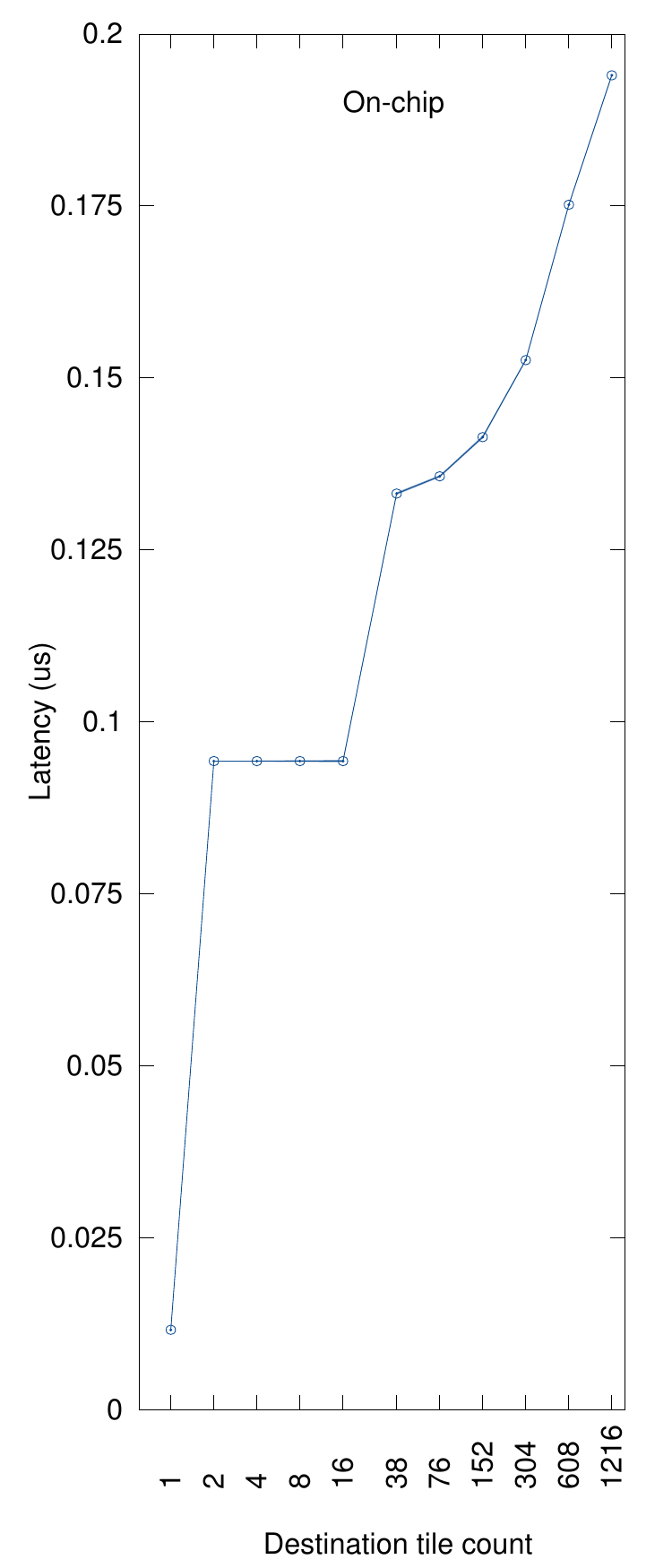}
  \includegraphics[width=0.49\columnwidth, trim=2mm 0mm 3mm 0mm]{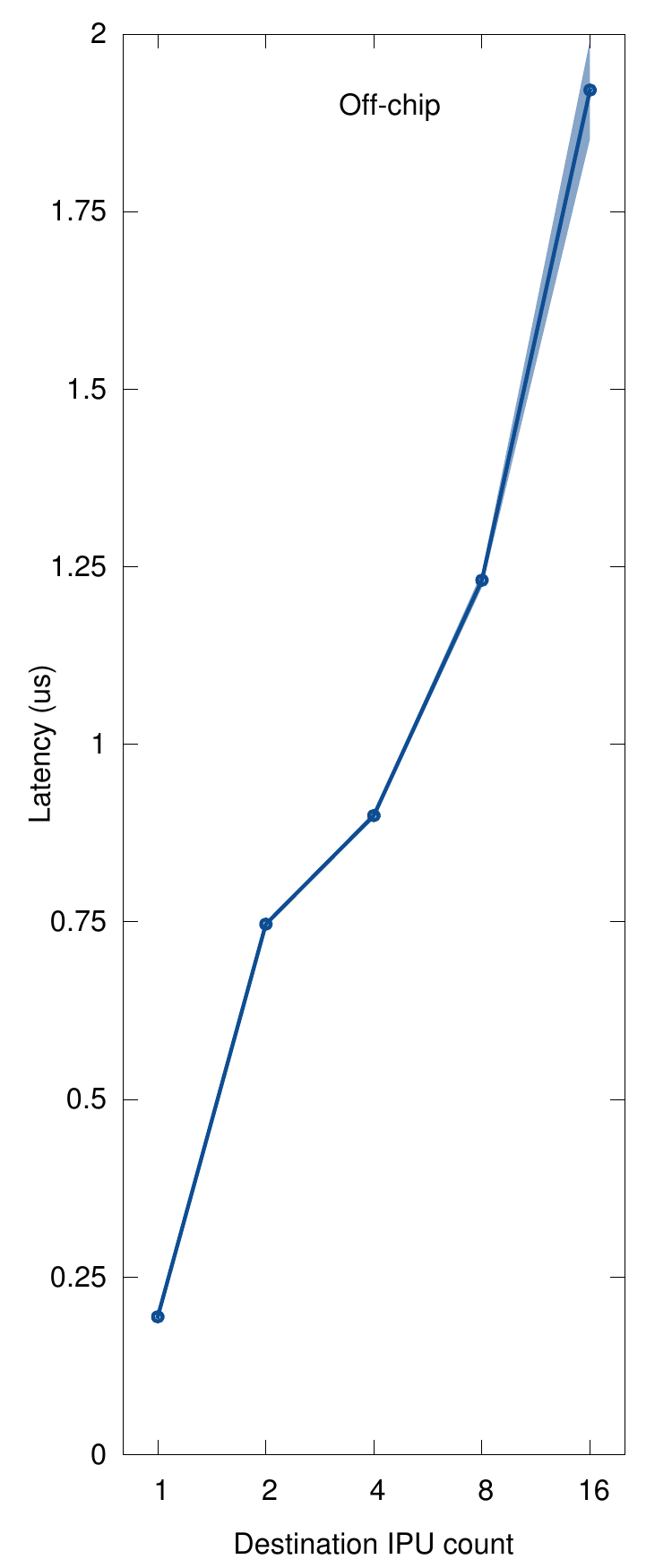}
  \caption{Broadcast latency scaling. Left: scaling within an IPU (on
    chip). Right: scaling and across multiple IPUs (off chip).}
  \label{fig:broadcast-latency-scaling}
\end{figure}

\noindent\textbf{Observations:}
\begin{itemize}
\item the latency of a broadcast operation involving the entire chip
  roughly grows logarithmically with tile count
  (Figure~\ref{fig:broadcast-latency-scaling}, left);
\item similarly, the latency of a broadcast operation spanning across
  multiple IPUs roughly grows logarithmically with IPU count
  (Figure~\ref{fig:broadcast-latency-scaling}, right);
\item performing a broadcast across 2 IPUs takes approximately 0.7
  microseconds; that is approximately 0.5 microseconds more expensive
  than a single-IPU broadcast;
\item the cost of performing an off-chip broadcast is dominated by the
  cost of traversing an IPU Link; the latency we measure in
  experiments (b) and (e) is identical.  This suggests that in
  experiment (e), the on-chip broadcast involving the source IPU
  occurs in parallel with the sequence of operations comprising the
  transfer to across the IPU link followed by the on-chip broadcast on
  the destination IPU.  The latter sequence takes longer than the
  local broadcast and determines the overall latency.  Intuitively,
  once the penalty to reach remote tiles across the IPU Link is paid,
  the marginal cost of also reaching local tiles is null.
\item same considerations apply between pair of experiments (c,f), and
  pair (d,g);
\item unexpectedly, a broadcast between two IPUs on the same board (e)
  takes marginally more time than between the IPUs located on
  different boards (f,g). This result is consistent with our gather
  results (see Section~\ref{sec:gather-latency}).
\end{itemize}


\subsection{Peak Broadcast Bandwidth}

In this section we study the peak bandwidth available to broadcast
operations of different scales. As in the previous section, we study
the effect of source-destination IPU proximity in the topologies
(a)...(j) of Figure~\ref{fig:broadcast-setup}. Our results are in
Table~\ref{tab:broadcast-bandwidth}.

The message size we employ in these experiments is 100 KiB. It is
sufficiently large to saturate the aggregate bandwidth per tile, and
at the same time, it is close enough to the maximum size that allows
for two copies (source and destination buffer) to exist simultaneously
in each tile's local memory (248 KiB are available to the user).
Fitting both buffers in local tile memory allows us to run experiments
where the source tile is also among the destinations.

No other traffic is occurring in the system except for that created by
the broadcast operation itself.


\begin{table}[h]
  \center
  \footnotesize
  \begin{tabular}{clcrrrr}
    \toprule
    Scale  & Experiment                                                     & \multicolumn{2}{c}{Concurrent} &  \multicolumn{1}{c}{Aggregate} & \multicolumn{1}{c}{Bandwidth} \\
    (IPUs) &                                                                & \multicolumn{2}{c}{transfers}  &  \multicolumn{1}{c}{bandwidth} & \multicolumn{1}{c}{per transfer} \\
           &                                                                &                 &              &  \multicolumn{1}{c}{(GB/s)}    & \multicolumn{1}{c}{(GB/s)} \\
    \midrule
      & \multicolumn{4}{l}{On-chip: source and destination tiles are on the same IPU} \\
      & Point to point                                                                   &     &     1   &      6.35  &  6.35   \\
      & One tile ... to 2 tiles                                                          &     &     2   &     12.70  &  6.35   \\
      & ... to 4 tiles                                                                   &     &     4   &     50.49  &  12.62  \\
      & ...                                                                              &     &     8   &    100.92  &  12.61  \\
      & ...                                                                              &     &    16   &    201.58  &  12.60  \\
      & ...                                                                              &     &    38   &    477.10  &  12.56  \\
      & ...                                                                              &     &    76   &    954.17  &  12.55  \\
  1/8 & ...                                                                              &     &   152   &  1,906.51  &  12.54  \\
  1/4 & ... to a quarter of one IPU                                                      &     &   304   &  3,808.50  &  12.53  \\
  1/2 & ... to half IPU                                                                  &     &   608   &  7,595.88  &  12.49  \\
  1   & ... to the entire IPU (same as source tile)                                      & (a) & 1,216   & 15,134.80  &  12.46  \\
  \midrule
  1  & \multicolumn{4}{l}{to an entire IPU, different than that of source tile} \\
     & ... both IPUs on the same board                                                   & (b) &  1,216 &    6,871  &  5.65  \\
     & ... across boards, direct IPU link                                                & (c) &  1,216 &    6,433  &  5.29  \\
     & ... across boards, no direct IPU link                                             & (d) &  1,216 &    6,427  &  5.29  \\
  \midrule
  2  & to 2 IPUs \\
     & ... both IPUs on the same board                                                   & (e) &  2,432 &    9,440  &  3.88  \\
     & ... across boards, direct IPU link                                                & (f) &  2,432 &    9,051  &  3.72  \\
     & ... across boards, no direct IPU link                                             & (g) &  2,432 &    9,034  &  3.71  \\
  \midrule
  4  & to 4 IPUs on two boards                                                           & (h) &  4,864 &   25,437  &  5.23  \\
  8  & to 8 IPUs on four boards                                                          & (i) &  9,728 &   36,756  &  3.78  \\
  16 & to 16 IPUs, entire system                                                         & (j) & 19,456 &   47,343  &  2.43  \\
  \bottomrule
  \end{tabular}
  \caption{Peak bandwidth available to broadcast operations of varying
    scale. Experiments labeled (a)-(j) correspond to the topologies
    illustrated in Figure~\ref{fig:broadcast-setup}; we describe them
    in detail.}
  \label{tab:broadcast-bandwidth}
\end{table}

\noindent\textbf{On-chip performance.}  As the scale of the broadcast
increases within an IPU, the bandwidth per tile achieved by the
operation saturates around 12 GB/s.  12 GB/s seems to be the on-chip
average per-transfer peak. Saturation occurs with small destination
sets: a destination set of 4 tiles is sufficient to achieve peak
per-tile bandwidth.

\noindent\textbf{Off-chip performance.} A direct comparison between
experiments (a) and (b), that have identical destination count,
reveals the penalty for extending the broadcast off chip: a drop in
average per-transfer bandwidth from 12.4 to 5.6 GB/s. The performance
difference between on-board and inter-board communication is
negligible.

\noindent\textbf{Scaling.} The results of experiments (h)-(j) show
that larger system scales benefit from a monotonically increasing
aggregate broadcast bandwidth. The growth trend is, however, not
linear, and not trivially explained.

\begin{table}[b]
  \center
  \footnotesize
  \begin{tabular}{clcrrrr}
    \toprule
    Scale  & Experiment                                                                  & \multicolumn{2}{c}{\% peak}     &  \multicolumn{1}{c}{90\% peak}  & \multicolumn{1}{c}{Peak} \\
    (IPUs) &                                                                             & \multicolumn{2}{c}{at 1}        &  \multicolumn{1}{c}{message}    & \multicolumn{1}{c}{bandwidth} \\
    &                                                                                    & \multicolumn{2}{c}{KiB }        &  \multicolumn{1}{c}{size}       & \multicolumn{1}{c}{(GiB/s)} \\
    &                                                                                    & \multicolumn{2}{c}{message}     &  \multicolumn{1}{c}{(KiB)}      &\\
    \midrule
    1   & ... the entire IPU (same as source tile)                                         & (a) &         30.3 \%           &        13.2                 & 15,134  \\
  \midrule
    1  & \multicolumn{4}{l}{to an entire IPU, different than that of source tile} \\
       & ... both IPUs on the same board                                                   & (b) &         18.9 \%           &          5.7                &  6,871  \\
       & ... across boards, direct IPU link                                                & (c) &         22.1 \%           &          5.4                &  6,433  \\
       & ... across boards, no direct IPU link                                             & (d) &         16.4 \%           &          4.8                &  6,427  \\
  \midrule
    2  & to 2 IPUs \\
       & ... both IPUs on the same board                                                   & (e) &         25.2 \%           &          7.3                &  9,440  \\
       & ... across boards, direct IPU link                                                & (f) &         31.4 \%           &          7.5                &  9,051  \\
       & ... across boards, no direct IPU link                                             & (g) &         29.5 \%           &          7.4                &  9,034  \\
  \midrule
    4  & to 4 IPUs on two boards                                                           & (h) &         16.5 \%           &         22.2                & 25,437  \\
    8  & to 8 IPUs on four boards                                                          & (i) &         15.2 \%           &         33.6                & 36,756  \\
    16 & to 16 IPUs, entire system                                                         & (j) &         14.6 \%           &         41.2                & 47,343  \\
  \bottomrule
  \end{tabular}
  \caption{Summary of aggregate broadcast bandwidth below peak.}
  \label{tab:broadcast-bandwidth-fraction}
\end{table}

\begin{figure}
  \center
  \includegraphics[width=\columnwidth]{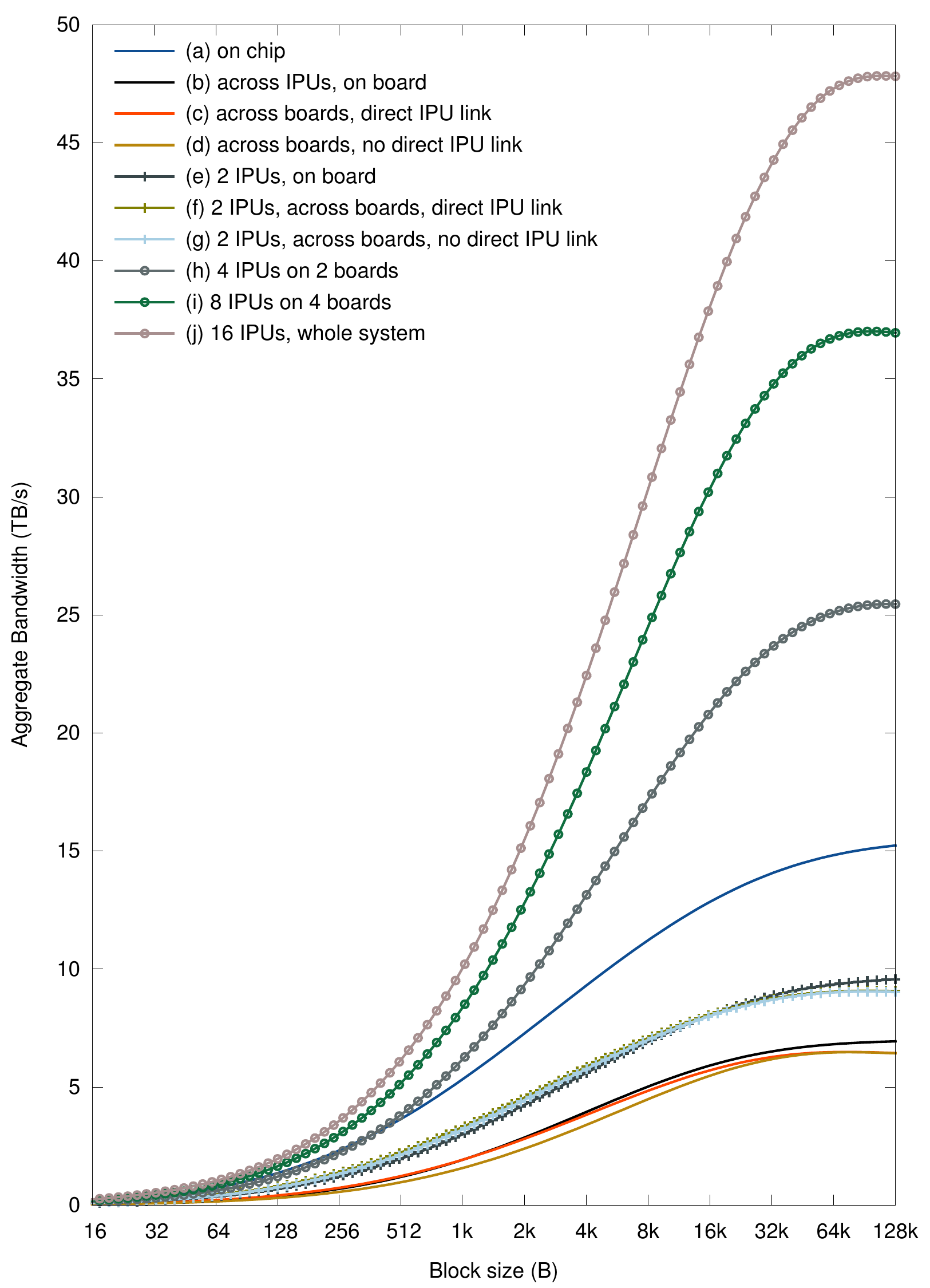}
  \caption{Impact of block size on aggregate broadcast bandwidth. Each line
    represents a different experiment topology (a)-(j) as illustrated in
    Figure~\ref{fig:broadcast-setup}.}
  \label{fig:broadcast-bandwidth-aggregate-blocksize}
\end{figure}

\begin{figure}
  \center
  \includegraphics[width=\columnwidth]{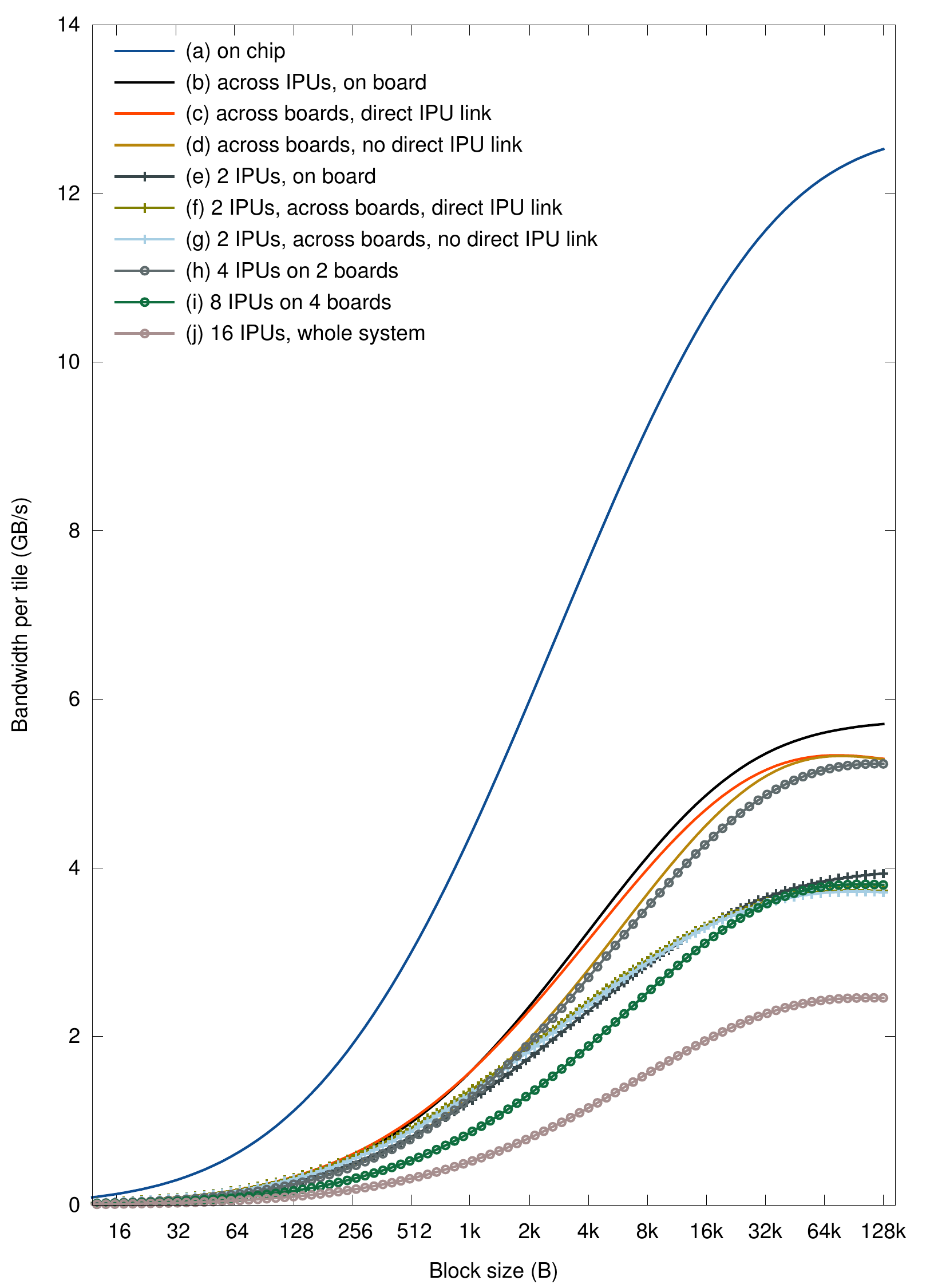}
  \caption{Impact of block size on broadcast bandwidth per tile. Each line
    represents a different experiment topology (a)-(j) as illustrated in
    Figure~\ref{fig:broadcast-setup}.}
  \label{fig:broadcast-bandwidth-pertile-blocksize}
\end{figure}

\subsection{Effect of Message Size on Broadcast Bandwidth}
\label{sec:bw-as-function-of-block-size}

We study broadcast bandwidth performance below peak, i.e., when
smaller messages are transferred.  We find that smaller messages
achieve a fraction of peak bandwidth depending on system scale
(e.g., 1 KiB messages achieve 15-30\% peak bandwidth).
Moreover, as system scale grows,
larger message sizes are needed to achieve close-to-peak
performance. These observations are not surprising and apply similarly
to most parallel systems. We summarize them quantitatively in
Table~\ref{tab:broadcast-bandwidth-fraction}.  We also chart in detail
the effects of message size on broadcast bandwidth in
Figures~\ref{fig:broadcast-bandwidth-aggregate-blocksize} and
\ref{fig:broadcast-bandwidth-pertile-blocksize}. The charts' data refer to
the same scales (1-16 IPUs) and experimental topologies (a)-(j)
already discussed.

In almost all experimental conditions, bandwidth saturates smoothly,
increasing monotonically with block size and suffering no degradation.
Minor exceptions are topologies (c) and (d) (broadcasting to a whole
IPU across boards), where block sizes larger than 64 KiB seem to
experience slightly lower bandwidths than at 64 KiB.

IPU systems behave well at scale, with bandwidth growing smoothly with
scale as the number IPUs increases to 2, 4, 8, and 16 (experiments
(g),(h),(i) and (j) respectively). We see no saturation at scale, and
larger scales benefit from higher aggregate bandwidths. Per-tile
bandwidths decrease smoothly, roughly with the logarithm of the system
size.


\section{Gather}

In this section we study the latency and bandwidth performance
available to gather operations.

\noindent\textbf{Gather.} For the complete avoidance of doubt, a
gather is a collective operation in which one tile receives one
message from multiple source tiles. The destination tile receives a
distinct message from each source, and must maintain a destination
buffer available to the operation, in its local memory, as large as
the product of message size and source count. Each source only needs
to maintain one destination buffer as large as the message size. In
benchmarks involving whole IPUs, the source tile is also among the
destinations, and must maintain in its local memory one source buffer
plus multiple destination buffers.

\noindent\textbf{Scale.} We benchmark the same scales as in our
broadcast experiments described in Section~\ref{sec:broadcast}.

\noindent\textbf{No load.} Experiments in this section study load
conditions in which no other operation is in progress. Each transfer's
performance is affected only by the load caused by the remainder of
the collective operation.

\noindent\textbf{Topologies.} We benchmark the topologies discussed in
the previous section and illustrated in
Figure~\ref{fig:broadcast-setup}, with the notable difference that
data flows are \textbf{reversed} with respect to those of a broadcast
operation, depicted in the figure and described in
Section~\ref{sec:broadcast}.

\noindent\textbf{Message size.}  The destination tile must dedicate to
the operation one buffer per source tile in its local memory.  This
constraints limits the maximum message size usable in a gather
operation. In turn, the limited message size is the primary factor
limiting the aggregate bandwidth available to the operation. A
complete discussion follows.

In benchmarks involving whole IPUs, the destination tile is also among
the sources, and maintains at the same time one output buffer and
multiple input buffers in its local memory.

\subsection{Congestion-free Gather Latency}
\label{sec:gather-latency}

\begin{table}[t]
  \center
  \footnotesize
  \begin{tabular}{clcrrrr}
    \toprule
  Scale  &  Experiment                               & \multicolumn{2}{c}{Source} & \multicolumn{1}{c}{Total}      & \multicolumn{1}{c}{Avg. latency} \\
  \,     &                                           & \multicolumn{2}{c}{tile}   & \multicolumn{1}{c}{latency}    & \multicolumn{1}{c}{per transfer} \\
  \,     &                                           & \multicolumn{2}{c}{count}  & \multicolumn{1}{c}{($\upmu$s)} & \multicolumn{1}{c}{(ns)} \\
  \midrule
         & Transfer to self (via local memory)         &      &         1      &  0.012  & 11.557  \\
  \midrule
       & \multicolumn{4}{l}{On-chip: source and destination tiles are on the same IPU} \\
       &   ... from 2 tiles                            &      &         2      &  0.094  &  47.142 \\
       &   ...                                         &      &         4      &  0.094  &  23.567 \\
       &   ...                                         &      &         8      &  0.094  &  11.784 \\
       &   ...                                         &      &        16      &  0.094  &  5.892 \\
       &   ...                                         &      &        38      &  0.097  &  2.563 \\
       &   ...                                         &      &        76      &  0.121  &  1.595 \\
  1/8  &   ...                                         &      &       152      &  0.169  &  1.111 \\
  1/4  &   ... from a quarter of one IPU               &      &       304      &  0.264  &  0.869 \\
  1/2  &   ... from half IPU                           &      &       608      &  0.455  &  0.748 \\
  1    &   ... from the entire IPU                     & (a)  &     1,216      &  0.835  &  0.687 \\
   \midrule
  1  & \multicolumn{4}{l}{from an entire IPU to a destination tile located on a different IPU} \\
     & ... both IPUs on the same board                 & (b) &      1,216      &  4.929  &  4.053 \\
     & ... across boards, direct IPU link              & (c) &      1,216      &  3.205  &  2.636 \\
     & ... across boards, no direct IPU link           & (d) &      1,216      &  6.693  &  5.504 \\
 \midrule
  \multicolumn{6}{l}{Multi-IPU experiments} \\
   2   & IPUs on the same board                          & (e) &     2,432      &  5.708 &  2.347 \\
   2   & IPUs on different boards, direct IPU link       & (f) &     2,432      &  3.983 &  1.638 \\
   2   & IPUs on different boards, no direct IPU link    & (g) &     2,432      &  7.472 &  3.072 \\
   4   & IPUs on two boards                              & (h) &     4,864      &  8.995 &  1.849 \\
   8   & IPUs on four boards                             & (i) &     9,728      & 12.377 &  1.272 \\
  16   & IPUs, entire system                             & (j) &    19,456      & 25.159 &  1.293 \\
  \bottomrule
  \end{tabular}
  \caption{Minimum gather latency in experiments of varying scale
    and topology. The topology of experiments (a)-(j) is as
    illustrated in Figure~\ref{fig:broadcast-setup}, but with reversed
    data flow directions.}
  \label{tab:gather-latency}
\end{table}

We study the minimum latency associated with a gather operation. To do
that, we transfer a message of minimum size, i.e., a word of 32 bits.
A gather involving an entire IPU completes in 0.8 microseconds.
Gather latencies in Multi-IPUs of increasing size grow sub-linearly,
with 16 IPUs completing a whole-system gather in 25 microseconds.  Our
results are in Table~\ref{tab:gather-latency}.

As we did for the broadcast operation, we first report the latency (12
nanoseconds) of a trivial gather operation involving the same tile as source
and destination. It results in a local memory copy, and we only report
it for comparison and latency breakdown.

\noindent\textbf{Latency off chip.} The direct comparison between
experiments that involve a pair of IPUs, connected directly vs.
indirectly by IPU links, reveal that gather latency depends directly
on the number of hops between source and destination IPUs in the ladder
network.

Specifically, experiment (d) results in approximately twice
as much latency as experiment (c).  Similarly, experiment (g) exhibits
roughly twice as much latency as (f).  This is intuitively consistent
with the fact that experiments (g) and (d) involve two hops in the IPU
link network, whereas experiments (c) and (f) only involve one hop
(see Figure~\ref{fig:broadcast-setup}, but assume the reverse data
flow direction).

A gather operation involving two IPUs on the same board exhibits
slightly longer latency when the IPUs are across boards and directly
connected.  This is evident by direct comparison between (c) and (b)
in the table and also by comparison between (f) and (e).  This result
is consistent with our results for the broadcast and scatter
operations, which we report in Tables~\ref{tab:broadcast-latency} and
\ref{tab:scatter-latency}, respectively.

\noindent\textbf{Latency at scale.}  Results show that a doubling of system
size roughly corresponds to a doubling of network diameter, which
causes in turn an approximate doubling of total latency.

\subsection{Peak Gather Bandwidth}

\begin{table}[tbh]
  \center \scriptsize
  \begin{tabular}{clcrrrr}
  \toprule
   Scale  & Experiment                                         & \multicolumn{2}{c}{Source    } &  Message &  Aggregate   &  Bandwidth \\
   \,     &                                                    & \multicolumn{2}{c}{tile}       &  size    &  bandwidth   &  per transfer \\
   \,     &                                                    & \multicolumn{2}{c}{count}      &  (bytes) &  (GB/s)      &  (MB/s) \\
  \midrule
         & Transfer to self (via local memory)               &      &       1     &  160  &  0.571   &  570.61  \\
  \midrule
        & \multicolumn{5}{l}{On-chip: source and destination tiles are on the same IPU} \\
        & To one tile ... from 2 tiles                       &      &       2     &  160  &  0.580   &  290.07  \\
        & ... from 4 tiles                                   &      &       4     &  160  &  1.258   &  314.47  \\
        & ...                                                &      &       8     &  160  &  2.105   &  263.16  \\
        & ...                                                &      &      16     &  160  &  3.148   &  196.75  \\
        & ...                                                &      &      38     &  160  &  4.465   &  117.51  \\
        & ...                                                &      &      76     &  160  &  5.260   &   69.22  \\
   1/8  & ...                                                &      &     152     &  160  &  5.759   &   37.89  \\
   1/4  & ... from a quarter of one IPU                      &      &     304     &  160  &  6.063   &   19.94  \\
   1/2  & ... from half IPU                                  &      &     608     &  160  &  6.219   &   10.23  \\
   1    & ... from the entire IPU                            & (a)  &   1,216     &  160  &  6.303   &    5.18  \\
  \midrule
   1    & \multicolumn{5}{l}{from an entire IPU, different than that of the destination tile} \\
        & ... both IPUs on the same board            & (b)  &   1,216     &  80   &  4.815   &    3.96   \\
        & ... across boards, direct IPU link         & (c)  &   1,216     &  80   &  4.805   &    3.95   \\
        & ... across boards, no direct IPU link      & (d)  &   1,216     &  80   &  4.561   &    3.75   \\
 \midrule
  \multicolumn{6}{l}{Multi-IPU experiments} \\
    2   & on the same board                          & (e)  &   2,432     &  80   &  5.829   &    2.40  \\
    2   & across boards, direct IPU link             & (f)  &   2,432     &  80   &  5.685   &    2.34  \\
    2   & across boards, no direct IPU link          & (g)  &   2,432     &  80   &  5.677   &    2.33  \\
    4   & two boards                                 & (h)  &   4,864     &  40   &  5.856   &    1.20  \\
    8   & four boards                                & (i)  &   9,728     &  20   &  5.743   &    0.59  \\
   16   & eight boards, entire system                & (j)  &  19,456     &  4    &  5.520   &    0.28  \\
  \bottomrule
  \end{tabular}
  \caption{Peak gather bandwidth: bandwidth available to
    gather operations that use the largest message size allowed by
    local memory capacity, in varying scales and topologies. We
    explicitly report the message used in each experiment, as
    different scales correspond to different maximum message
    sizes. Experiments labeled (a)-(j) use topologies corresponding to
    these illustrated in Figure~\ref{fig:broadcast-setup}, but with
    reversed data flow directions.}
  \label{tab:gather-bandwidth}
\end{table}

In this section we study the peak bandwidth available to gather
operations at different scales and in different topologies.  As we did
in the previous sections, we study the effects of source-destination
IPU proximity in topologies (a)...(j) of
Figure~\ref{fig:broadcast-setup}, with the caveat that gather
operations involve data flows of the reverse direction than that depicted
in the figure. At each scale we use the largest message size allowed
by local memory capacity. Results show that the operation's
performance is primarily limited by the small message size and is
less sensitive to scale. Aggregate gather bandwidth degrades very
gracefully with scale, with 16-IPU systems experiencing an aggregate
bandwidth that is only 5.3\% lower than that available on a 2-IPU
board.  Our results are in Table~\ref{tab:gather-bandwidth}.

\noindent\textbf{Small messages.} Local memory capacity on the source
tile is \emph{the} limiting factor for message size. As scale
increases, the maximum usable message size decreases from 160 to 4
bytes. Small message size limits, in turn, available aggregate
bandwidth. The bandwidth decrease is consistent with that already
benchmarked in Section~\ref{sec:bw-as-function-of-block-size}.

\noindent\textbf{Off chip.} Scatter operations exhibit minimal
degradation of performance when moving from on chip to off chip:
aggregate bandwidth decreases from 6.3 to 5.6...5.2 GB/s.
The aggregate bandwidth at scales of 2...16 IPUs is virtually identical
for gather and scatter operations.

\section{Scatter}

In this section we study the latency and bandwidth performance
available to scatter operations.

\noindent\textbf{Scatter.} A scatter operation is similar to a
broadcast in the sense that both operations involve one source and
multiple destination tiles.  The two operations differ in the scatter
sending a distinct message to each destination, whereas a broadcast
sends the same message to all destination. Moreover, a scatter is the
reverse operation of a gather.

\noindent\textbf{Scale.} We benchmark the same scales as in our
broadcast and gather experiments. We described it in
Section~\ref{sec:broadcast}.

\noindent\textbf{No load.} Experiments in this section study load
conditions in which no other operations are in progress. Each transfer's
performance is affected only by the load caused by the remainder of
the collective operation.

\noindent\textbf{Topologies.} We benchmark the same topologies
discussed in the previous sections, and illustrated in
Figure~\ref{fig:broadcast-setup}. Scatter data flow directions are
correctly depicted in the figure (thin blue arrows) and also match
those described in Section~\ref{sec:broadcast} for broadcast
operations.

\noindent\textbf{Message size.}  The source tile must dedicate to the
operation one buffer per destination in its local memory.  This
constraints limits the maximum message size usable in a scatter
operation. In turn, the limited message size is the primary factor
limiting the aggregate bandwidth available to the operation. A
complete discussion follows.

In benchmarks involving whole IPUs, the source tile is also among the
destinations, and maintains at the same time one input buffer
and multiple output buffers in its local memory.

\subsection{Congestion-free Scatter Latency}

\begin{table}[tb]
  \center
  \footnotesize
  \begin{tabular}{clcrrrr}
  \toprule
 Scale  &  Experiment                                        & \multicolumn{2}{c}{Destination    } & \multicolumn{1}{c}{Total}      & Avg. latency   \\
 \,     &                                                    & \multicolumn{2}{c}{tile}            & \multicolumn{1}{c}{latency}    & per transfer \\
 \,     &                                                    & \multicolumn{2}{c}{count}           & \multicolumn{1}{c}{($\upmu$s)} & \multicolumn{1}{c}{(ns)} \\
  \midrule
        &  Transfer to self (via local memory)               &     &         1      &  0.012   &  11.542 \\
  \midrule
        & \multicolumn{4}{l}{On-chip: source and destination tiles are on the same IPU} \\
        &  ... to 2 tiles                                    &     &         2      &  0.181   &  90.713 \\
        &  ...                                               &     &         4      &  0.183   &  45.823 \\
        &  ...                                               &     &         8      &  0.186   &  23.237 \\
        &  ...                                               &     &        16      &  0.191   &  11.927 \\
        &  ...                                               &     &        38      &  0.205   &   5.384 \\
        &  ...                                               &     &        76      &  0.226   &   2.973 \\
   1/8  &  ...                                               &     &       152      &  0.273   &   1.795 \\
   1/4  &  ... to a quarter of one IPU                       &     &       304      &  0.367   &   1.207 \\
   1/2  &  ... to half IPU                                   &     &       608      &  0.555   &   0.913 \\
   1    &  ... to the entire IPU                             & (a) &     1,216      &  0.927   &   0.762 \\
  \midrule
     1  & \multicolumn{4}{l}{to an entire IPU, different than that of the source tile} \\
        & ... both IPUs on the same board                    & (b) &     1,216      &  1.361   &   1.361 \\
        & ... across boards, direct IPU link                 & (c) &     1,216      &  1.275   &   1.048 \\
        & ... across boards, no direct IPU link              & (d) &     1,216      &  1.404   &   1.155 \\
  \midrule
  \multicolumn{6}{l}{Multi-IPU experiments} \\
    2   & IPUs on the same board                             & (e) &     2,432      &  2.268   &   0.932 \\
    2   & IPUs on different boards, direct IPU link          & (f) &     2,432      &  2.181   &   0.897 \\
    2   & IPUs on different boards, no direct IPU link\hspace{-3ex}& (g) & 2,432    &  2.181   &   0.897 \\
    4   & IPUs on two boards                                 & (h) &     4,864      &  3.707   &   0.762 \\
    8   & IPUs on four boards                                & (i) &     9,728      &  7.115   &   0.731 \\
   16   & IPUs, entire system                                & (j) &    19,456      & 13.729   &   0.706 \\
  \bottomrule
  \end{tabular}
  \caption{Minimum scatter latency in experiments of varying scale
    and topology. The topology of experiments (a)-(j) is as
    illustrated in Figure~\ref{fig:broadcast-setup}.}
  \label{tab:scatter-latency}
\end{table}

We study the minimum latency associated with scatter operations of
varying scale and in different topologies. We transfer a message of
minimum size, i.e., a word of 32 bits.  We vary the experiment's scale
from one tile to the whole system.  We find that a whole-IPU scatter
completes in 0.9 microseconds, only marginally slower than the gather
operation of equal scale.  Scatter operations show remarkable
scalability, with latencies in Multi-IPUs growing sub-linearly as a
function of the number of hops in the ladder network.  A whole-system
gather completes in 14 microseconds. Interestingly, off-chip scatter
operations are significantly faster than gathers of equal
scale. Quantitative results are in Table~\ref{tab:gather-latency}.

As we did in earlier tests, we report first the latency (12 nanoseconds) of a
trivial operation involving the same tile as source and
destination. It results in a local memory copy, and its latency is
only intended for comparison and contribution breakdown.

\noindent\textbf{Latency off chip.} Results collected over experiments
at scale show that scatter latencies also depend on the number of hops
between sources and destination in the ladder network, but the
off-chip penalty going from a single-IPU to a 2-IPU test is much
smaller for scatter than for gather operations (0.4 vs. 3.1
microseconds).

Comparisons between 2-IPU, direct-vs-indirect topologies yield
somewhat similar considerations as gathers: directly connected pairs
of IPUs on different boards see a marginally better latency pairs of
IPUs located on the same board which, in turn, see a better latency
than pairs of IPU not directly connected via IPU links.  The latency
differences between the respective topologies are, however, much
smaller than those seen in the gather experiments.

\noindent\textbf{At scale.}  Similarly to what we found for gather
operations, a doubling of system size roughly corresponds to a
doubling of network diameter, which causes in turn an approximate
doubling of total latency. Scatter operations complete, however,
almost twice as quickly as gathers of equal scale.

\subsection{Peak Scatter Bandwidth}

\begin{table}[tbh]
  \center
  \scriptsize
  \begin{tabular}{clcrrrr}
  \toprule
  Scale  &  Experiment                                         & \multicolumn{2}{c}{Destination} &  Message &  Aggregate    &  Bandw. \\
         &                                                     & \multicolumn{2}{c}{tile }       &  size    &  bandwidth    &  per tile  \\
         &                                                     & \multicolumn{2}{c}{count }      &  (bytes) &  (GB/s)       &  (MB/s)    \\
  \midrule
        &  Transfer to self (via local memory)                &      &       1     &  160  &  13.884  &  13883.50  \\
  \midrule
        & \multicolumn{4}{l}{On-chip: source and destination tiles are on the same IPU} \\
        &  One tile ... to 2 tiles                            &      &       2     &  160  &   1.728   &  864.10    \\
        &  ... to 4 tiles                                     &      &       4     &  160  &   2.706   &  676.54    \\
        &  ...                                                &      &       8     &  160  &   3.812   &  476.55    \\
        &  ...                                                &      &      16     &  160  &   4.770   &  298.12    \\
        &  ...                                                &      &      38     &  160  &   5.592   &  147.15    \\
        &  ...                                                &      &      76     &  160  &   5.971   &  78.56     \\
   1/8  &  ...                                                &      &     152     &  160  &   6.175   &  40.62     \\
   1/4  &  ... a quarter of one IPU                           &      &     304     &  160  &   6.279   &  20.65     \\
   1/2  &  ... half IPU                                       &      &     608     &  160  &   6.332   &  10.41     \\
   1    &  ... the entire IPU                                 & (a)  &   1,216     &  160  &   6.360   &  5.23      \\
  \midrule
   1    & \multicolumn{6}{l}{to an entire IPU, different than that of source tile} \\
        & ... both IPUs on the same board                     & (b)  &   1,216     &  80   &   5.633   &  4.63      \\
        & ... across boards, direct IPU link                  & (c)  &   1,216     &  80   &   5.229   &  4.30      \\
        & ... across boards, no direct IPU link               & (d)  &   1,216     &  80   &   5.217   &  4.29      \\
  \midrule
  \multicolumn{6}{l}{Multi-IPU experiments} \\
    2   & IPUs on the same board                              & (e)  &    2,432    &  80  &   5.899   &  2.43      \\
    2   & IPUs on different boards - direct IPU link\hspace{-3ex} & (f) & 2,432    &  80  &   5.685   &  2.34      \\
    2   & IPUs on different boards - no direct IPU link\hspace{-3ex} & (g) & 2,432 &  80  &   5.677   &  2.33      \\
    4   & IPUs on two boards                                  & (h)  &    4,864    &  40  &   5.855   &  1.20      \\
    8   & IPUs on four boards                                 & (i)  &    9,728    &  20  &   5.743   &  0.59      \\
   16   & IPUs, entire system                                 & (j)  &   19,456    &  4   &   5.514   &  0.28      \\
  \bottomrule
  \end{tabular}
  \caption{Peak scatter bandwidth: bandwidth available to
    scatter operations that use the largest message size allowed by
    local memory capacity, in varying scales and topologies. We
    explicitly report the message used in each experiment, as
    different scales correspond to different maximum message sizes.
    Experiments labeled (a)-(j) use the corresponding topologies
    illustrated in Figure~\ref{fig:broadcast-setup} and described in
    Section~\ref{sec:broadcast}.}
  \label{tab:scatter-bandwidth}
\end{table}

In this section we study the peak bandwidth available to scatter
operations of different scales. As we did in the previous sections, we
study the effect of source-destination IPU proximity in topologies
(a)...(j) of Figure~\ref{fig:broadcast-setup}. At each scale we use
the largest message size allowed by local memory capacity.  Results
show that the operation's performance is primarily limited by the
small message size, and is less sensitive to scale. Aggregate scatter
bandwidth degrades very gracefully with scale, with 16-IPU systems
experiencing an aggregate bandwidth that is only 6.6\% lower than that
available on a 2-IPU board.  Our results are in
Table~\ref{tab:scatter-bandwidth}.

\noindent\textbf{Small messages.} We remark that local memory capacity
on the source tile is \emph{the} limiting factor for message size. As
scale increases, the maximum usable message size decreases from 160 to
4 bytes. Small message size is, in turn, the primary factor for the
low aggregate available bandwidth. Results are consistent with the
findings already presented in
Section~\ref{sec:bw-as-function-of-block-size}.

\noindent\textbf{Off chip.} Scatter operations exhibit minimal
degradation of performance when moving from on chip to off chip:
aggregate bandwidth decreases from 6.3 to 5.6...5.2 GB/s.
The aggregate bandwidth at scales 2...16 IPUs is virtually identical
for gather and scatter operations.

\section{All to all}

In this section we study the performance of all-to-all collective
operations on the IPU. Because of the number of distinct personalized
transfers involved in this operation and the associated local memory
footprint, we were unable to scale our benchmark to multiple IPUs, or
even to an entire IPU.  For that reason, results in this section will
be incomplete. We report latency results up to half IPU (609
tiles). Half an IPU completes an all-to-all in 0.55 microseconds.
Our results are in Table~\ref{tab:all-to-all-latency}.

\noindent\textbf{All to all.} In an all-to-all operation, each tile in
a group sends one distinct message to each tile in the group, itself
included.  If the group contains $n$ members, each participant
concurrently sends $n$ distinct messages. The whole operation involves
transferring $n^2$ distinct messages.

\noindent\textbf{Message size.}  Each tile must dedicate to the
operation $n$ output buffers and $n$ input buffers.  This constraint
limits the maximum message size usable in the operation. Specifically,
we only consider messages consisting of a single word, because they
allow us to benchmark the largest scale (1/2 IPU).

\noindent\textbf{Scale.} Because of the aforementioned considerations,
we were unable to benchmark a whole-IPU all-to-all operation, or one
involving multiple IPUs.

\begin{table}
  \center
  \footnotesize
  \begin{tabular}{clcrrrr}
  \toprule
 Scale  &  Experiment                    & \multicolumn{2}{r}{Total} & Concurrent  & Total      & Avg. latency   \\
 \,     &                                & \multicolumn{2}{r}{tile}  & transfers   & latency    & per transfer \\
 \,     &                                & \multicolumn{2}{r}{count} &             & ($\upmu$s) & \multicolumn{1}{c}{(ns)} \\
  \midrule
        & Transfer to self (via local memory) & &        1           &  1          & 0.012      & 11.542   \\
  \midrule
        & \multicolumn{4}{l}{On-chip: all tiles are on the same IPU} \\
        &  between 2 tiles               &     &         2           &  4          & 0.131      & 32.816  \\
        &  ...                           &     &         4           &  16         & 0.125      & 7.812   \\
        &  ...                           &     &         8           &  64         & 0.143      & 2.227   \\
        &  ...                           &     &        16           &  256        & 0.163      & 0.637   \\
        &  ...                           &     &        38           &  1,444      & 0.182      & 0.126   \\
        &  ...                           &     &        76           &  5,776      & 0.215      & 0.037   \\
   1/8  &  ...                           &     &       152           &  23,104     & 0.256      & 0.011   \\
   1/4  &  ... a quarter of one IPU      &     &       304           &  92,416     & 0.355      & 0.004   \\
   1/2  &  ... half IPU                  &     &       608           &  369,664    & 0.552      & 0.001   \\
   1    &  ... the entire IPU            &     &     1,216           &  1,478,656  & -          & -       \\
  \bottomrule
  \end{tabular}
  \caption{Minimum all-to-all latency: latency of all-to-all
    communications in different scales.}
  \label{tab:all-to-all-latency}
\end{table}

\section{Reduction}

We investigate the performance of reduction operations at different
scales and in different topologies.

\noindent\textbf{Reduction.} A reduction involves a variable number of
source tiles (depending on scale) plus one destination tile. The
operation takes one input numerical array on each source tile, and
returns one output numerical array on the destination tile.  (For the
avoidance of doubt, we are describing the equivalent of an
\lstinline|MPI_Reduce| primitive, not an \lstinline|MPI_Allreduce|.)

\noindent\textbf{Weak vs. Strong Scaling.} As we vary the scale of the
experiment, we consider two scenarios: one in which the total amount
of input operands grows linearly with parallelism (weak scaling) and
one in which it remains constant (strong scaling).

\noindent\textbf{Benchmark}. The Poplar SDK offers
\lstinline|popops::reduce|, a library primitive that performs the
reduction. Our experiments simply benchmark this primitive's
performance.  We use sum as a reduction operation.

\noindent\textbf{No load.} Experiments in this section study load
conditions in which no other operation is in progress. Each transfer's
performance is affected only by the load caused by the remainder of
the collective operation.

\noindent\textbf{SDK version}. Actual performance depends on the
library function implementation provided with the specific version of
SDK employed. Our results are only representative of SDK 1.0.49.

\subsection{Minimum Reduction Latency - Weak Scaling}

\begin{table}[p]
  \center
  \footnotesize
  \begin{tabular}{lrrrr}
  \toprule
  Experiment                                       & Operands & Tiles    & Total      & Avg. latency \\
  \,                                               &  per     & involved & latency    & per input \\
  \,                                               &  tile    &          & ($\upmu$s) & operand (ns)\\
  \midrule
  \textbf{Baseline sequential}\\
  1 tile                                           &  1,216 &        1 &  1.69  & 1.39  \\
  1 tile                                           &  2,432 &        1 &  2.44  & 1.00  \\
  1 tile                                           &  4,864 &        1 &  3.97  & 0.82  \\
  1 tile                                           &  9,728 &        1 &  7.02  & 0.72  \\
  1 tile                                           & 19,456 &        1 & 13.10  & 0.67  \\
  \midrule
  \textbf{Increasing system diameter} \\
  1 IPU, 1 tile                                    &      1 &        1 &  0.44  & 0.44  \\
  2 IPUs on the same board                         &      1 &        2 &  1.13  & 0.57  \\
  4 IPUs on two PCI boards                         &      1 &        4 &  1.48  & 0.37  \\
  8 IPUs on four PCI boards                        &      1 &        8 &  2.02  & 0.25  \\
  16 IPUs, entire system                           &      1 &       16 &  3.16  & 0.20  \\
  \midrule
  \textbf{Increasing scale} \\
   1 IPU, all 1,216 tiles                          &      1 &    1,216 &  1.97  & 1.62  \\
   2 IPUs on the same board                        &      1 &    2,432 &  5.32  & 2.19  \\
   2 IPUs on different boards, direct IPU link     &      1 &    2,432 &  3.60  & 1.48  \\
   2 IPUs on different boards, no direct IPU link  &      1 &    2,432 &  7.09  & 2.92  \\
   4 IPUs on two PCI boards                        &      1 &    4,864 &  7.18  & 1.48  \\
   8 IPUs on four PCI boards                       &      1 &    9,728 &  7.68  & 0.79  \\
   16 IPUs, entire system                          &      1 &   19,456 & 14.51  & 0.74  \\
  \bottomrule
  \end{tabular}
  \caption{Reduction, weak scaling: Latency of one reduction as it
    scales up from a single tile to an entire 16-IPU system.}
  \label{tab:reduction-weak-scaling}
\end{table}

We study how the latency of one reduction scales (first weakly, then
strongly) as it extends across tiles, then across IPUs on one board,
then across boards, finally spanning an entire 16-IPU system.
Interestingly, we find the latency associated with completely
distributed reduction is comparable (11...16\% more expensive) than a
sequential operation of equal input size entirely performed on one
tile. Detailed results are in Table~\ref{tab:reduction-weak-scaling}.

\noindent\textbf{Intent} The purpose of this experiment is to
understand how larger scale and parallelism affect the overall latency
and the per-operand latency of the operation.  In this context, we are
especially interested in the data transfer component, and not in the
arithmetics involved in the reduction.  Specifically, the larger is
the scale, the higher is the fraction of messages occurring over
slower interconnects, i.e., IPU link vs. exchange, or more vs. fewer
hops.

\noindent\textbf{Input size}: in this section, all parallel experiments
intentionally use the smallest inputs, i.e., one single-precision
floating point value (32 bits) per tile.  This choice exposes
primarily the cost of data transfers, as opposed to that of arithmetic
computation or local memory access.

For comparison against the parallel experiments, we also perform
single-tile baseline experiments of equal input size as the parallel
ones (i.e., 1,216, then 2,432, ... finally 19,456 operands).  These
measurements serve as a baseline for the cost of a reduction when no
parallelism is involved.

\clearpage

\subsection{Minimum Reduction Latency - Strong Scaling}

\begin{table}[tb]
  \center
  \footnotesize
  \begin{tabular}{lrrrr}
  \toprule
  Experiment                                       & Operands & Tiles    & Total      & Avg. latency \\
  \,                                               & per      & involved & latency    & per input \\
  \,                                               & tile     &          & ($\upmu$s)  & operand (ns) \\
  \midrule
  Baseline - one tile                              &  19,456  &        1 &  13.1  & 0.67   \\
  \midrule
   1 IPU, all 1,216 tiles                          &      16  &    1,216 &   2.1  & 0.11   \\
   2 IPUs on the same board                        &       8  &    2,432 &   5.4  & 0.28   \\
   2 IPUs on different boards, direct IPU link     &       8  &    2,432 &   3.7  & 0.19   \\
   2 IPUs on different boards, no direct IPU link  &       8  &    2,432 &  7.20  & 0.37   \\
   4 IPUs on two PCI boards                        &       4  &    4,864 &  7.28  & 0.37   \\
   8 IPUs on four PCI boards                       &       2  &    9,728 &  7.78  & 0.40   \\
  16 IPUs, entire system                           &       1  &   19,456 & 14.51  & 0.75   \\
  \bottomrule
  \end{tabular}
  \caption{Reduction, strong scaling: Latency of one reduction as it
    scales up from a single tile to an entire 16-IPU system.}
  \label{tab:reduction-strong-scaling}
\end{table}

We benchmark reduction latency once more, but in a strong scaling
scenario: problem size is kept constant across experiments at
different scales, in order to expose the cost of parallelization.
We choose the smallest input size that would place one operand per
tile on the largest system considered, i.e., 19,456 operands.
Reducing 19,456 single-precision values on one IPU takes 2.1
microsecond. Performing the same reduction on a 16-IPU system costs
14.51 microseconds.

\subsection{Peak Reduction Bandwidth}

\begin{table}[tbh]
  \center \footnotesize
  \begin{tabular}{lrrrr}
  \toprule
  Experiment                                       & Operands & Tiles      & Aggregate    & Bandw. \\
  \,                                               & per      & involved   & bandwidth    & per tile \\
  \,                                               & tile     &            & (GB/s)       & (GB/s) \\
  \midrule
  \multicolumn{4}{l}{\textbf{Single-IPU baseline performance}} \\
  Baseline - 1,000 tiles                           &   1,216 &    1,000    &  1,671  &  1.67  \\
  Baseline - 1,000 tiles                           &   2,432 &    1,000    &  2,657  &  2.68  \\
  Baseline - 1,000 tiles                           &   4,864 &    1,000    &  3,747  &  3.75  \\
  Baseline - 1,000 tiles                           &   9,728 &    1,000    &  4,729  &  4.73  \\
  Baseline - 1,000 tiles                           &  19,456 &    1,000    &  5,432  &  5.43  \\
  \midrule
  \multicolumn{4}{l}{\textbf{Large operands - Single-IPU Weak scaling}}\\
   Running on ... one tile                                  & 25,000  &       1    &      6  &  6.03   \\
   ... 2 tiles                                              & 25,000  &       2    &     12  &  5.97   \\
   ... 4 tiles                                              & 25,000  &       4    &     24  &  5.91   \\
   ...                                                      & 25,000  &       8    &     47  &  5.86   \\
   ...                                                      & 25,000  &      16    &     92  &  5.72   \\
   ...                                                      & 25,000  &      38    &    224  &  5.90   \\
   ...                                                      & 25,000  &      76    &    448  &  5.89   \\
   ...                                                      & 25,000  &     152    &    890  &  5.85   \\
   ... a quarter of one IPU                                 & 25,000  &     304    &  1,769  &  5.82   \\
   ... half IPU                                             & 25,000  &     608    &  3,481  &  5.73   \\
   ... the entire IPU                                       & 25,000  &    1,216   &  6,756  &  5.56   \\
   \midrule
   \multicolumn{4}{l}{\textbf{Smaller operands - Multi-IPU Weak scaling}}\\
   1 IPU, all 1,216 tiles                                      &  1,000  &    1,216   &  1,636  &  1.34  \\
   2 IPUs on the same board                                    &  1,000  &    2,432   &  1,540  &  0.63  \\
   2 IPUs on different boards, direct IPU link\hspace{-5ex}    &  1,000  &    2,432   &  2,090  &  0.86  \\
   2 IPUs on different boards, no direct IPU link\hspace{-5ex} &  1,000  &    2,432   &  1,230  &  0.51  \\
   4 IPUs on two PCI boards                                    &  1,000  &    4,864   &  2,381  &  0.49  \\
   8 IPUs on two PCI boards                                    &  1,000  &    9,728   &  4,512  &  0.46  \\
  16 IPUs, entire system                                       &  1,000  &   19,456   &  5,016  &  0.26  \\
  \midrule
  \multicolumn{4}{l}{\textbf{Larger operands - Multi-IPU Weak scaling}}\\
   1 IPU, all 1,216 tiles                                        & 25,000  &    1,216   &   6,756 &  5.56 \\
   2 IPUs on the same board                                      & 25,000  &    2,432   &  11,399 &  4.69 \\
   2 IPUs on different boards, direct IPU link \hspace{-5ex}     & 25,000  &    2,432   &  12,400 &  2.55 \\
   2 IPUs on different boards, no direct IPU link\hspace{-5ex}   & 25,000  &    2,432   &  10,485 &  2.16 \\
   4 IPUs on two PCI boards                                      & 25,000  &    4,864   &  20,965 &  4.31 \\
   8 IPUs on two PCI boards                                      & 25,000  &    9,728   &  41,128 &  4.23 \\
  16 IPUs, entire system                                         & 25,000  &   19,456   &  63,724 &  3.28 \\
  \bottomrule
  \end{tabular}
  \caption{Reduction bandwidth available at different scales
    (sequential, 1-16 IPUs) and for different input sizes, in weak
    scaling conditions.}
    \label{tab:reduction-bandwidth}
\end{table}

\begin{figure}[tb]
  \includegraphics[width=\columnwidth]{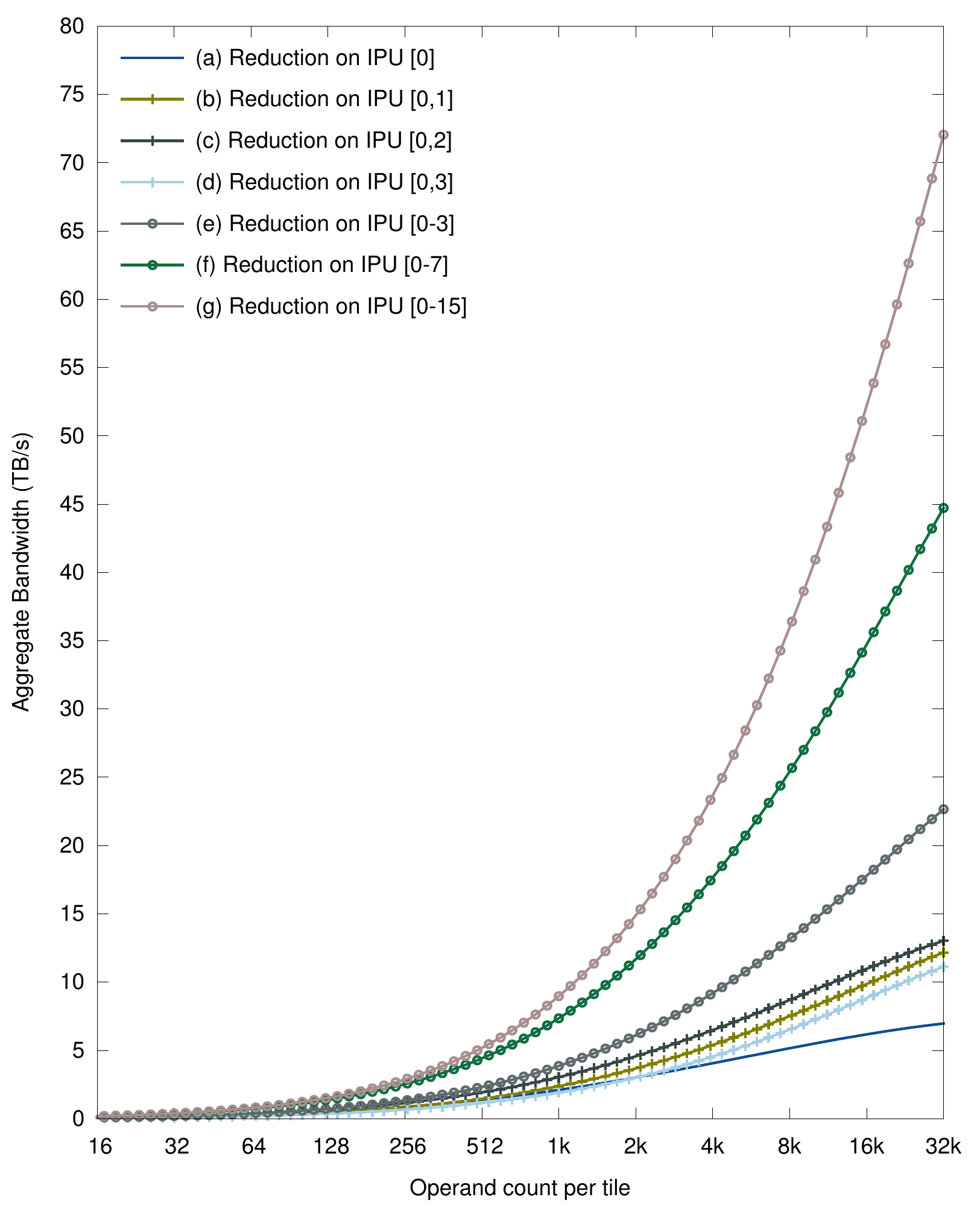}
  \caption{Effect of operand size on reduction bandwidth in weak
    scaling conditions. IPU identifiers are DNC IDs.}
  \label{fig:red_bandwidth}
\end{figure}

We measure the throughput achieved by reductions of different sizes
and scale.  As a measure of reduction throughput, we adopt the input
operand size consumed in the unit of time. In this section, all
experiments study weak scaling.  For simplicity, we report
quantitative results for only two choices (1,000 and 25,000 operands
per tile) of input size, in Table~\ref{tab:reduction-bandwidth}.  In
Figure~\ref{fig:red_bandwidth} we chart the effect of operand size on
reduction bandwidth in a broader and more detailed spectrum of input
sizes.

\noindent\textbf{On-chip scaling.} In weak scaling conditions, on-chip
aggregate reduction bandwidth grows remarkably well with system
size. The aggregate per-chip throughput reaches 6.76 TB/s, with a
parallel efficiency equal to 92.2\%.

\noindent\textbf{Off-chip scaling.}  For larger inputs (25,000 values
per tile), the aggregate throughput grows by a factor of 9.4$\times$
as the system scales up from 1 to 16 IPUs. The peak aggregate
throughput of a 16-IPU system reaches an impressive 64.7 TB/s.  The
per-tile bandwidth degradation is approximately 41\%; the parallel
efficiency is 59\%.

\noindent\textbf{Operand size.} The effect of operand size on
performance is evident by direct comparison between the third and
fourth band of the table, which report results of experiments with
1,000 and 25,000 values per tiles, respectively.  Larger operands
achieve a substantially higher bandwidth at any scale, and better
scalability overall: the large-vs-small operand speedup is 4.1$\times$
on a single IPU, but it grows to 12.7$\times$ on a 16-IPU system.  We
chart a broader set of results describing the same trends in
Figure~\ref{fig:red_bandwidth}.

\clearpage

\section{Host Connectivity}

In this section we present bandwidth and latency associated with
host-to-IPU data transfers. These metrics are relevant to evaluate
potential bottleneck in achieving high performance in hybrid CPU/IPU
applications where the PCI-express bus lies in the critical path.
Bandwidth is important for streaming applications where CPU data are
continuously offloaded to the IPU, and latency is important for
inference applications where the output is produced in a ping-pong
fashion across CPU and IPU. The reader should pay consideration to
fact that PCI-express is an industry-standard I/O technology; the
results presented here are meant to be compared against the
theoretical limit.

\noindent\textbf{SDK version}. Performance depends on the version of
SDK and drivers employed. Our results are only representative of SDK
1.0.49. We expect Graphcore to further optimize host-to-IPU
connectivity performance in future releases.

\subsection{Minimum Host-to-IPU Latency}

\textbf{Benchmark description:} we use a benchmark that sends data from the
host to a tensor on IPUs using the \lstinline|graph::DataStream| mechanism
with an empty callback to minimize overhead associated with the software stack.
Depending on the benchmark scale and IPU count, the tensor is distributed across
local memories such as each tile receives exactly 4 byte data. The latency
measurements are taken on the host side as average over multiple multiple
synchronous transfers.

\noindent\textbf{Observations:} as shown in
Table~\ref{tab:host-to-ipu-latency}, the latency to transfer data to
an IPU is at least 8.81 $\upmu$s and is stable across the PCI-express
topology. Sending data to different destination IPUs doesn't show
obvious differences in latency. Data transfer latency also doesn't
increase when communicating with multiple IPUs. We only notice a
slight increase in latency as destination tile count increases. This,
however, is a consequence of moving a tensor large enough to fit the
maximum PCI-express payload size, requiring multiple transactions to
complete the transfer.

\begin{table}
  \footnotesize
  \begin{tabular}{clcr}
  \toprule
  Scale  & Experiment                                      & Destination&        Total   \\
  (IPUs) &                                                 & tiles      &        latency \\
         &                                                 &            &     ($\upmu$s) \\
  \midrule
        & \multicolumn{3}{l}{Destination tiles are on IPU 0}\\
        & From CPU ... to 1 tile                           &         1  &      8.81  \\
        & ... to 2 tiles                                   &         2  &      8.84  \\
        & ...                                              &         4  &      8.89  \\
        & ...                                              &         8  &      8.31  \\
        & ...                                              &        16  &      8.81  \\
        & ...                                              &        38  &      8.81  \\
        & ...                                              &        76  &      9.01  \\
   1/8  & ...                                              &       152  &      8.80  \\
   1/4  & ... to a quarter of one IPU                      &       304  &      8.95  \\
   1/2  & ... to half IPU                                  &       608  &      9.56  \\
   1    & ... to the entire IPU                            &     1,216  &      9.87  \\

   \midrule
   1    & \multicolumn{3}{l}{Single destination IPU}  \\
        & ... to 1 tile on IPU 0                           &         1  &      8.87  \\
        & ... to 1 tile on IPU 1                           &         1  &      8.89  \\
        & ... to 1 tile on IPU 2                           &         1  &      8.83  \\
        & ... to 1 tile on IPU 3                           &         1  &      8.80  \\
        & ... to 1 tile on IPU 4                           &         1  &      8.82  \\
        & ... to 1 tile on IPU 5                           &         1  &      8.81  \\
        & ... to 1 tile on IPU 6                           &         1  &      8.83  \\
        & ... to 1 tile on IPU 7                           &         1  &      8.85  \\
        & ... to 1 tile on IPU 8                           &         1  &      8.84  \\
        & ... to 1 tile on IPU 9                           &         1  &      8.80  \\
        & ... to 1 tile on IPU 10                          &         1  &      8.82  \\
        & ... to 1 tile on IPU 11                          &         1  &      8.81  \\
        & ... to 1 tile on IPU 12                          &         1  &      8.80  \\
        & ... to 1 tile on IPU 13                          &         1  &      8.81  \\
        & ... to 1 tile on IPU 14                          &         1  &      8.88  \\
        & ... to 1 tile on IPU 15                          &         1  &      8.82  \\
  \midrule
        & \multicolumn{3}{l}{Multiple destination IPUs, 1 tile each IPU} \\
   2    & ... to 2 IPUs on the same board                  &         2  &      8.88  \\
   2    & ... to 2 IPUs across boards, direct IPU link     &         2  &      8.83  \\
   2    & ... to 2 IPUs across boards, no direct IPU link  &         2  &      8.84  \\
   4    & ... to 4 IPUs on two boards                      &         4  &      8.83  \\
   8    & ... to 8 IPUs on four boards                     &         8  &      8.82  \\
   16   & ... to 16 IPUs, entire system                    &        16  &      8.83  \\
  \bottomrule
  \end{tabular}
  \caption{Minimum host-to-IPU latency in experiments of different scale and topology.}
  \label{tab:host-to-ipu-latency}
\end{table}

\subsection{Peak Host-to-IPU Bandwidth}

\textbf{Benchmark description:} we measure aggregate bandwidth with a
benchmark that transfers data from the host to a tensor on IPUs via
\lstinline|graph::DataStream|.  The destination tensor is partitioned
linearly across all involved tiles and IPUs. The benchmark sends 40 KB
data to each tile, and up to 778.24 MB to all 16 IPUs.  We report
results in Table~\ref{tab:host-to-ipu-bandwidth}.

\noindent\textbf{Observations:} the achieved aggregate bandwidth
increases with the growth of involved PCIe lane count. Since each IPU
is connected to the PCIe switch via 8 PCIe Gen3 lanes, transfers to
one IPU reaches up to $\sim6$ GB/s.  Every four IPUs share the same 16
PCIe lanes, thus the transfer to 4 IPUs (h) can only achieve 13.78
GB/s. The transfer to all 16 IPUs (j) enjoys a bandwidth of 55.04
GB/s.

\begin{table}
  \center
  \footnotesize
  \begin{tabular}{clrrrr}
  \toprule
  Scale  & Experiment                                          & Destination & PCIe    & Aggregate     & Bandw.   \\
  (IPUs) &                                                     & tiles       & lanes   & bandwidth     & per tile \\
         &                                                     &             &         & (GB/s)        & (MB/s) \\
  \midrule
        & \multicolumn{4}{l}{Destination tiles are on IPU 0}\\
        & From CPU to 1 tile                                   &         1   &     8   &  2.57         & 2,565.29 \\
        & ... to 2 tile                                        &         2   &     8   &  3.54         & 1,768.85 \\
        &                                                      &         4   &     8   &  4.41         & 1,103.05 \\
        &                                                      &         8   &     8   &  5.07         &  634.30 \\
        & ...                                                  &        16   &     8   &  5.43         &  339.50 \\
        & ...                                                  &        38   &     8   &  5.85         &  154.04 \\
        & ...                                                  &        76   &     8   &  5.74         &  75.50 \\
   1/8  & ...                                                  &       152   &     8   &  5.80         &  38.17 \\
   1/4  & ... to a quarter of one IPU                          &       304   &     8   &  5.82         &  19.15 \\
   1/2  & ... to half IPU                                      &       608   &     8   &  5.87         &   9.66 \\
   1    & ... to the entire IPU                                &     1,216   &     8   &  5.86         &   4.82 \\
   \midrule
   1    & \multicolumn{5}{l}{Single destination IPU}  \\
        & ... to all tiles on IPU 1                            &    1,216    &     8   &  5.84         & 4.80     \\
        & ... to all tiles on IPU 2                            &    1,216    &     8   &  5.85         & 4.81     \\
        & ... to all tiles on IPU 3                            &    1,216    &     8   &  5.87         & 4.82     \\
        & ... to all tiles on IPU 4                            &    1,216    &     8   &  5.84         & 4.80     \\
        & ... to all tiles on IPU 5                            &    1,216    &     8   &  5.84         & 4.80     \\
        & ... to all tiles on IPU 6                            &    1,216    &     8   &  5.84         & 4.80     \\
        & ... to all tiles on IPU 7                            &    1,216    &     8   &  5.84         & 4.80     \\
        & ... to all tiles on IPU 8                            &    1,216    &     8   &  5.97         & 4.91     \\
        & ... to all tiles on IPU 9                            &    1,216    &     8   &  6.02         & 4.95     \\
        & ... to all tiles on IPU 10                           &    1,216    &     8   &  6.03         & 4.96     \\
        & ... to all tiles on IPU 11                           &    1,216    &     8   &  6.02         & 4.95     \\
        & ... to all tiles on IPU 12                           &    1,216    &     8   &  6.03         & 4.96     \\
        & ... to all tiles on IPU 13                           &    1,216    &     8   &  6.03         & 4.96     \\
        & ... to all tiles on IPU 14                           &    1,216    &     8   &  6.03         & 4.96     \\
        & ... to all tiles on IPU 15                           &    1,216    &     8   &  6.03         & 4.96     \\
  \midrule
        & \multicolumn{5}{l}{Multiple destination IPUs} \\
   2    & ... to 2 IPUs on the same board                      &     2,432   &    16   &  11.35        & 4.67     \\
   2    & ... to 2 IPUs across boards, direct IPU link\hspace{-5ex}    & 2,432 &  16   &  11.36        & 4.67     \\
   2    & ... to 2 IPUs across boards, no direct IPU link\hspace{-5ex} & 2,432 &  16   &  11.35        & 4.67     \\
   4    & ... to 4 IPUs on two boards                          &     4,864   &    16   &  13.78        & 2.83     \\
   8    & ... to 8 IPUs on four boards                         &     9,728   &    32   &  27.55        & 2.83     \\
   16   & ... to 16 IPUs, entire system                        &    19,456   &    64   &  55.04        & 2.83     \\
  \bottomrule
  \end{tabular}
  \caption{Peak host-to-IPU bandwidth available to concurrent
    transfers in different transfer topologies.}
  \label{tab:host-to-ipu-bandwidth}
\end{table}

\chapter{Notable Arithmetic Primitives}
\label{chap:workloads}

\section{Matrix Multiplication}
\label{sec:matmul}

Dense matrix multiplication (matmul) is a workload of such pervasive
presence in HPC and AI/ML applications that its performance on a
computing architecture is frequently used (and sometimes abused) as a
singular proxy for an architecture's overall performance. In this
section, we report and discuss IPU dense matrix multiplication
performance as offered by poplin, Graphcore's linear algebra library,
at the time of writing. We compare the IPU's matmul performance with
contemporary GPUs in terms of aggregate throughout and energy
efficiency.  We find that the IPU offers an impressive arithmetic
throughput, up to 31.1 TFlops/s in single precision and 124.5 TFlops/s
in mixed precision per chip, surpassing the GPU's theoretical
limits. Actual performance measured on GEMM benchmarks show the IPU as
a clear winner in single precision against NVidia's V100 GPU.  In
mixed precision, the comparison does not yield a clear winner and
requires a more nuanced discussion that follows.

\begin{table}[tb]
  \center
  \small
  \begin{tabular}{lllrlr}
    \toprule
    Arithmetic        &                & \multicolumn{2}{c}{Graphcore C2 IPU} & \multicolumn{2}{c}{NVidia Volta V100 (SXM2)} \\
    \cmidrule(lr){3-4}
    \cmidrule(lr){5-6}
    Precision         &                & Units       & 1 IPU           & Units          & 1 GPU     \\
    \midrule
    Single            & Theoretical    & AMP         & 31.1            & TensorCores    & N/A    \\
    (FP32)            & Theoretical    & Vector      &  7.8            & FP cores       & 15.7   \\
    \cmidrule(lr){2-6}
    \,                & Actual GEMM    & AMP         & 18.9            & FP cores       & 15.5   \\
    \,                & \hspace{5ex} \% Theor. &       & 60.7\%        &                & 98.7\% \\
    \midrule
    Mixed             & Theoretical    & AMP        & 124.5           & TensorCores    & 125.0   \\
    (FP16.32)         & Theoretical    & Vector     &  15.6           & FP cores       & 31.4    \\
    \cmidrule(lr){2-6}
    \,                & Actual GEMM    & AMP        &  58.9           & TensorCores    & 90.0    \\
    \,                & \hspace{5ex} \% Theor. &    &  47.3\%         &                & 72.0\% \\
    \bottomrule
  \end{tabular}
  \caption{Arithmetic throughput per-chip comparison between IPUs and
    GPUs. We offer theoretical upper bounds and actual peak
    performance that we benchmarked on matrix-matrix multiplication
    (GEMM). Both platforms offer specialized units; we offer
    independent theoretical upper bounds for specialized and
    non-specialized units on both platforms. Benchmarks used the
    respective vendors' optimized GEMM functions.}
  \label{tab:arithmetic-throughput-comparison}
\end{table}

\noindent\textbf{IPU-GPU comparisons.}  The IPU offers impressive
theoretical compute power thanks to its specialized hardware called
\emph{Accumulating Matrix Product} (AMP) units. These units are
similar in purpose to the GPU's TensorCore units.  In the comparisons
that follow, the reader should pay consideration to the fact that
while one C2 IPU board contains two IPU processors, a Volta GPU board
(either PCI or SXM2) only contains one GPU processor; we focus on the
per-chip comparisons as of interest to architecture designers.

\noindent\textbf{Theoretical limits.} Each IPU tile contains one AMP
unit. An AMP unit can finish 64 mixed-precision or 16 single-precision
floating point operations per clock cycle. At a 1.6-GHz clock rate,
the 1,216 tiles on one IPU deliver 31.1 and 124.5 TFlops/s in single
and mixed precision, respectively. In single precision, one IPU
processor offers almost twice as much single-precision theoretical
throughput as one V100 GPU: 31.1 vs. 15.7 TFlops/s. This result also
reflects the fact that on the GPU, TensorCores do not support single
precision, and single-precision computation uses regular FP cores.  In
mixed precision, one IPU processor roughly matches one V100 GPU (124.5
vs. 125.0 TFlops/s). Rows labeled as ``Theoretical'' in
Table~\ref{tab:arithmetic-throughput-comparison} report these numbers
side to side, offering separate limits for specialized and
non-specialized units.

\noindent\textbf{Benchmark.}  Our experiments benchmark each device's
performance as all of the chip's resources are used to perform one,
large, matrix-matrix multiplication. We study performance sensitivity
to input size.  For simplicity, we only consider square input
matrices, i.e., matmul of $A$ and $B$, both of size $n\times n$.  We
vary $n$ from 16 to the largest size fitting matrices.

\noindent\textbf{Version sensitivity.}  Because we are benchmarking
library functions, performance is also a factor of code optimization,
not just hardware compute power. We report actual GEMM performance
delivered on each device by the respective manufacturers' optimized
linear algebra libraries: Graphcore's poplin and NVidia's cuBlas.  Our
IPU results only describe SDK version 1.0.49, and may differ
significantly from future, more optimized poplin versions. Same
considerations apply to the GPU, where we used cuBLAS version
10.1.0.105.

\noindent\textbf{Single precision.}  We measured 18.9 TFlops/s per IPU
at peak, which is 60\% of the theoretical limit. The two IPUs on a C2
board can deliver twice as much performance (37.8 TFlops/s) if used in
parallel on independent matrix operands. The IPU outperforms the GPU
in a per-chip comparison. This result reflects the fact that the GPU's
specialized TensorCore units do not support pure single precision.
In Table~\ref{tab:arithmetic-throughput-comparison},
``Actual GEMM'' rows report these results, also specifying what hardware
units the library's GEMM implementation uses at each precision and what
fraction of theoretical throughput they achieve.


While the IPU delivers higher throughput, the GPU supports larger
operands thanks to its higher device memory capacity.  In our
experiments, the largest square matrix operands fitting one IPU are
2,944$\times$2,944, while on a 32-GB GPU they are roughly
$\sim$50,000$\times\sim$50,000.


\noindent\textbf{Mixed precision.}  On both devices, specialized
hardware (TensorCores and AMP units) supports matrix multiplication in
mixed precision. Despite one IPU delivering roughly the same
theoretical throughput as one GPU, in GEMM benchmarks the IPU
yields lower performance than a V100 GPU: 58.9 TFlops/s vs. 90.0
TFlops/s, respectively.  The IPU uses a lower fraction of its
theoretical limit (47.3\%) than the GPU (72.0\%). See the lower half
of Table~\ref{tab:arithmetic-throughput-comparison}.

In mixed precision as well, the GPU supports larger operands.  The
largest square matrix operands fitting an IPU are 2,688$\times$2,688,
while on a 32-GB GPU they are roughly $\sim$72,000$\times\sim$72,000.

\begin{table}[tb]
  \center
  \begin{tabular}{lrr}
    \toprule
    \,                                & \multicolumn{2}{c}{Precision} \\ \cmidrule(lr){2-3}
    \,                                & \multicolumn{1}{c}{Single} &  \multicolumn{1}{c}{Mixed} \\
    \midrule
    Theoretical arithmetic throughput & 31.1 TFlops/s      & 124.5 TFlops/s \\
    Actual, Peak                      & 18.9 TFlops/s      & 58.2 TFlops/s \\
    Actual/theoretical fraction       & 60.7 \%            & 46.7\% \\
    \bottomrule
  \end{tabular}
  \caption{Theoretical arithmetic throughput compared to the actual
    peak throughput we measured on dense matrix multiplication on one
    IPU.}
  \label{tab:matmul-throughput}
\end{table}




\noindent\textbf{Tile mapping.}  We benchmark two mappings of input
operands to tile memory:

\begin{itemize}
\item \textbf{basic}: this benchmark invokes function
  \lstinline|poplin::matMulAcc|, mapping matrices to tiles linearly via
  function \lstinline|poputil::mapTensorLinearly|. Specifically, input
  matrices are spreads evenly over tiles in a linear lexicographic
  manner, with the indices of the flattened matrix mapped across
  increasing tile IDs;
\item \textbf{optimized}: this benchmark includes an optimization that
  preconditions the two input operands via functions
  \lstinline|poplin::createMatMulInputLHS| and \lstinline|...RHS|. It
  then uses \lstinline|poplin::matMul| and
  \lstinline|popops::scaledAddTo|.
\end{itemize}




\noindent\textbf{Comparison with GPUs.}  We compare IPUs and GPUs in
terms of throughput and energy efficiency in single and mixed
precision. This is a measurement of how well a platform can minimize
the latency to complete matmuls when the entirety of the processor is
available for a single task.  We offer comparisons on a per-chip and
per-board basis. Since the IPU board has two chips, we consider a
library-based matmul benchmark that extends the matmul operation to
both IPUs on a C2 board.  The GPU we consider is NVidia's V100.
Table~\ref{tab:matmul-throughput} summarizes our findings.

\begin{figure}
  \includegraphics[width=\columnwidth]{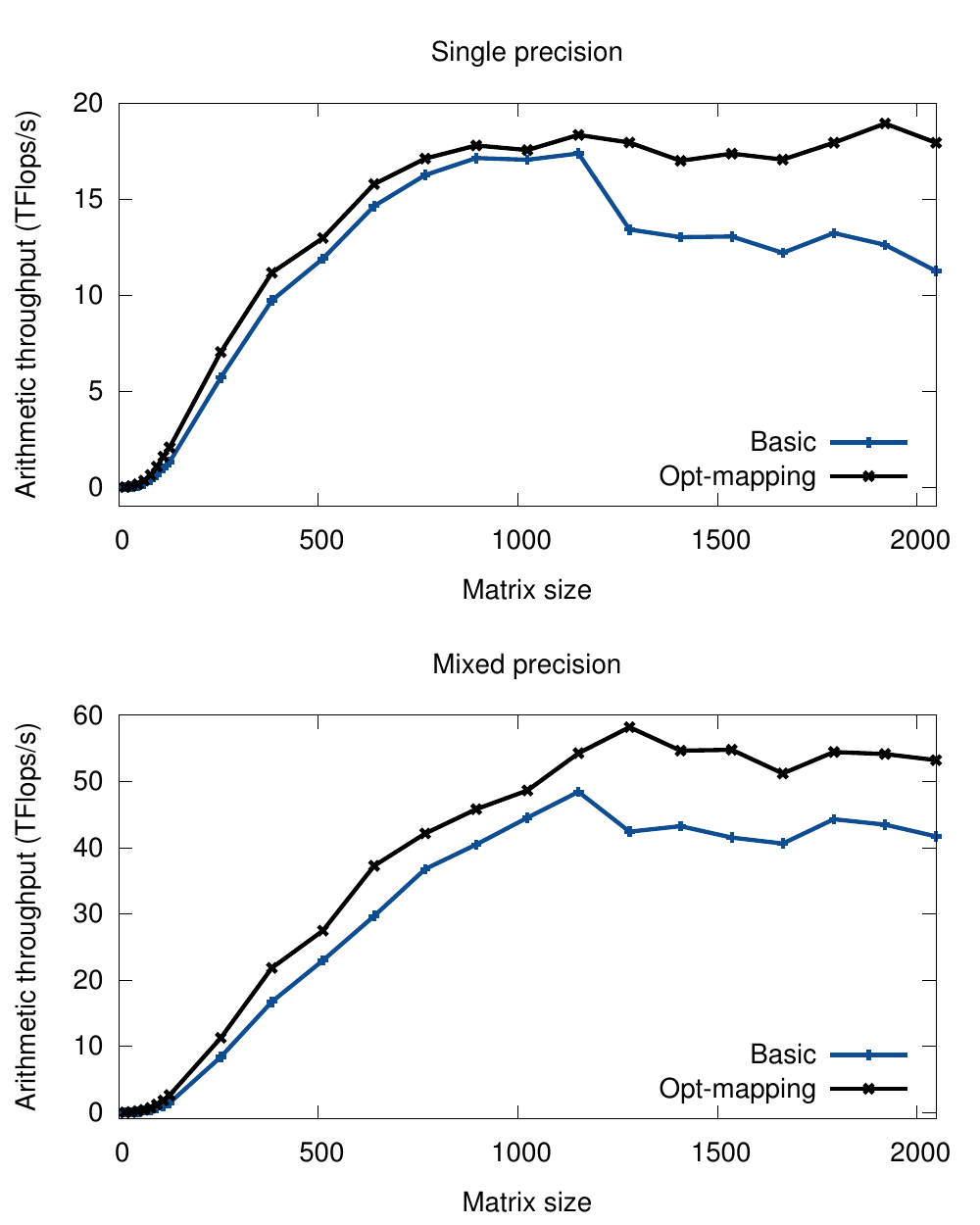}
  \caption{Floating-point arithmetic throughput achieved by dense
    matrix multiplication on one IPU. For simplicity, we use square
    inputs $(n\times n) \times (n\times n)$; the horizontal axis
    varies $n$.}
  \label{fig:matmul-IPU}
\end{figure}

We chart results in Figure~\ref{fig:matmul-IPU}, where the horizontal
axis represents $n$, one side of each square input operand matrix.
Performance varies with input size, with smaller matrices unable to
achieve sufficient occupancy of the device. This is a common
occurrence among massively parallel computing platform and is of
little consequence in most cases.  In our benchmark using optimized
tile mapping, performance saturates around 18 and 55~...~58 TFlops/s
in single and mixed precision, respectively.

\begin{figure}
  \center
  \includegraphics[width=\columnwidth]{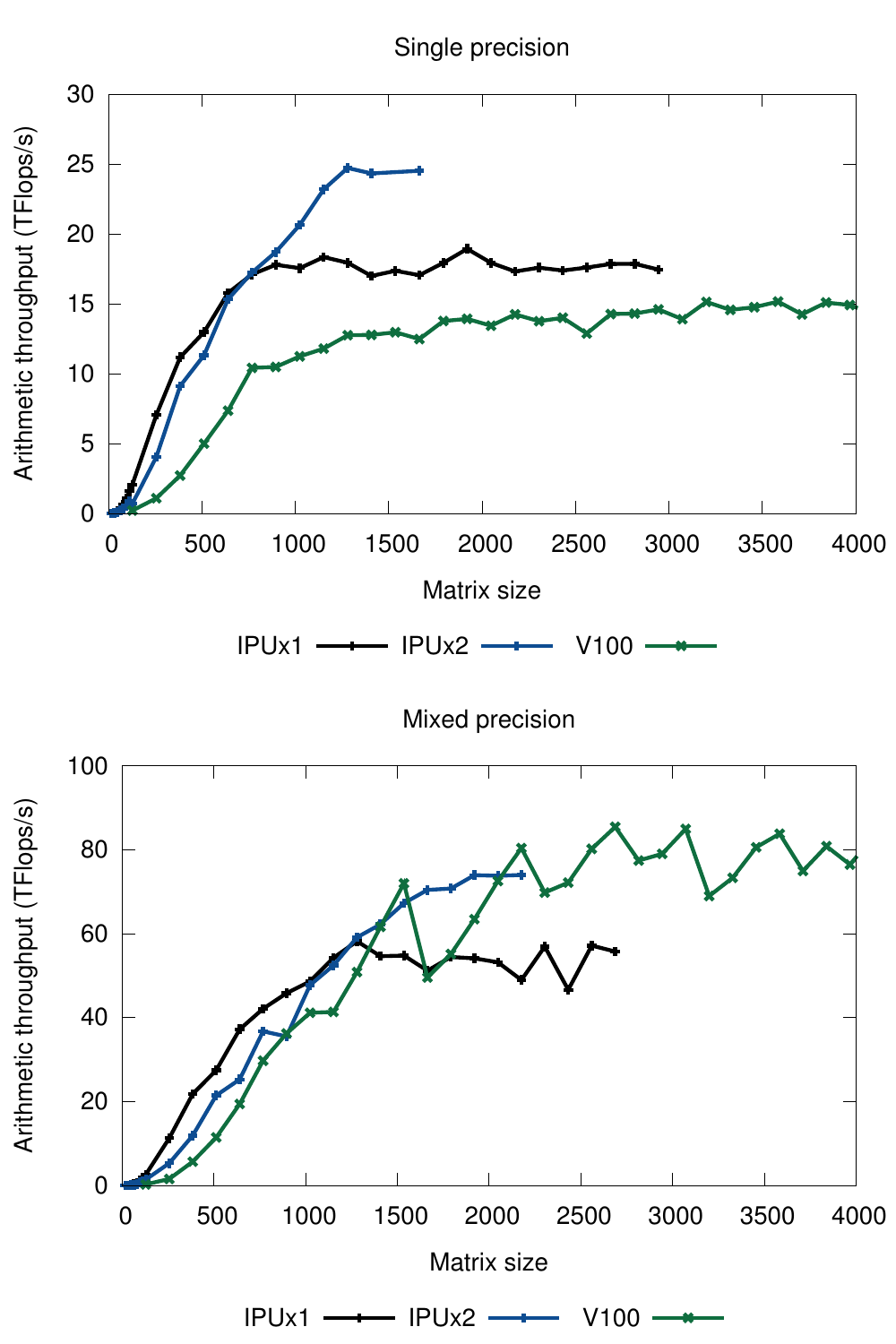}
        \caption{Matrix multiplication performance on IPUs and GPUs.}
  \label{fig:matmul-flops_one_board}
\end{figure}

\begin{figure}
  \center
  \includegraphics[width=\columnwidth]{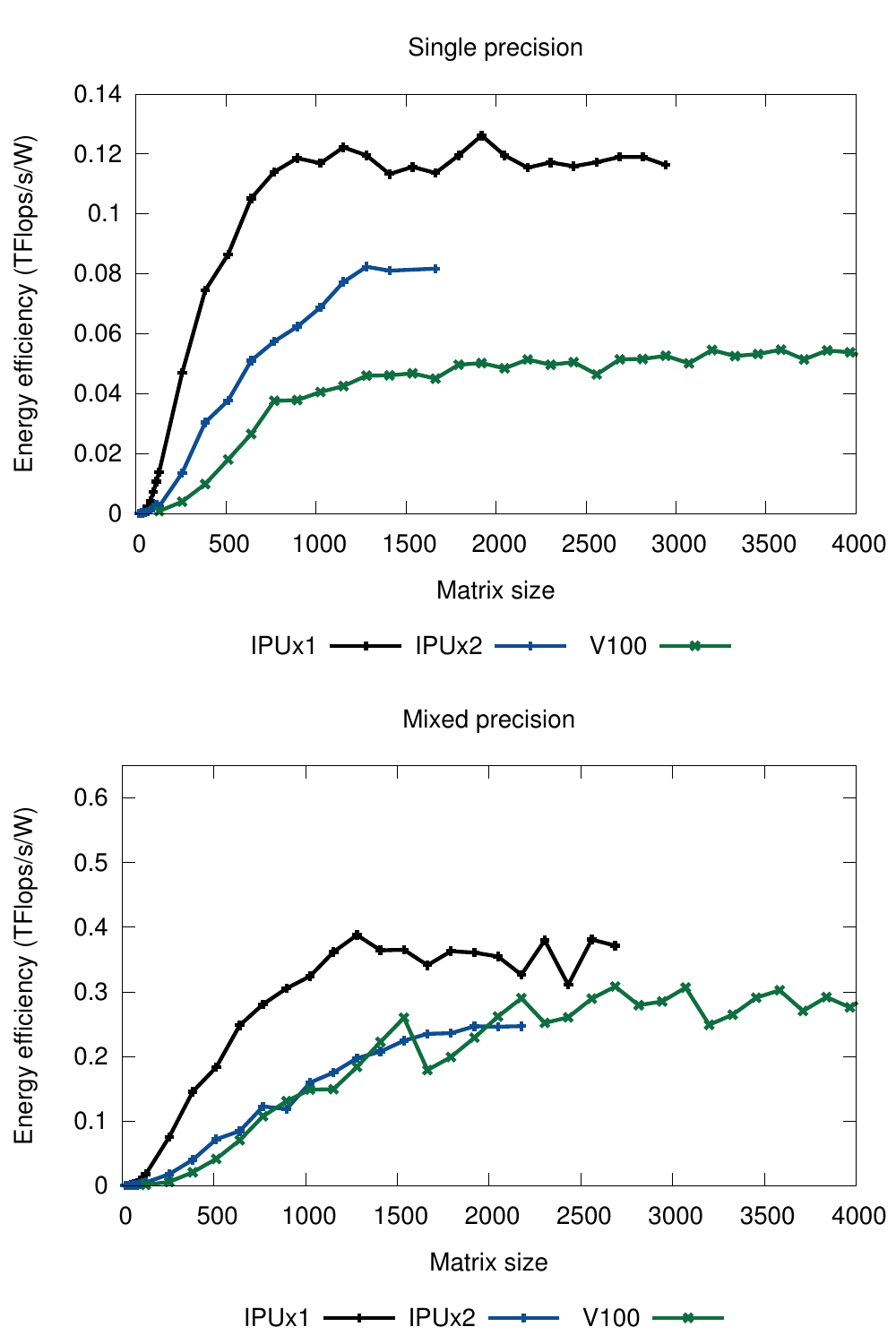}
  \caption{Energy efficiency while running matrix multiplication on
    IPUs and GPUs.}
  \label{fig:matmul-flops_per_watt}
\end{figure}

We also compare the devices in terms of energy efficiency by charting
throughput values divided by nominal power absorbed by each board in
Figure~\ref{fig:matmul-flops_per_watt}. Results are expressed in units
of TFlops/s per Watt (TFlops/s/W).  For the V100 GPU, we use nominal
power, 250 W.

In both single and mixed precision cases, a single IPU delivers higher
efficiency than two IPUs and higher efficiency than the V100 GPU.

\section{Convolution}

We dedicate this section to studying the IPU's convolution
performance.  For our benchmarking suite, we selected a basket of 6
commonly used CNNs in image classification according to our survey of
recent
literature~\cite{xie2016,he2015,szegedy2015,simonyan2014,liu2016,krizhevsky2012}.
We compare the IPU's and the V100 GPU's forward-pass performances, on a
per-chip basis, first in terms of arithmetic throughput (TFlops/s),
and then in terms of throughput normalized per nominal power
consumption (TFlops/s/W).  In
our results, IPUs tend to outperform GPUs at smaller batch sizes. While
the GPU supports larger batches thanks to its larger device memory, it
also needs to use larger batches for its compute resources to achieve
sufficient occupancy.
For CNNs whose architectures were designed without GPU efficiency in mind
(such as ResNeXt), we observe speedups upward of 700x.

\noindent\textbf{Benchmarking.} On each device, we use the
manufacturer's respective optimized convolution primitives:
\lstinline|poplin::convolution| from the Poplar SDK version 1.0.49,
and \lstinline|cudnnConvolutionForward| from cuDNN 7.5.0.
All benchmarks use half precision floating point math.

\noindent\textbf{Batch size.}  On both platforms, we vary batch size
over a meaningful range.  Note that sufficiently large batches exceed
the memory capacity of both IPUs and GPUs. Because of the larger GPU
device capacity, it supports larger batch sizes for most layers.  Our
experiments sweep the range 1 ... 2,048 on the GPU and 1 ... 128 on
the IPU.

\noindent\textbf{Raw results.} In Tables~\ref{tab:half-conv-v100} and
\ref{tab:half-conv-IPU} we report in detail the arithmetic throughput
achieved by each layer of each CNN, for each batch size considered, on
the GPU and IPU respectively. Results show that the GPU prefers
larger batch sizes, with most network layers reaching their peak
performances at batch size 512 or 1,024. In contrast, the IPU supports
smaller batch sizes and tends to achieve peak performance with batch
sizes in the 8...32 range. Moreover, for sizes below peak, the IPU
still delivers a good fraction of peak throughput.

\noindent\textbf{Comparison at peak.} We now present a simpler and
more concise analysis that abstracts away from batch size.
Specifically, we compare the IPU's and the GPU's peak performance at
each CNN layer (at the respective peak, whichever batch sizes they
correspond to). Table~\ref{tab:half-conv-pwatt} shows both arithmetic
throughput comparisons and energy efficiency comparisons.  We use
nominal power figures (250 W for the GPU, 150 W for the IPU).  Columns
labeled \textbf{Ratio} show the IPU/GPU throughput and energy
efficiency ratios, with over-unity values indicating an IPU advantage.

\noindent\textbf{Discussion}.
Results in Table~\ref{tab:half-conv-pwatt} support the expectation
that the highly parallel fine-grained MIMD architecture of the IPU
would provide an advantage in models using group or separable
convolutions.  When G=32 for the ResNeXt model, the IPU shows $>$100x advantage
for certain convolutions.  This advantage is not
available in legacy CNN models that were originally optimized for the
GPU architecture.

{\scriptsize
\begin{longtable}{lrrrrrrrrrrrr}
  \caption{Comparison between one V100 GPU and one IPU (per chip) in terms of arithmetic throughput and energy efficiency (i.e., arithmetic throughput normalized over nominal power).
    Benchmark: reference basket of convolutional neural networks~\cite{xie2016,he2015,szegedy2015,simonyan2014,liu2016,krizhevsky2012}.
    The \textbf{ratio} column reports IPU-over-GPU ratios; over-unity values mean that the IPU is performing better than the GPU; below-unity, vice versa.
    Parameters: $F_s$: filter size (height = width); $I_s$: input size (height = width); $I_c$: input channel; $O_c$: output channel; $S$: stride; $G$: group.} \\
  \label{tab:half-conv-pwatt} \\
  \toprule
  \,            &        &         &       &        &       &      & \multicolumn{3}{c}{Arithmetic Throughput}  & \multicolumn{3}{c}{Energy Efficiency}  \\
  Layer         & \multicolumn{6}{c}{Parameters}                   & \multicolumn{3}{c}{ (TFlops/s) }           & \multicolumn{3}{c}{(TFlops/s/W)} \\
  \cmidrule(lr){2-7}                                                          \cmidrule(lr){8-10}                        \cmidrule(lr){11-13}
  \,            & $F_s$  & $I_s$   & $I_c$  & $O_c$  & $S$   & $G$  &              IPU    &   V100  & Ratio      & IPU      & V100 & Ratio   \\
  \midrule
  \endfirsthead
  \toprule
  \,            &          &         &       &        &       &    & \multicolumn{3}{c}{Arithmetic Throughput}  & \multicolumn{3}{c}{Energy Efficiency}  \\
  Layer         & \multicolumn{6}{c}{Parameters}                   & \multicolumn{3}{c}{ (TFlops/s) }           & \multicolumn{3}{c}{(TFlops/s/W)} \\
  \cmidrule(lr){2-7}                                                          \cmidrule(lr){8-10}                        \cmidrule(lr){11-13}
  \,            & $F_s$  & $I_s$   & $I_c$  & $O_c$  & $S$   & $G$  &              IPU    &   V100  & Ratio    & IPU      & V100 & Ratio   \\
  \midrule
  \endhead
  \endfoot
  \endlastfoot
     \multicolumn{8}{l}{\textbf{ResNeXt}}          \\
  Conv1    & 7           & 224        & 3          & 64       & 2  & 1     & 15.04  & 6.26  &  2.40  &  0.10  &  0.03  &  4.00                      \\
  Conv2\_1 & 3           & 112        & 64         & 64       & 2  & 1     & 62.07  & 22.85  &  2.72  &  0.41  &  0.09  &  4.53                     \\
  Conv2\_2 & 1           & 56         & 64         & 128      & 1  & 32    & 0.32   & 0.04  &  8.11  &  0.00  &  0.00  &  13.51                     \\
  Conv2\_3 & 3           & 56         & 128        & 128      & 1  & 32    & 2.58   & 0.44  &  5.86  &  0.02  &  0.00  &  9.77                      \\
  Conv2\_4 & 1           & 56         & 128        & 256      & 1  & 32    & 1.14   & 0.15  &  7.59  &  0.01  &  0.00  &  12.65                     \\
  Conv3\_1 & 1           & 28         & 256        & 256      & 1  & 32    & 6.90   & 0.02  &  344.96  &  0.05  &  0.00  &  574.94                  \\
  Conv3\_2 & 3           & 28         & 256        & 256      & 1  & 32    & 36.13  & 0.75  &  48.17  &  0.24  &  0.00  &  80.29                    \\
  Conv3\_3 & 1           & 28         & 256        & 512      & 1  & 32    & 10.22  & 0.03  &  340.59  &  0.07  &  0.00  &  567.66                  \\
  Conv4\_1 & 1           & 14         & 512        & 512      & 1  & 32    & 28.21  & 0.04  &  705.23  &  0.19  &  0.00  &  1175.38                 \\
  Conv4\_2 & 3           & 14         & 512        & 512      & 1  & 32    & 45.69  & 1.38  &  33.11  &  0.30  &  0.01  &  55.18                    \\
  Conv4\_3 & 1           & 14         & 512        & 1024     & 1  & 32    & 31.57  & 0.06  &  526.23  &  0.21  &  0.00  &  877.05                  \\
  Conv5\_1 & 1           & 7          & 1024       & 1024     & 1  & 32    & 28.05  & 0.08  &  350.61  &  0.19  &  0.00  &  584.35                  \\
  Conv5\_2 & 3           & 7          & 1024       & 1024     & 1  & 32    & 37.14  & 4.74  &  7.84  &  0.25  &  0.02  &  13.06                     \\
  Conv5\_3 & 1           & 7          & 1024       & 2048     & 1  & 32    & 33.41  & 0.11  &  303.76  &  0.22  &  0.00  &  506.27                  \\
           &             &            &            &          &    &       &        &   &    &    &    &                                            \\
    \multicolumn{8}{l}{\textbf{ResNet-50 v1.5}}          \\
  Conv1    & 7           & 224        & 3          & 64       & 2  & 1     & 15.04  & 6.13  &  2.45  &  0.10  &  0.02  &  4.09                    \\
  Conv2\_1 & 3           & 112        & 64         & 64       & 2  & 1     & 62.32  & 23.27  &  2.68  &  0.42  &  0.09  &  4.46                   \\
  Conv2\_2 & 1           & 56         & 64         & 64       & 1  & 1     & 52.39  & 8.48  &  6.18  &  0.35  &  0.03  &  10.30                   \\
  Conv2\_3 & 3           & 56         & 64         & 64       & 1  & 1     & 73.43  & 47.74  &  1.54  &  0.49  &  0.19  &  2.56                   \\
  Conv2\_4 & 1           & 56         & 64         & 256      & 1  & 1     & 58.55  & 12.09  &  4.84  &  0.39  &  0.05  &  8.07                   \\
  Conv3\_1 & 1           & 28         & 256        & 128      & 1  & 1     & 54.87  & 20.71  &  2.65  &  0.37  &  0.08  &  4.42                   \\
  Conv3\_2 & 3           & 28         & 128        & 128      & 1  & 1     & 58.13  & 64.07  &  0.91  &  0.39  &  0.26  &  1.51                   \\
  Conv3\_3 & 1           & 28         & 128        & 512      & 1  & 1     & 62.39  & 21.38  &  2.92  &  0.42  &  0.09  &  4.86                   \\
  Conv4\_1 & 1           & 14         & 256        & 256      & 1  & 1     & 44.41  & 26.61  &  1.67  &  0.30  &  0.11  &  2.78                   \\
  Conv4\_2 & 3           & 14         & 256        & 256      & 1  & 1     & 52.78  & 78.29  &  0.67  &  0.35  &  0.31  &  1.12                   \\
  Conv4\_3 & 1           & 14         & 256        & 1024     & 1  & 1     & 60.92  & 33.67  &  1.81  &  0.41  &  0.13  &  3.02                   \\
  Conv5\_1 & 1           & 7          & 512        & 512      & 1  & 1     & 41.43  & 37.73  &  1.10  &  0.28  &  0.15  &  1.83                   \\
  Conv5\_2 & 3           & 7          & 512        & 512      & 1  & 1     & 52.71  & 85.48  &  0.62  &  0.35  &  0.34  &  1.03                   \\
  Conv5\_3 & 1           & 7          & 512        & 2048     & 1  & 1     & 57.72  & 48.69  &  1.19  &  0.38  &  0.19  &  1.98                   \\
           &             &            &            &          &    &       &        &   &    &    &    &                                          \\
    \multicolumn{8}{l}{\textbf{Inception v3}}          \\
  Conv1    & 3           & 299        &3           &32        & 2  & 1     & 7.08   & 3.91  &  1.81  &  0.05  &  0.02  &  3.02                    \\
  Conv2    & 3           & 149        &32          &32        & 1  & 1     & 71.59  & 24.05  &  2.98  &  0.48  &  0.10  &  4.96                   \\
  Conv3    & 3           & 147        &32          &64        & 1  & 1     & 70.25  & 38.33  &  1.83  &  0.47  &  0.15  &  3.05                   \\
  Conv4    & 3           & 147        &64          &64        & 2  & 1     & 57.36  & 23.39  &  2.45  &  0.38  &  0.09  &  4.09                   \\
  Conv5    & 3           & 73         &64          &80        & 1  & 1     & 65.66  & 37.49  &  1.75  &  0.44  &  0.15  &  2.92                   \\
  Conv6    & 3           & 71         &80          &192       & 2  & 1     & 61.90  & 32.77  &  1.89  &  0.41  &  0.13  &  3.15                   \\
  Conv7    & 3           & 35         &192         &288       & 1  & 1     & 62.12  & 61.78  &  1.01  &  0.41  &  0.25  &  1.68                   \\
  Conv8    & 3           & 35         &288         &768       & 2  & 1     & 50.85  & 59.17  &  0.86  &  0.34  &  0.24  &  1.43                   \\
  Conv9    & 3           & 17         &768         &1280      & 2  & 1     & 45.56  & 72.17  &  0.63  &  0.30  &  0.29  &  1.05                   \\
  Conv10   & 3           & 8          &1280        &2048      & 1  & 1     & 49.27  & 87.92  &  0.56  &  0.33  &  0.35  &  0.93                   \\
           &             &            &            &          &    &       &        &   &    &    &    &                                          \\
    \multicolumn{8}{l}{\textbf{VGG16}}          \\
  Conv1\_1 &  3           & 224        & 3          & 64      & 1  & 1     & 7.20   & 3.83  &  1.88  &  0.05  &  0.02  &  3.13                    \\
  Conv1\_2 &  3           & 224        & 64         & 128     & 1  & 1     & 75.04  & 52.01  &  1.44  &  0.50  &  0.21  &  2.40                   \\
  Conv2\_1 &  3           & 112        & 128        & 128     & 1  & 1     & 72.60  & 64.59  &  1.12  &  0.48  &  0.26  &  1.87                   \\
  Conv2\_2 &  3           & 112        & 128        & 256     & 1  & 1     & 71.79  & 70.46  &  1.02  &  0.48  &  0.28  &  1.70                   \\
  Conv3\_1 &  3           & 56         & 256        & 256     & 1  & 1     & 70.20  & 78.75  &  0.89  &  0.47  &  0.32  &  1.49                   \\
  Conv3\_2 &  3           & 56         & 256        & 512     & 1  & 1     & 72.10  & 81.35  &  0.89  &  0.48  &  0.33  &  1.48                   \\
  Conv4\_1 &  3           & 28         & 512        & 512     & 1  & 1     & 58.53  & 87.22  &  0.67  &  0.39  &  0.35  &  1.12                   \\
  Conv4\_2 &  3           & 28         & 512        & 512     & 1  & 1     & 58.54  & 87.28  &  0.67  &  0.39  &  0.35  &  1.12                   \\
  Conv5\_1 &  3           & 14         & 512        & 512     & 1  & 1     & 49.92  & 87.02  &  0.57  &  0.33  &  0.35  &  0.96                   \\
           &              &            &            &         &    &       &        &   &    &    &    &                                          \\
     \multicolumn{8}{l}{\textbf{SSD v1.1}}          \\
  Conv1    & 3           & 38         & 512        & 1024     & 2  & 1     & 53.35  & 71.51  &  0.75  &  0.36  &  0.29  &  1.24                   \\
  Conv2    & 1           & 19         & 1024       & 1024     & 1  & 1     & 54.83  & 53.46  &  1.03  &  0.37  &  0.21  &  1.71                   \\
  Conv3    & 1           & 19         & 1024       & 512      & 1  & 1     & 57.01  & 47.1  &  1.21  &  0.38  &  0.19  &  2.02                    \\
  Conv4    & 3           & 10         & 512        & 512      & 2  & 1     & 46.76  & 63.02  &  0.74  &  0.31  &  0.25  &  1.24                   \\
  Conv5    & 1           & 10         & 512        & 128      & 1  & 1     & 35.19  & 24.09  &  1.46  &  0.23  &  0.10  &  2.43                   \\
  Conv6    & 3           & 10         & 128        & 256      & 2  & 1     & 31.83  & 52.52  &  0.61  &  0.21  &  0.21  &  1.01                   \\
  Conv7    & 1           & 5          & 256        & 128      & 1  & 1     & 16.21  & 16.07  &  1.01  &  0.11  &  0.06  &  1.68                   \\
  Conv8    & 3           & 5          & 128        & 256      & 1  & 1     & 30.16  & 63.86  &  0.47  &  0.20  &  0.26  &  0.79                   \\
  Conv9    & 1           & 3          & 256        & 128      & 1  & 1     & 8.81   & 10.12  &  0.87  &  0.06  &  0.04  &  1.45                   \\
  Conv10   & 3           & 3          & 128        & 256      & 1  & 1     & 17.56  & 45.51  &  0.39  &  0.12  &  0.18  &  0.64                   \\
           &             &            &            &          &    &       &        &   &    &    &    &                                          \\
    \multicolumn{8}{l}{\textbf{AlexNet}}          \\
  Conv1    & 11          & 227        & 3          & 96       & 4  & 1     & 20.04  & 14.7  &  1.36  &  0.13  &  0.06  &  2.27                    \\
  MaxPool1 & 3           & 55         & 96         & 96       & 2  & 1     & 51.73  & 33.12  &  1.56  &  0.34  &  0.13  &  2.60                   \\
  Conv2    & 5           & 27         & 96         & 256      & 1  & 1     & 64.34  & 86.46  &  0.74  &  0.43  &  0.35  &  1.24                   \\
  MaxPool2 & 3           & 27         & 256        & 256      & 2  & 1     & 49.21  & 50.37  &  0.98  &  0.33  &  0.20  &  1.63                   \\
  Conv3    & 3           & 13         & 256        & 384      & 1  & 1     & 53.70  & 82.07  &  0.65  &  0.36  &  0.33  &  1.09                   \\
  Conv4    & 3           & 13         & 384        & 256      & 1  & 1     & 52.91  & 82.02  &  0.65  &  0.35  &  0.33  &  1.08                   \\
  Conv5    & 3           & 13         & 256        & 256      & 2  & 1     & 41.77  & 54.93  &  0.76  &  0.28  &  0.22  &  1.27                   \\
           &             &            &            &          &    &       &  \\
  \bottomrule
\end{longtable}
}

%
%

%
%
%
%

{ \scriptsize
\begin{longtable}{lrrrrrrrrrrrr}
  \caption{Convolution performance (arithmetic throughput, TFlops/s)
    of one NVidia V100 GPU, in half precision, over our reference
    basket of convolutional neural network architectures. Missing data
    represents experiments that did not fit in device memory.}  \\
  \label{tab:half-conv-v100} \\
  \toprule
    Layer name & \multicolumn{12}{c}{Batch size}                                                                    \\
               &  1    & 2    & 4 & 8    & 16    & 32    & 64    & 128   & 256   & 512   & 1024 & 2048     \\
  \midrule
  \endfirsthead
  \toprule
    Layer name & \multicolumn{12}{c}{Batch size}                                                                    \\
               &  1    & 2    & 4 & 8    & 16    & 32    & 64    & 128   & 256   & 512   & 1024 & 2048     \\
  \midrule
  \endhead
  \endfoot
  \endlastfoot
    \multicolumn{13}{l}{\textbf{ResNeXt}}          \\
   Conv1    &  2.71  & 3.75  & 4.05  & 5.01  & 5.89  & 6.02  & 6.10  & 6.14  & 6.21  & 6.24  & 6.26  & 6.25  \\
   Conv2\_1 &  5.89  & 9.92  & 15.04 & 17.91 & 18.42 & 21.02 & 22.25 & 22.68 & 22.85 & 22.79 & 22.72 & 22.63 \\
   Conv2\_2 &  0.00  & 0.01  & 0.01  & 0.03  & 0.04  & 0.04  & 0.04  & 0.04  & 0.04  & 0.04  & 0.04  & 0.04  \\
   Conv2\_3 &  0.11  & 0.20  & 0.32  & 0.40  & 0.43  & 0.44  & 0.44  & 0.44  & 0.44  & 0.44  & 0.44  & 0.44  \\
   Conv2\_4 &  0.01  & 0.03  & 0.05  & 0.11  & 0.14  & 0.15  & 0.15  & 0.15  & 0.15  & 0.15  & 0.15  & 0.15  \\
   Conv3\_1 &  0.00  & 0.01  & 0.02  & 0.02  & 0.02  & 0.02  & 0.02  & 0.02  & 0.02  & 0.02  & 0.02  & 0.02  \\
   Conv3\_2 &  0.10  & 0.20  & 0.39  & 0.75  & 0.17  & 0.18  & 0.18  & 0.18  & 0.18  & 0.18  & 0.17  & 0.18  \\
   Conv3\_3 &  0.01  & 0.02  & 0.02  & 0.02  & 0.03  & 0.03  & 0.03  & 0.03  & 0.03  & 0.03  & 0.03  & 0.03  \\
   Conv4\_1 &  0.00  & 0.01  & 0.02  & 0.04  & 0.04  & 0.04  & 0.04  & 0.04  & 0.04  & 0.04  & 0.04  & 0.04  \\
   Conv4\_2 &  0.10  & 0.20  & 0.36  & 0.73  & 1.38  & 0.34  & 0.36  & 0.35  & 0.36  & 0.35  & 0.36  & 0.35  \\
   Conv4\_3 &  0.01  & 0.02  & 0.03  & 0.04  & 0.05  & 0.05  & 0.06  & 0.06  & 0.05  & 0.06  & 0.06  & 0.06  \\
   Conv5\_1 &  0.00  & 0.01  & 0.02  & 0.04  & 0.07  & 0.07  & 0.08  & 0.08  & 0.08  & 0.08  & 0.08  & 0.08  \\
   Conv5\_2 &  0.09  & 0.19  & 0.37  & 0.74  & 1.44  & 2.80  & 0.70  & 4.74  & 0.72  & 0.72  & 0.73  & 0.73  \\
   Conv5\_3 &  0.01  & 0.02  & 0.03  & 0.07  & 0.09  & 0.10  & 0.11  & 0.11  & 0.11  & 0.11  & 0.11  & 0.11  \\
&         &        &        &        &        &        &        &        &        &        &        &        \\
    \multicolumn{13}{l}{\textbf{ResNet v1.5}}          \\
   Conv1    &  4.66 & 3.70  &4.10   &4.94   &5.78   &5.91   &5.98   &6.02   &6.09   &6.13   &6.13   &6.13   \\
   Conv2\_1 &  4.71 & 10.00 &15.45  &18.15  &18.70  &21.38  &22.72  &23.13  &23.27  &23.23  &23.14  &23.07  \\
   Conv2\_2 &  1.00 & 2.00  &3.99   &6.40   &6.30   &7.12   &7.79   &8.15   &8.32   &8.42   &8.46   &8.48   \\
   Conv2\_3 &  6.02 & 11.79 &17.47  &29.06  &32.83  &37.18  &42.50  &44.11  &45.32  &46.58  &47.41  &47.74  \\
   Conv2\_4 &  4.00 & 7.92  &8.29   &9.76   &10.65  &11.22  &11.69  &11.91  &12.01  &12.07  &12.08  &12.09  \\
   Conv3\_1 &  2.11 & 4.16  &7.77   &12.23  &15.95  &16.02  &18.36  &19.93  &20.33  &20.53  &20.61  &20.71  \\
   Conv3\_2 &  3.97 & 7.94  &15.51  &28.79  &40.58  &42.01  &53.38  &61.81  &62.70  &63.32  &63.44  &64.07  \\
   Conv3\_3 &  4.07 & 8.28  &13.90  &14.69  &17.34  &19.04  &19.88  &20.59  &20.97  &21.21  &21.33  &21.38  \\
   Conv4\_1 &  1.03 & 2.01  &4.00   &7.94   &13.61  &18.80  &19.62  &22.74  &25.24  &25.85  &26.22  &26.61  \\
   Conv4\_2 &  2.38 & 4.67  &9.18   &18.04  &34.50  &48.05  &50.21  &65.12  &76.76  &77.85  &78.10  &78.29  \\
   Conv4\_3 &  3.56 & 8.04  &14.87  &21.85  &23.42  &28.51  &32.23  &31.85  &32.38  &32.94  &33.36  &33.67  \\
   Conv5\_1 &  0.85 & 1.68  &3.15   &6.10   &11.46  &20.43  &28.45  &29.07  &34.73  &36.68  &37.27  &37.73  \\
   Conv5\_2 &  1.06 & 2.10  &4.12   &8.35   &16.80  &32.97  &51.04  &54.24  &70.48  &82.64  &84.69  &85.48  \\
   Conv5\_3 &  3.28 & 6.51  &12.10  &21.96  &31.54  &33.78  &42.23  &48.69  &47.44  &43.98  &44.33  &44.76  \\
            &      &        &        &        &        &        &        &        &        &        &        \\
    \multicolumn{13}{l}{\textbf{Inception v3}}          \\
   Conv1    &  1.43  &1.02  &1.15  &1.30  &1.42  &1.46  &1.50  &1.52  &1.53  &1.54  &1.54  &3.91   \\
   Conv2    &  10.47  &14.26  &16.71  &19.13  &21.09  &22.43  &23.29  &23.67  &23.89  &23.99  &24.04  &24.05  \\
   Conv3    &  18.47  &23.19  &26.73  &30.21  &33.03  &34.69  &36.28  &37.27  &37.94  &38.05  &38.33  &  \\
   Conv4    &  9.02  &14.25  &17.35  &18.12  &20.40  &22.21  &23.21  &23.28  &23.39  &23.34  &23.24  &  \\
   Conv5    &  12.14  &18.11  &20.81  &26.57  &31.35  &34.52  &36.46  &36.68  &37.23  &37.25  &37.49  &37.49  \\
   Conv6    &  7.10  &13.68  &19.25  &21.16  &24.51  &28.09  &30.42  &31.59  &32.39  &32.77  &32.55  &32.72  \\
   Conv7    &  14.95  &28.98  &40.86  &43.01  &56.34  &57.85  &58.77  &61.21  &61.52  &61.67  &61.75  &61.78  \\
   Conv8    &  9.21  &18.02  &28.60  &53.40  &57.41  &59.17  &52.52  &55.04  &56.74  &57.43  &57.19  &57.56  \\
   Conv9    &  3.68  &7.29  &14.62  &24.16  &47.03  &52.68  &71.49  &72.17  &59.45  &61.66  &62.97  &62.78  \\
   Conv10   &  4.11  &8.15  &16.04  &32.20  &55.52  &65.78  &72.01  &84.09  &87.92  &73.27  &67.99  &64.01  \\
            &         &        &        &        &        &        &        &        &        &        &        &        \\
    \multicolumn{13}{l}{\textbf{VGG16}}          \\
   Conv1\_1 &  2.30   &2.72   &3.22   &3.43   &3.60   &3.72   &3.79   &3.81   &3.83   &3.83   &      &       \\
   Conv1\_2 &  41.87  &48.67  &49.97  &50.68  &51.28  &51.58  &52.01  &51.78  &51.96  &       &      &       \\
   Conv2\_1 &  40.21  &42.06  &53.15  &62.04  &62.83  &63.37  &63.57  &64.28  &64.59  &64.59  &64.58 &       \\
   Conv2\_2 &  43.70  &53.64  &66.60  &68.17  &68.78  &69.18  &69.96  &70.40  &70.43  &70.46  &      &       \\
   Conv3\_1 &  34.34  &48.55  &50.16  &65.04  &76.28  &76.92  &77.02  &77.34  &78.27  &78.73  &78.71 & 78.75 \\
   Conv3\_2 &  45.16  &50.56  &66.90  &78.72  &79.38  &79.92  &79.88  &80.80  &81.29  &81.35  &81.28 &       \\
   Conv4\_1 &  16.99  &33.27  &51.54  &54.48  &70.34  &82.23  &84.11  &85.05  &85.60  &86.61  &87.22 & 87.14 \\
   Conv4\_2 &  16.94  &33.23  &51.53  &54.48  &70.33  &82.30  &84.10  &85.03  &85.59  &86.59  &87.17 & 87.28 \\
   Conv5\_1 &  4.13   &8.37   &16.86  &33.07  &51.28  &54.29  &70.73  &82.92  &84.59  &85.56  &86.00 & 87.02 \\
            &         &        &        &        &        &        &        &        &        &        &        &        \\
    \multicolumn{13}{l}{\textbf{SSD v1.1}}          \\
   Conv1    &  13.35  &26.03  &40.11  &46.52  &61.36  &71.51  &60.11  &62.62  &64.01  &64.19  &64.71  &64.86 \\
   Conv2    &  11.12  &20.92  &37.38  &37.73  &46.92  &52.85  &49.97  &51.73  &52.71  &53.00  &53.25  &53.46 \\
   Conv3    &  7.49   &14.27  &25.34  &32.57  &31.91  &41.13  &43.66  &44.40  &45.70  &46.48  &46.89  &47.10 \\
   Conv4    &  0.54   &1.07   &2.14   &4.17   &8.33   &13.23  &25.46  &47.36  &49.94  &63.00  &62.44  &63.02 \\
   Conv5    &  0.43   &0.82   &1.62   &3.11   &5.91   &9.94   &12.75  &19.01  &19.58  &22.23  &23.87  &24.09 \\
   Conv6    &  0.26   &0.51   &1.03   &2.04   &3.91   &7.57   &11.21  &20.49  &35.59  &37.91  &46.05  &52.52 \\
   Conv7    &  0.07   &0.13   &0.26   &0.52   &1.06   &1.75   &3.96   &7.19   &11.24  &14.82  &15.02  &16.07 \\
   Conv8    &  0.26   &0.52   &1.03   &2.02   &3.98   &7.84   &15.22  &28.90  &39.35  &48.53  &54.89  &63.86 \\
   Conv9    &  0.02   &0.05   &0.10   &0.18   &0.38   &0.63   &1.31   &2.63   &4.50   &6.70   &8.91   &10.12 \\
   Conv10   &  0.09   &0.19   &0.38   &0.75   &1.47   &2.84   &5.52   &10.56  &19.66  &33.61  &42.98  &45.51 \\
            &         &        &        &        &        &        &        &        &        &        &        &        \\
    \multicolumn{13}{l}{\textbf{AlexNet}}          \\
   Conv1    &  3.65  &7.32   &8.62   &11.78  &12.19  &13.42  &14.13  &14.51  &14.57  &14.67  &14.68  &14.70 \\
   MaxPool1 &  2.71  &5.32   &10.07  &13.88  &23.14  &24.09  &28.85  &32.18  &32.43  &32.83  &32.87  &33.12 \\
   Conv2    &  9.04  &17.87  &34.83  &48.31  &50.03  &63.49  &78.25  &79.18  &83.13  &85.21  &86.36  &86.46 \\
   MaxPool2 &  2.39  &4.59   &8.96   &16.13  &29.04  &41.23  &42.44  &45.37  &48.62  &49.93  &49.59  &50.37 \\
   Conv3    &  2.97  &5.81   &11.57  &22.43  &42.98  &59.82  &63.38  &65.26  &74.03  &79.92  &80.39  &82.07 \\
   Conv4    &  2.18  &4.29   &8.52   &16.72  &31.74  &44.33  &46.29  &59.59  &70.69  &77.62  &81.80  &82.02 \\
   Conv5    &  0.60  &1.19   &2.37   &4.61   &9.03   &16.13  &29.33  &41.50  &42.93  &48.96  &54.35  &54.93 \\
            &         &        &        &        &        &        &        &        &        &        &        &        \\
  \bottomrule
\end{longtable}
}
{
\scriptsize
\begin{longtable}{lrrrrrrrr}
  \caption{Convolution performance (arithmetic throughput, TFlops/s)
  of one IPU, in half precision, over our reference basket of
  convolutional neural network architectures.  Missing data
  represents experiments that did not fit in device memory.  } \\

  \label{tab:half-conv-IPU} \\
  \toprule
    Layer name & \multicolumn{6}{c}{Batch size}                                                                    \\
               & 1    & 2    & 4 & 8    & 16    & 32  &64 &128 \\
  \midrule
  \endfirsthead
  \toprule
    Layer name & \multicolumn{6}{c}{Batch size}                                                                    \\
               & 1    & 2    & 4 & 8    & 16    & 32  &64 &128 \\
  \midrule
  \endhead
  \endfoot
  \endlastfoot
  \multicolumn{7}{l}                    {\textbf{ResNeXt}}                                 \\
   Conv1    & 7.45   &  14.23  &  15.04  &  11.95  &         &  7.75   &  17.82  &         \\
   Conv2\_1 & 30.60  &  39.87  &  47.74  &  55.50  &  62.07  &  55.93  &         &         \\
   Conv2\_2 & 0.18   &  0.23   &  0.28   &  0.30   &  0.32   &  0.32   &  0.49   &  0.49   \\
   Conv2\_3 & 0.87   &  1.32   &  1.62   &  1.86   &  2.10   &  2.58   &  0.71   &  0.51   \\
   Conv2\_4 & 0.37   &  0.75   &  0.92   &  1.03   &  1.10   &  1.14   &  0.46   &         \\
   Conv3\_1 & 1.57   &  2.62   &  3.90   &  5.19   &  6.23   &  6.90   &  2.17   &  1.83   \\
   Conv3\_2 & 5.81   &  10.97  &  16.30  &  21.39  &  27.70  &  36.13  &  10.78  &  1.87   \\
   Conv3\_3 & 2.54   &  4.27   &  6.19   &  7.99   &  9.33   &  10.22  &  3.91   &  1.72   \\
   Conv4\_1 & 3.22   &  5.85   &  9.88   &  15.12  &  22.02  &  28.21  &  33.75  &  37.18  \\
   Conv4\_2 & 10.28  &  15.78  &  20.59  &  33.03  &  41.61  &  45.69  &  45.26  &  50.12  \\
   Conv4\_3 & 5.91   &  9.66   &  14.12  &  20.20  &  25.63  &  31.57  &  35.93  &  37.87  \\
   Conv5\_1 & 2.74   &  4.82   &  8.74   &  14.41  &  20.22  &  28.05  &  35.76  &  45.81  \\
   Conv5\_2 & 7.77   &  11.16  &  15.91  &  25.60  &  31.31  &  37.14  &  45.06  &  51.85  \\
   Conv5\_3 & 4.84   &  8.45   &  13.36  &  18.98  &  27.61  &  33.41  &  41.51  &  47.44  \\
            &        &         &         &         &         &         &         &         \\
   \multicolumn{7}{l}                   {\textbf{ResNet-50 v1.5}}                          \\
   Conv1    & 7.45   &  14.23  &  15.04  &  11.95  &         &  7.75   &  17.82  &        \\
   Conv2\_1 & 30.58  &  39.91  &  47.74  &  53.98  &  62.32  &  55.82  &         &        \\
   Conv2\_2 & 11.85  &  18.92  &  26.73  &  36.84  &  46.03  &  52.39  &  55.93  &  59.29 \\
   Conv2\_3 & 34.44  &  45.79  &  54.30  &  65.21  &  70.79  &  73.43  &  72.00  &  66.37 \\
   Conv2\_4 & 27.48  &  35.51  &  44.54  &  51.39  &  56.99  &  58.55  &  43.21  &        \\
   Conv3\_1 & 16.23  &  22.59  &  27.89  &  35.41  &  43.79  &  54.87  &  60.86  &  57.06 \\
   Conv3\_2 & 29.82  &  38.20  &  47.03  &  53.89  &  56.63  &  58.13  &  55.88  &  62.66 \\
   Conv3\_3 & 23.16  &  32.72  &  41.29  &  51.86  &  59.07  &  62.39  &  45.49  &  45.58 \\
   Conv4\_1 & 10.92  &  15.88  &  21.75  &  28.29  &  36.09  &  44.41  &  55.60  &  58.85 \\
   Conv4\_2 & 26.53  &  32.99  &  41.86  &  47.94  &  52.66  &  52.78  &  50.36  &  51.51 \\
   Conv4\_3 & 20.65  &  27.09  &  34.94  &  45.20  &  56.45  &  60.92  &  57.94  &  54.69 \\
   Conv5\_1 & 10.30  &  15.56  &  20.84  &  26.60  &  34.44  &  41.43  &  51.27  &  55.72 \\
   Conv5\_2 & 22.52  &  29.48  &  32.88  &  44.85  &         &  52.71  &  50.35  &  49.28 \\
   Conv5\_3 & 21.39  &  27.91  &  32.84  &  42.66  &  49.81  &  57.72  &  55.14  &  53.44 \\
            &        &         &         &         &         &         &         &        \\
   \multicolumn{7}{l}                   {\textbf{Inceptionv3}}                            \\
   Conv1    & 2.90   &  2.66   &  2.82   &  3.17   &  5.70   &  7.08   &         &        \\
   Conv2    & 44.61  &  58.19  &  70.92  &  71.59  &         &         &         &        \\
   Conv3    & 55.98  &  69.61  &  70.25  &  67.43  &         &         &         &        \\
   Conv4    & 37.05  &  45.06  &  51.92  &  57.36  &         &  49.85  &         &        \\
   Conv5    & 43.30  &  52.82  &  61.51  &  65.65  &  65.66  &  63.77  &  63.55  &        \\
   Conv6    & 32.79  &  41.10  &  46.42  &  55.94  &  61.90  &  50.70  &  35.80  &        \\
   Conv7    & 51.80  &  58.89  &  59.99  &  62.12  &         &  58.10  &  61.71  &        \\
   Conv8    & 44.93  &  50.85  &         &         &         &  49.13  &  49.05  &        \\
   Conv9    & 36.64  &  45.56  &         &         &         &  43.23  &  43.58  &        \\
   Conv10   &        &         &  49.27  &  41.35  &         &  42.81  &  45.57  &        \\
            &        &         &         &         &         &         &         &        \\
   \multicolumn{7}{l}                   {\textbf{VGG16}} \\
   Conv1\_1 & 6.33   &  5.36   &  7.20   &  7.08   &         &         &         &        \\
   Conv1\_2 & 75.04  &  74.89  &  69.21  &         &         &         &         &        \\
   Conv2\_1 & 68.90  &  72.60  &  70.71  &  70.41  &         &         &         &        \\
   Conv2\_2 & 71.79  &  70.87  &  69.25  &         &         &         &         &        \\
   Conv3\_1 & 64.05  &  69.31  &  70.20  &  69.03  &         &  66.61  &         &        \\
   Conv3\_2 & 65.29  &  72.10  &  69.13  &  69.07  &         &         &         &        \\
   Conv4\_1 & 57.55  &  58.53  &         &         &  56.39  &  56.07  &         &        \\
   Conv4\_2 & 57.56  &  58.54  &         &         &  56.39  &  55.88  &         &        \\
   Conv5\_1 & 41.40  &  46.76  &         &         &         &  49.92  &  46.93  &        \\
            &        &         &         &         &         &         &         &        \\
   \multicolumn{7}{l}                   {\textbf{SSD v1.1}} \\
   Conv1    &        &         &         &  47.72  &  47.58  &  53.35  &         &        \\
   Conv2    & 40.53  &  47.34  &         &         &         &  54.83  &  56.07  &        \\
   Conv3    & 32.90  &  40.36  &  48.38  &  57.01  &         &  55.19  &  56.07  &        \\
   Conv4    & 14.78  &  22.21  &  30.26  &  35.26  &         &  46.76  &  42.77  &        \\
   Conv5    & 6.62   &  10.49  &  16.08  &  22.04  &  28.92  &  35.19  &  43.25  &        \\
   Conv6    & 5.30   &  8.37   &  12.45  &  18.04  &  26.01  &  31.83  &  38.45  &        \\
   Conv7    & 1.08   &  2.09   &  3.88   &  7.13   &  11.23  &  16.21  &  22.95  &        \\
   Conv8    & 4.51   &  7.07   &  10.51  &  14.83  &  23.64  &  30.16  &  37.25  &        \\
   Conv9    & 0.43   &  0.81   &  1.53   &  2.94   &  5.51   &  8.81   &  13.24  &        \\
   Conv10   & 1.47   &  2.77   &  4.62   &  6.68   &  11.68  &  17.56  &  23.35  &        \\
            &        &         &         &         &         &         &         &        \\
   \multicolumn{7}{l}                   {\textbf{AlexNet}} \\
   Conv1    & 8.40   &  9.75   &  11.14  &  11.15  &         &  20.04  &  17.07  &        \\
   MaxPool1 & 20.22  &  28.43  &  37.38  &  43.72  &  51.12  &  51.73  &  47.83  &        \\
   Conv2    & 44.03  &  57.30  &  62.83  &  64.34  &         &  61.69  &  60.52  &        \\
   MaxPool2 & 24.40  &  28.69  &  36.60  &  44.66  &  49.21  &  46.32  &  42.26  &        \\
   Conv3    & 28.05  &  37.17  &  45.68  &  52.25  &         &  53.70  &  51.04  &        \\
   Conv4    & 28.62  &  34.72  &  42.61  &  51.43  &         &  52.91  &  50.10  &        \\
   Conv5    & 12.20  &  16.15  &  20.35  &  26.10  &  33.43  &  41.77  &  44.11  &        \\
            &        &         &         &         &         &         &         &        \\
  \bottomrule
\end{longtable}
}

%
%

\clearpage


\begin{figure}[b]
  \includegraphics[width=\columnwidth]{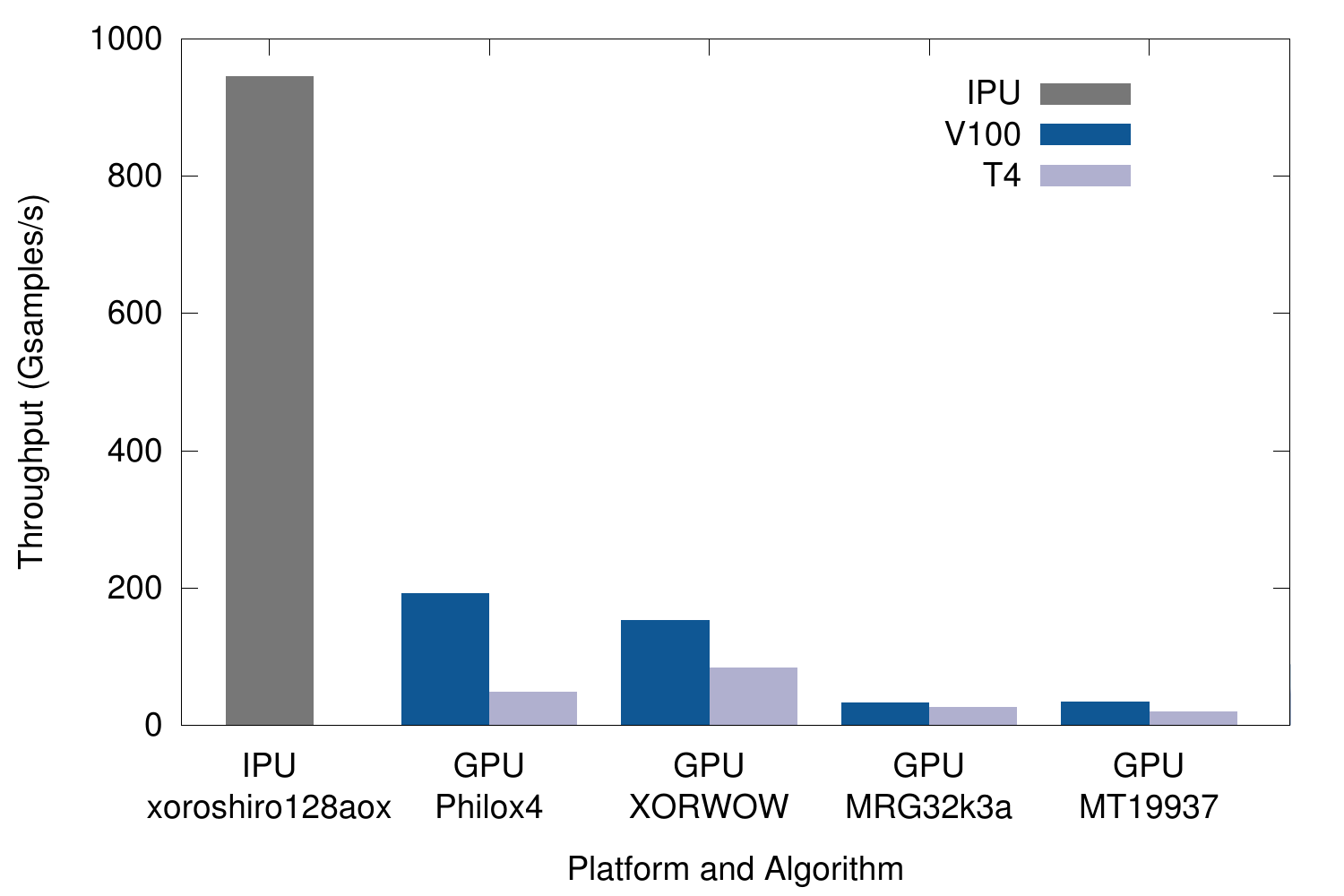}
  \caption{Bulk random number generation performance on IPU and GPUs:
    Peak generation throughput in billions random samples per second,
    uniform distribution. Per-chip comparison.}
  \label{fig:rng_barchart}
\end{figure}

\section{Pseudo-Random Number Generation}

We compare IPU and GPU performance at bulk pseudo-random number
generation (PRNG) in terms of aggregate throughput on a per-chip
basis. We find that, thanks to its dedicated, in-core PRNG hardware
units, the IPU generates up to 944 billion random samples per second
(Gsamples/s), whereas a V100 GPU generates up to 192 Gsamples/s. While
the IPU offers 4.9$\times$ more aggregate throughput than a V100 GPU,
it employs a PRNG algorithm that delivers a lower quality of randomness
than the fastest algorithm we benchmarked on the GPU. The IPU's
performance advantage over the GPU doubles in a per-board comparison.
We are not qualified to judge which platform the performance-quality
trade-off favors.  All details follow.

For simplicity, our benchmarks focus only on the generation of
pseudo-random numbers extracted from a uniform distribution.

\begin{figure}
  \includegraphics[width=\columnwidth]{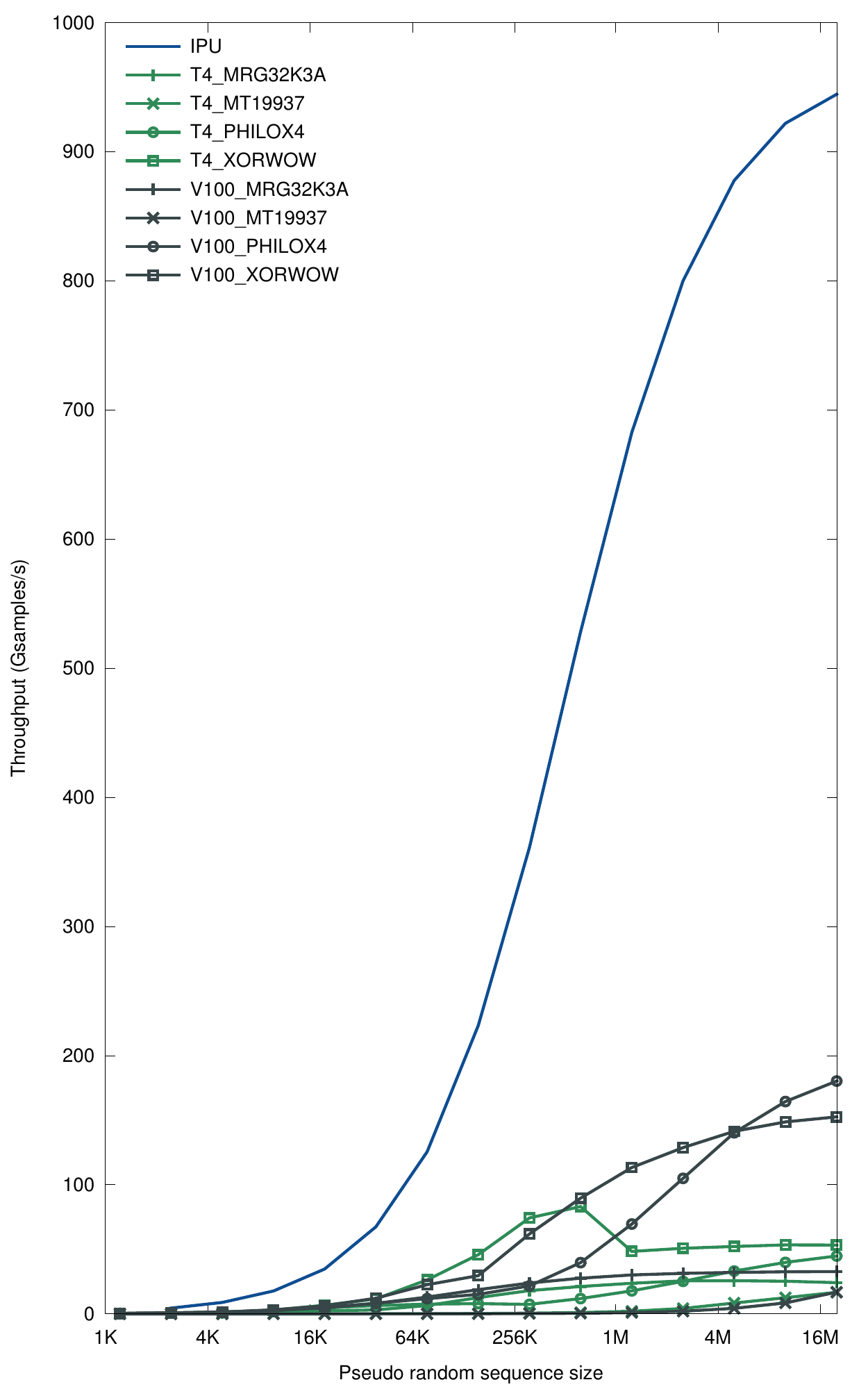}
  \caption{Bulk random generation throughput on IPUs and GPUs as a
    function of block size. Per-chip comparison.}
  \label{fig:rng_uniform_block_size}
\end{figure}

\textbf{PRNG on the IPU}. On the IPU, each tile includes PRNG
acceleration circuitry that implements a variant of the
\emph{xoroshiro128+} algorithm by David Blackman and Sebastiano
Vigna~\cite{vigna2019}. Details on the algorithm are available upon
request from Graphcore.  Our benchmark invokes the
\lstinline|poprand::uniform| function from the Poplar SDK.  For
sufficiently large blocks, the aggregate IPU throughput offered by the
function reaches 944 Gsamples/s; see Figure~\ref{fig:rng_barchart}.

\textbf{PRNG on the GPU}. On the GPUs, our benchmarks exercise PRNG
functions offered with NVidia's cuRand\cite{cuRand} library. cuRand
supports multiple PRNG algorithms; we benchmark XORWOW, MRG32K3A,
MTGP32, MT19937, Philox4. The fastest algorithm at volume on the V100
GPU is Philox4\cite{salmon} by John Salmon. We run the
experiment on each GPU (V100 and T4) at the maximum clock frequency
supported by each device.

A comparison on the quality of randomness provided by PRNG algorithms
(e.g., TestU01\cite{testU01} performance) is beyond the scope of this
paper.

\textbf{Output size.} We study how bulk PRNG performance varies as a
function of requested output block size. Our results show that both on
IPUs and GPUs, the generation of larger output blocks achieves higher
throughputs.  On the two platforms, performance grows and saturates
similarly, as a function of output size
(Figure~\ref{fig:rng_uniform_block_size})

\clearpage
\begin{small}
\listoffigures
\end{small}

\bibliographystyle{IEEEtran}
\raggedright
\bibliography{tech_report}

\begin{thebibliography}{10}
\providecommand{\url}[1]{#1}
\csname url@samestyle\endcsname
\providecommand{\newblock}{\relax}
\providecommand{\bibinfo}[2]{#2}
\providecommand{\BIBentrySTDinterwordspacing}{\spaceskip=0pt\relax}
\providecommand{\BIBentryALTinterwordstretchfactor}{4}
\providecommand{\BIBentryALTinterwordspacing}{\spaceskip=\fontdimen2\font plus
\BIBentryALTinterwordstretchfactor\fontdimen3\font minus
  \fontdimen4\font\relax}
\providecommand{\BIBforeignlanguage}[2]{{%
\expandafter\ifx\csname l@#1\endcsname\relax
\typeout{** WARNING: IEEEtran.bst: No hyphenation pattern has been}%
\typeout{** loaded for the language `#1'. Using the pattern for}%
\typeout{** the default language instead.}%
\else
\language=\csname l@#1\endcsname
\fi
#2}}
\providecommand{\BIBdecl}{\relax}
\BIBdecl

\bibitem{zhe2018}
\BIBentryALTinterwordspacing
Z.~Jia, M.~Maggioni, B.~Staiger, and D.~P. Scarpazza, ``Dissecting the {NVidia}
  {Volta} {GPU} architecture via microbenchmarking,'' 2018. [Online].
  Available: \url{https://arxiv.org/abs/1804.06826}
\BIBentrySTDinterwordspacing

\bibitem{zhe2019}
\BIBentryALTinterwordspacing
Z.~Jia, M.~Maggioni, J.~Smith, and D.~P. Scarpazza, ``Dissecting the {NVidia}
  {Turing} {T4} {GPU} architecture via microbenchmarking,'' 2019. [Online].
  Available: \url{https://arxiv.org/abs/1903.07486}
\BIBentrySTDinterwordspacing

\bibitem{culler1993}
\BIBentryALTinterwordspacing
D.~Culler, R.~Karp, D.~Patterson, A.~Sahay, K.~E. Schauser, E.~Santos,
  R.~Subramonian, and T.~von Eicken, ``Logp: Towards a realistic model of
  parallel computation,'' in \emph{Proceedings of the Fourth ACM SIGPLAN
  Symposium on Principles and Practice of Parallel Programming}, ser. PPOPP
  '93.\hskip 1em plus 0.5em minus 0.4em\relax New York, NY, USA: ACM, 1993, pp.
  1--12. [Online]. Available: \url{http://doi.acm.org/10.1145/155332.155333}
\BIBentrySTDinterwordspacing

\bibitem{alexandrov1995}
A.~Alexandrov, M.~F. Ionescu, K.~E. Schauser, and C.~Scheiman, ``Loggp:
  Incorporating long messages into the logp model --- one step closer towards a
  realistic model for parallel computation,'' Santa Barbara, CA, USA, Tech.
  Rep., 1995.

\bibitem{panda2004}
J.~Liu, B.~Chandrasekaran, W.~Yu, J.~Wu, D.~Buntinas, S.~Kini, D.~Panda, and
  P.~Wyckoff, ``Microbenchmark performance comparison of high-speed cluster
  interconnects,'' \emph{IEEE Micro}, vol.~24, pp. 42 -- 51, 02 2004.

\bibitem{kistler2006}
\BIBentryALTinterwordspacing
M.~Kistler, M.~Perrone, and F.~Petrini, ``Cell multiprocessor communication
  network: Built for speed,'' \emph{IEEE Micro}, vol.~26, no.~3, pp. 10--23,
  May 2006. [Online]. Available: \url{http://dx.doi.org/10.1109/MM.2006.49}
\BIBentrySTDinterwordspacing

\bibitem{bsp1990}
\BIBentryALTinterwordspacing
L.~G. Valiant, ``A bridging model for parallel computation,''
  \emph{Communications of the ACM}, vol.~33, no.~8, pp. 103--111, Aug. 1990.
  [Online]. Available: \url{http://doi.acm.org/10.1145/79173.79181}
\BIBentrySTDinterwordspacing

\bibitem{omb2019}
\BIBentryALTinterwordspacing
{The Ohio State University's Network-Based Computing Laboratory}, ``Osu micro
  benchmarks,'' 2019. [Online]. Available:
  \url{http://mvapich.cse.ohio-state.edu/benchmarks/}
\BIBentrySTDinterwordspacing

\bibitem{xie2016}
S.~Xie, R.~Girshick, P.~Dollar, Z.~Tu, and K.~He, ``Aggregated residual
  transformations for deep neural networks,'' 2016.

\bibitem{he2015}
K.~He, X.~Zhang, S.~Ren, and J.~Sun, ``Deep residual learning for image
  recognition,'' 2015.

\bibitem{szegedy2015}
C.~Szegedy, V.~Vanhoucke, S.~Ioffe, J.~Shlens, and Z.~Wojna, ``Rethinking the
  inception architecture for computer vision,'' 2015.

\bibitem{simonyan2014}
K.~Simonyan and A.~Zisserman, ``Very deep convolutional networks for
  large-scale image recognition,'' 2014.

\bibitem{liu2016}
\BIBentryALTinterwordspacing
W.~Liu, D.~Anguelov, D.~Erhan, C.~Szegedy, S.~Reed, C.-Y. Fu, and A.~C. Berg,
  ``Ssd: Single shot multibox detector,'' \emph{Lecture Notes in Computer
  Science}, pp. 21--37, 2016. [Online]. Available:
  \url{http://dx.doi.org/10.1007/978-3-319-46448-0_2}
\BIBentrySTDinterwordspacing

\bibitem{krizhevsky2012}
A.~Krizhevsky, I.~Sutskever, and G.~E. Hinton, ``Imagenet classification with
  deep convolutional neural networks,'' pp. 1097--1105, 2012.

\bibitem{vigna2019}
\BIBentryALTinterwordspacing
D.~Blackman and S.~Vigna, ``Scrambled linear pseudorandom number generators,''
  \emph{preprint article}, 2019. [Online]. Available:
  \url{http://vigna.di.unimi.it/ftp/papers/ScrambledLinear.pdf}
\BIBentrySTDinterwordspacing

\bibitem{cuRand}
\BIBentryALTinterwordspacing
NVidia, ``{cuRand} -- the {API} reference guide for {cuRand}, the {CUDA} random
  number generation library,'' 2019. [Online]. Available:
  \url{https://docs.nvidia.com/cuda/curand/index.html}
\BIBentrySTDinterwordspacing

\bibitem{salmon}
\BIBentryALTinterwordspacing
J.~Salmon, M.~Moraes, R.~Dror, and D.~Shaw, ``Parallel random numbers: As easy
  as 1, 2, 3,'' 11 2011, p.~16. [Online]. Available:
  \url{https://thesalmons.org/john/random123/papers/random123sc11.pdf}
\BIBentrySTDinterwordspacing

\bibitem{testU01}
P.~L'Ecuyer and R.~Simard, ``{TestU01}: A c library for empirical testing of
  random number generators,'' \emph{{ACM} Transactions on Mathematical
  Software}, vol.~33, 2007.

\end{thebibliography}

\chapter*{The Authors}
\newlength{\picsize}
\newlength{\biosize}
\setlength{\picsize}{0.8in}
\setlength{\biosize}{\textwidth-\picsize}

{
\begin{tabular}{p{\picsize}p{\biosize}}
  \raisebox{-\height}{\includegraphics[width=\picsize]{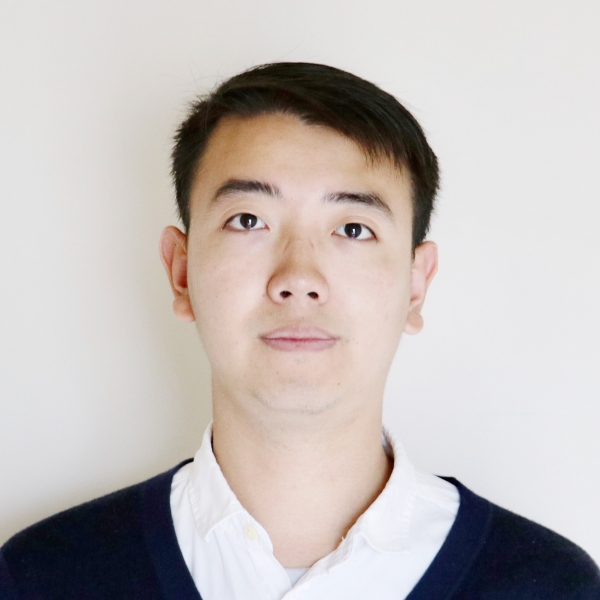}}
  & \textbf{Zhe Jia} is a senior R\&D engineer with the
  High-Performance Computing group at Citadel.  Prior to this
  position, he was a senior R\&D engineer with Alicloud, and a
  software engineering intern at Microsoft, Beijing.  He received his
  B.S. degree in Physics and his M.S. degree in Meteorology from
  Peking University, China.  His interests include the performance
  optimization of deep learning, numerical modeling, and atmospheric
  simulation workloads. \\
  & \\
  \raisebox{-\totalheight}{\includegraphics[width=\picsize]{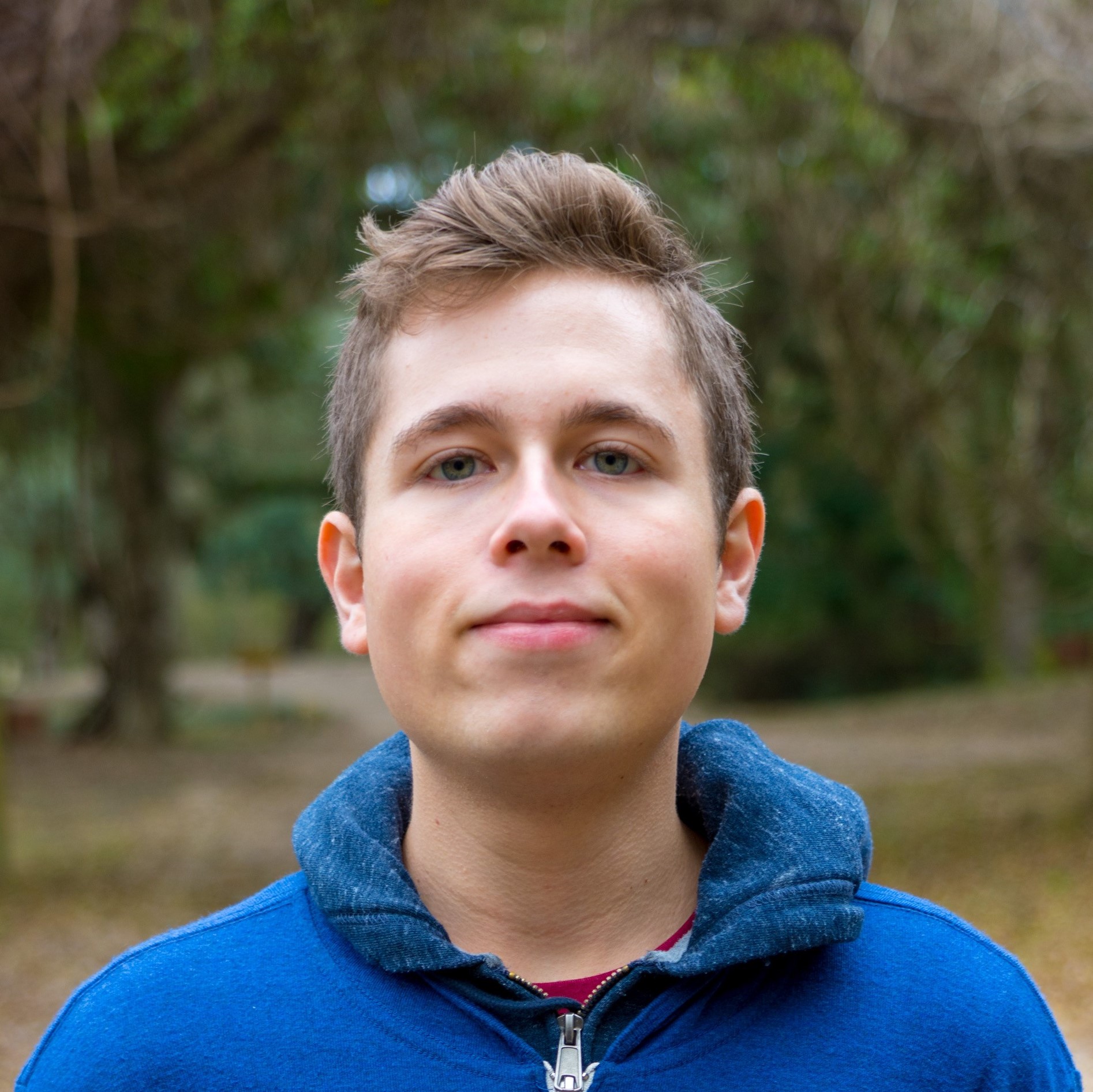}}
  & \textbf{Blake Tillman} is a Software Engineer with the
  High-Performance Computing group at Citadel. He joined the firm in
  July 2019, after receiving a B.S. degree in Computer Science and a
  B.S. degree in Mathematics. Prior to joining Citadel, Blake
  completed internships at Google and Apple. \\
  & \\
  \raisebox{-\totalheight}{\includegraphics[width=\picsize]{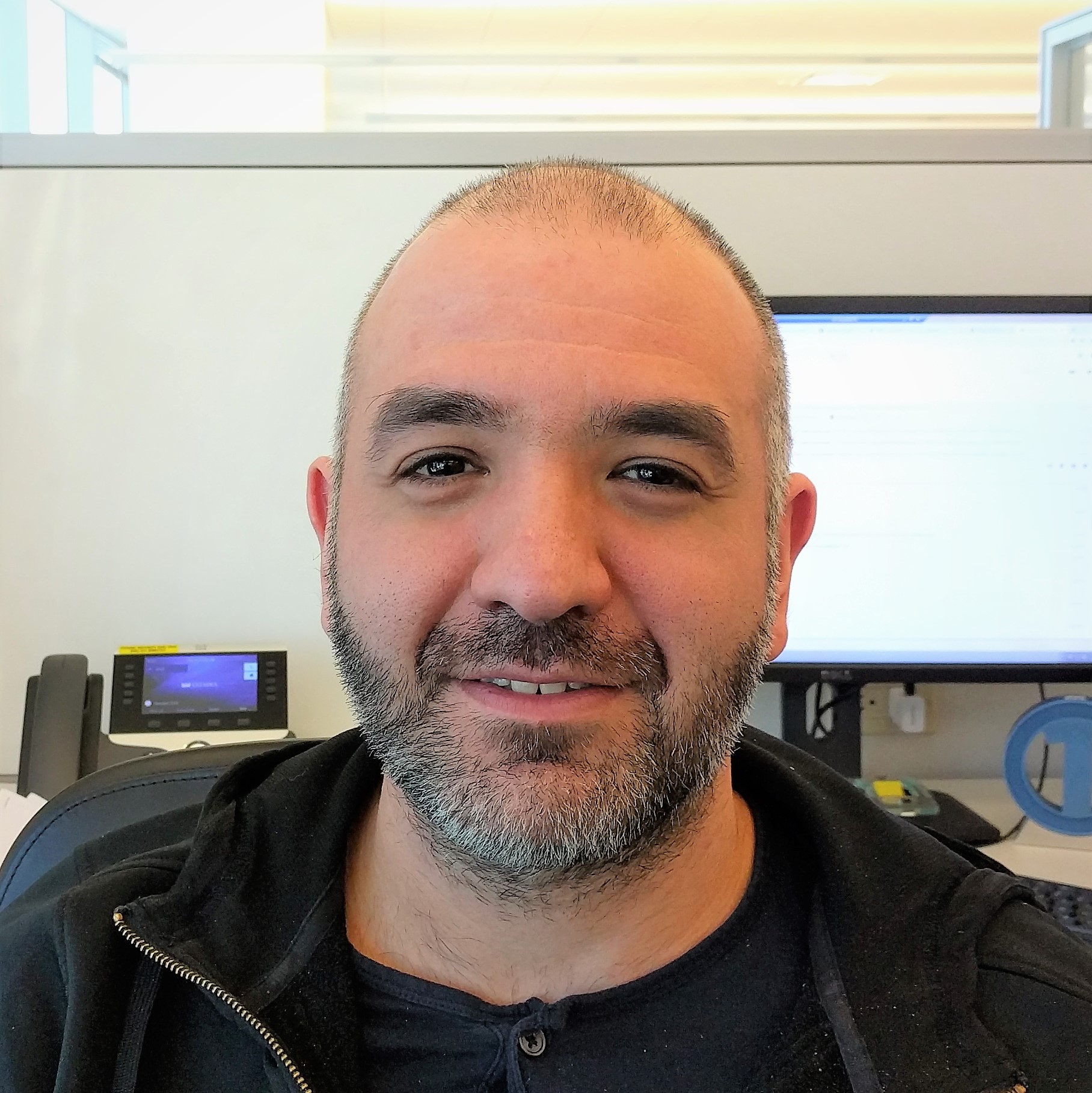}}
  & \textbf{Marco Maggioni} is a senior R\&D engineer with the
  High-Performance Computing group at Citadel, Chicago.  He received
  his Ph.D. in Computer Science from the University of Illinois at
  Chicago, where he focused on sparse linear algebra and convex
  optimization on GPUs.  His research established records for the fastest
  sparse matrix-vector multiplication GPU kernel in research and
  industry. \\
  & \\
  \raisebox{-\totalheight}{\includegraphics[width=\picsize]{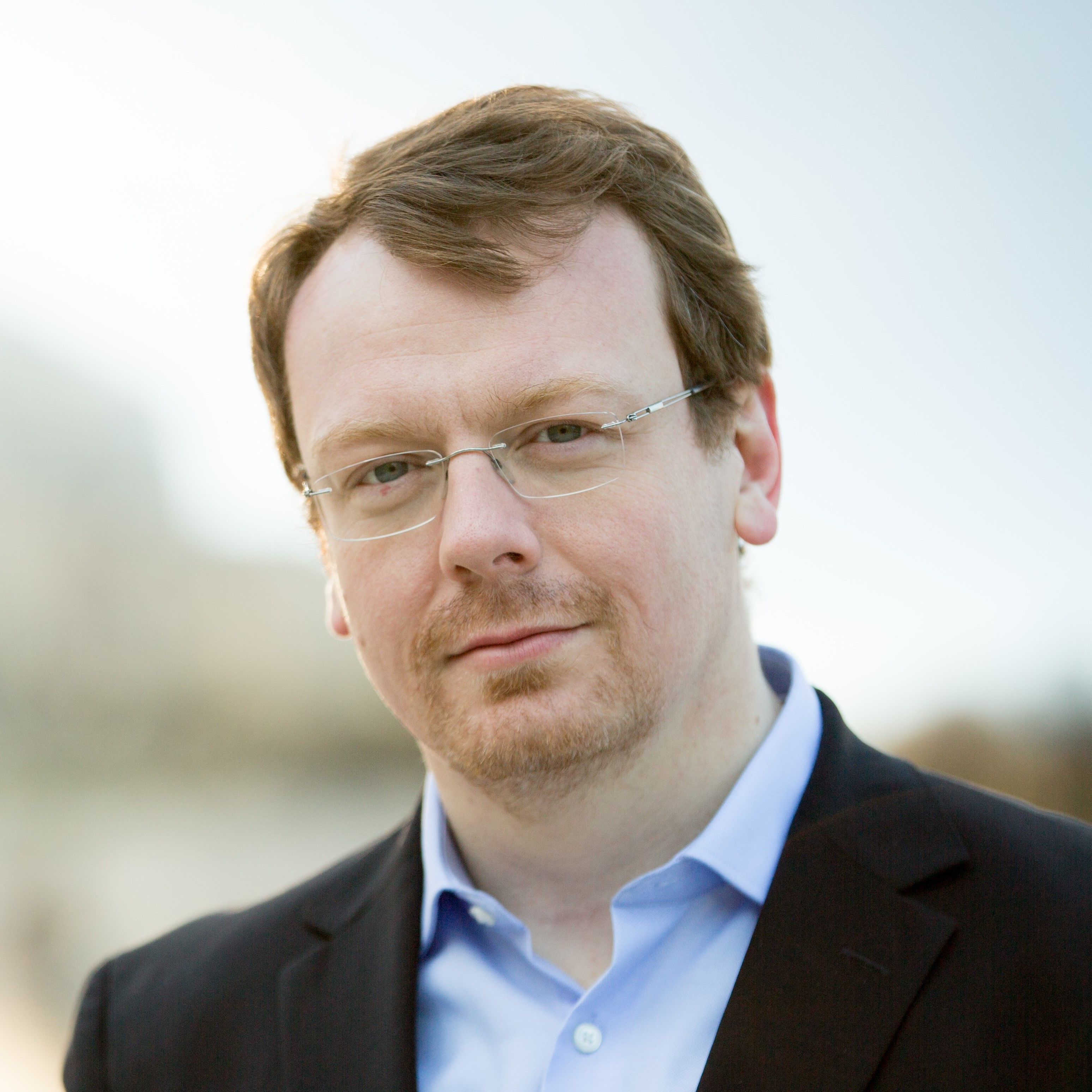}}
  & \textbf{Daniele Paolo Scarpazza}  leads the
  High-Performance Computing group at Citadel, Chicago.  Prior to this
  position, he was a Research Scientist with D. E. Shaw Research, a
  Research Staff Member with the IBM T. J. Watson Research Center, and
  a Post-Doc with the Pacific Northwest National Laboratory.  He
  received his Ph.D. in Information Engineering from Politecnico di
  Milano, Italy.  He is the co-recipient of a Gordon Bell prize.  He
  focuses on quantitative performance analysis and optimization of
  algorithms on parallel architectures.
\end{tabular}
}

\end{document}